\documentclass[twocolumn,floatfix,prb,aps,showpacs]{revtex4}
\usepackage{graphicx,amsmath,amssymb,color}
\usepackage{nicefrac}
\usepackage[titletoc,title]{appendix}

\newcommand{\be}{\begin{equation}}
\newcommand{\ee}{\end{equation}}

\newcommand{\ba}{\begin{eqnarray}}
\newcommand{\ea}{\end{eqnarray}}

\newcommand{\ecf}{$^{\rm e}$CF}
\newcommand{\hcf}{$^{\rm h}$CF}
\newcommand{\nues}{\nu^*}
\newcommand{\nue}{\nu}
\newcommand{\nuhs}{\nu^*}

\renewcommand{\vec}[1]{{\textbf{\textit{#1}}}}
\renewcommand{\vr}{\vec{r}}

\begin{document}

\title{Luttinger theorem for the strongly correlated Fermi liquid of composite fermions}
\author{Ajit C. Balram,$^1$ Csaba T\H oke,$^2$ and J. K. Jain,$^1$}
\affiliation{
   $^{1}$Department of Physics, 104 Davey Lab, Pennsylvania State University, University Park, Pennsylvania 16802, USA}
\affiliation{
   $^{2}$BME-MTA Exotic Quantum Phases ``Lend\"ulet" Research Group, Budapest University of Technology and Economics,
Institute of Physics, Budafoki \'ut 8, H-1111 Budapest, Hungary}

\begin{abstract} While an ordinary Fermi sea is perturbatively robust to interactions, the paradigmatic composite-fermion (CF) Fermi sea arises as a non-perturbative consequence of emergent gauge fields in a system where there was no Fermi sea to begin with. A mean-field picture suggests two Fermi seas, of composite fermions made from electrons or holes in the lowest Landau level, which occupy different areas away from half filling and thus appear to represent distinct states. We show that in the microscopic theory of composite fermions, which satisfies particle-hole symmetry in the lowest Landau level to an excellent degree, the Fermi wave vectors at filling factors $\nu$ and $1-\nu$ are the same, and are generally consistent with the experimental findings of Kamburov {\em et al.} [Phys. Rev. Lett. {\bf 113}, 196801 (2014)]. Our calculations suggest that the area of the CF Fermi sea may slightly violate the Luttinger area rule.
\end{abstract}
\maketitle

A fundamental property of a Landau Fermi liquid is captured by Luttinger's theorem\cite{Luttinger60b}, according to which the volume occupied by the Fermi sea, appropriately defined, remains invariant when the interaction is switched on, so long as no phase boundary is crossed. A violation of this theorem signifies non-Fermi liquid behavior, which has motivated investigations\cite{Schulz95,Schofield99,Yamanaka97,Oshikawa00,Varma02,Kiaran13} of its applicability for various strongly correlated systems, such as high temperature superconductors. This article investigates the Luttinger theorem for an exotic emergent Fermi sea. 

When two-dimensional electrons are subjected to a strong magnetic field, they exhibit the phenomenon of the fractional quantum Hall effect (FQHE)\cite{Tsui82}, which is understood in terms of topological particles called composite fermions\cite{Jain89,Lopez91,Halperin93,Jain07,Jain15}. 
Halperin, Lee and Read\cite{Halperin93} and Kalmeyer and Zhang \cite{Kalmeyer92} theoretically predicted that at Landau level (LL) filling factor $\nu=1/2$, the external magnetic field is canceled, in a mean field (MF) approximation, by the emergent gauge field carried by composite fermions, and they form a Fermi sea.  The composite fermion (CF) Fermi sea is special in the following sense. Ordinarily, we begin with the perfect Fermi sea of non-interacting fermions and then ask how interactions degrade or destroy it. In contrast, interactions are fully responsible for creating the CF Fermi sea (CFFS) in a system of electrons confined to the lowest LL (LLL) where, originally, there was no Fermi sea, and, in fact, no kinetic energy and no composite fermions. The very existence of the CFFS thus is a manifestation of strong correlations. 
The essential validity of the CFFS has been confirmed in extensive detail in many experiments\cite{Willett93,Kang93,Goldman94,Smet96,Smet98,Willett99,Smet99,Freytag02,Kamburov12}, and it also dovetails with the prominently observed sequences of fractions at $\nu=n/(2n\pm 1)$\cite{Jain89}.  

Kamburov {\em et al.}\cite{Kamburov14b} have recently made accurate measurements of the CF Fermi wave vector through commensurability effects in the presence of a periodic modulation. They have observed more commensurability oscillations than before, and thus provided the most detailed confirmation of the CFFS state to date. Furthermore, the unprecedented accuracy of their measurement has revealed an intriguing puzzle. For electrons confined to the LLL, one can take two exactly equivalent starting points: One can define the problem in terms of either electrons at $\nu$ or holes at $1-\nu$. One can then go ahead and composite fermionize either electrons or holes to produce what we will label \ecf s or \hcf s, which experience an effective magnetic field given by $B^*=B-2\rho\phi_0$, where $\phi_0=hc/e$ and $\rho$ is the density of composite fermions.  (All CF quantities are marked by an asterisk.) For $\nu\neq 1/2$ the \ecf s and \hcf s have different densities, producing, {\em in the MF description}, different Fermi wave vectors for the fully spin polarized CFFS state:
\begin{eqnarray}
{\rm MF}\;\;{\rm for}\;\; ^{\rm e}{\rm CFFS}:\;\;k^*_{\rm F}&=&\sqrt{4\pi \rho_{\rm e}}\;\; \leftrightarrow \;\; k^{\rm *}_{\rm F}\ell=
\sqrt{2\nu}\nonumber \\
{\rm MF}\;\;{\rm for}\;\; ^{\rm h}{\rm CFFS}:\;\;k^*_{\rm F}&=&\sqrt{4\pi \rho_{\rm h}}\;\; \leftrightarrow \;\; k^{\rm *}_{\rm F}\ell
=\sqrt{2(1-\nu)} \nonumber
\end{eqnarray}
where $\ell=\sqrt{\hbar c/eB}$ is the magnetic length, and the electron and hole densities are given by $\rho_{\rm e}=\rho_{\nu}=\nu/(2\pi \ell^2)$ and $\rho_{\rm h}=\rho_{1-\nu}=(1-\nu)/(2\pi \ell^2)$.  The CFFS thus appears to have a split personality. This raises many interesting conceptual questions. At a given $\nu$, do the $^{\rm e}$CFFS and $^{\rm h}$CFFSs represent two distinct states, or are they dual descriptions of the same state? If the former is true, then which of these two, if either, occurs in real systems? If the latter is true, how does one reconcile the seemingly incompatible consequences of the MF picture, and how does one understand the violation of the Luttinger theorem for at least one of the two descriptions? In either case, what is the role of particle-hole (p-h) symmetry in the LLL? Finally, how do we understand the remarkable finding of Kamburov {\em et al.}\cite{Kamburov14b} that the measured value of the CF Fermi wave vector is consistent with that expected from the smaller Fermi sea, namely $k^*_{\rm F}\ell=\min[\sqrt{2\nu},\sqrt{2(1-\nu)}]$?

These observations have motivated two striking theoretical proposals that lead to experimentally testable predictions. Son has proposed\cite{Son15} that viewing the composite fermion as a Dirac fermion allows a p-h symmetric description of the FQHE and the CFFS. 
Barkeshli, Mulligen and Fisher \cite{Barkeshli15} have taken the experimental observations to imply that the $^{\rm e}$CFFS and the $^{\rm h}$CFFS are distinct states of matter and a spontaneous breakdown of p-h symmetry within the LLL selects one of them. 
Within the Chern-Simons (CS) formulation of composite fermions\cite{Lopez91,Halperin93}, the MF Fermi wave vector $k_{\rm F}^{\rm *MF}\ell$ is not expected to change to all orders in a perturbative treatment of the Coulomb and the gauge interactions, suggesting that the $^{\rm e}$CFFS and $^{\rm h}$CFFS are perturbatively disconnected, i.e. are topologically distinct, for any $\nu\neq 1/2$, and, by extension, also for $\nu=1/2$. The CS formulation, however, does not implement the LLL constraint
, and hence does not satisfy the p-h symmetry, as has been stressed elsewhere in the literature\cite{Kivelson97,Son15}. 

We determine CFFS area using a different theoretical formulation of the CF paradigm, namely the microscopic wave functions of composite fermions\cite{Jain89,Jain89b,Jain07,Jain15}. This theory (i) is explicitly restricted to the LLL; (ii) satisfies p-h symmetry; and (iii) does not assume, a priori, any specific value for $k_{\rm F}^*\ell$.  We show that $k^*_{\rm F}\ell$ defined from Friedel oscillations in the pair-correlation function has the same value for states at $\nu$ and $1-\nu$ related by p-h symmetry. We explicitly calculate $k^*_{\rm F}\ell$ for filling factors in the vicinity of $\nu=1/2$.

We define the Fermi wave vector through the Friedel oscillations in the pair correlation function, for which we take the form\cite{Kamilla97b}
\be
g(\vec{r})=1+A (r\sqrt{4\pi \rho_{\rm e}})^{-\alpha} \sin(2k^*_{\rm F} r+\theta)
\label{pair_correlation_fitting_form}
\ee
where $A$, $\alpha$, $k^*_{\rm F}$ and $\theta$ are fitting parameters. We denote the particle coordinates by either $\vr_j$ or $z_j=x_j-iy_j$, and set $\ell=1$. This form is motivated by the observation that for noninteracting fully-spin polarized fermions in two dimensions at $B=0$, the oscillatory part of $g(r)$ for large $rk_{\rm F}$ is given by $(4/\pi r^3 k_{\rm F}^3)\sin(2k_{\rm F}r)$. Let us consider a wave function $\phi_\nu$ for a uniform density state at filling factor $\nu$. Its pair correlation function is given by
\be
g_\nu(\vec{r},\vec{r}')=\rho_{\nu}^{-2}\langle \phi_\nu | \hat{\Psi}^{\dagger}(\vec{r}) \hat{\Psi}^{\dagger}(\vec{r}')  \hat{\Psi}(\vec{r}')   \hat{\Psi}(\vec{r}) |\phi_\nu \rangle
\ee
\be
\phi_\nu={1\over N!} \int \prod_{j=1}^N d^2\vec{r}_j \phi_\nu(\vr_1, \cdots \vr_N) \prod_{k=1}^N\hat{\Psi}^{\dagger}(\vr_{k}) |0\rangle
\ee
where $\phi_\nu(\vr_1, \cdots \vr_N)$ is the real space wave function, $\hat{\Psi}(\vr)=\sum_{m=0}^\infty\eta_m(\vr) c_m$ is the electron annihilation operator in the LLL and $\hat{\Psi}^\dagger(\vr)$ is the corresponding electron creation operator, with the single particle LLL wave function defined as $\eta_m=(2\pi 2^m m!)^{-1/2} z^m \exp[-|z|^2/4]$.
We can similarly define the pair correlation function for electrons at $1-\nu$, with \
\be
\phi_{1-\nu}={1\over N!} \int \prod_{j=1}^N d^2\vec{r}_j \phi^*_\nu(\vr_1, \cdots \vr_N) \prod_{k=1}^N\hat{\Psi}(\vr_{k}) |1\rangle
\ee
where $|1\rangle$ is the state with the LLL fully occupied.  
Substituting into the expression for the pair correlation function and noting $\langle 1|f(c_m, c^\dagger_m)|1\rangle=\langle 0|f(c_m\rightarrow c^\dagger_m, c^\dagger_m\rightarrow c_m)|0\rangle$ produces the relation
\be
g_{1-\nu}(\vec{r},\vec{r}')=\rho_{1-\nu}^{-2}\langle \phi_\nu |   \hat{\Psi}(\vec{r})   \hat{\Psi}(\vec{r}') \hat{\Psi}^{\dagger}(\vec{r}') \hat{\Psi}^{\dagger}(\vec{r})|\phi_\nu \rangle
\ee
In terms of the LLL projected delta function\cite{Girvin84,Stone92} 
\be
 \bar{\delta}(\vec{r},\vec{r}')={1\over 2\pi} \exp\left[ -{1\over 4} (|\vec{r}-\vec{r}'|^2-z z'^{*}+z' z^* )\right]
\ee
which satisfies $ \bar{\delta}(\vec{r},\vec{r}')=[ \bar{\delta}(\vec{r}',\vec{r})]^*$,  
we have 
$\{  \hat{\Psi}(\vec{r}), \hat{\Psi}^{\dagger}(\vec{r}') \} \equiv \bar{\delta}(\vec{r},\vec{r}')
$, 
$
\langle \phi_\nu| \hat{\Psi}^{\dagger}(\vec{r})\hat{\Psi}(\vec{r}')|\phi_\nu\rangle=\nu\bar{\delta}(\vec{r}',\vec{r})
$, and 
$
\langle \phi_\nu| \hat{\Psi}(\vec{r})\hat{\Psi}^{\dagger}(\vec{r}')|\phi_\nu\rangle=(1-\nu)\bar{\delta}(\vec{r},\vec{r}').
$
Straightforward algebra gives the relation (assuming thermodynamic limit and translational invariance, and setting $\vec{r}'=0$)
\be
g_{1-\nu}(r)={ (1-2\nu) (1-e^{-r^2/2}) + \nu^2 g_{\nu}(r)  \over (1-\nu)^2  }
\label{gr-ph}
\ee
where we have assumed the same magnetic lengths for $\nu$ and $1-\nu$. For $r\gg 1$, this reduces to $g_{1-\nu}(r)=1+(\nu/(1-\nu))^2(g_\nu(r)-1)$. The important point is that an oscillatory term $\sin(2k^*_{\rm F} r)$ in $g_\nu$ implies identical oscillatory behavior for $g_{1-\nu}$, indicating that the states at $\nu$ and $1-\nu$ have the same $k^*_{\rm F} \ell$.  The ``exact" $k^*_{\rm F} \ell$ is thus the same at $\nu$ and $1-\nu$, and is independent of whether the problem is formulated in terms of electrons or holes.

We next determine the value of $k^*_{\rm F} \ell$ in a microscopic calculation from the oscillations in $g(r)$ following Refs.~\onlinecite{Kamilla97b,Park98b}. Being an equal time correlation function, $g(r)$ can be evaluated from the knowledge of the microscopic wave functions for the ground state in the vicinity of $\nu=1/2$. We concentrate on the fractions $\nu=n/(2n\pm 1)$ which approach $\nu=1/2$ in the limit of sufficiently large $n$. The microscopic Jain wave functions for these states are given by\cite{Jain89}
\be 
\Psi_{n/(2n\pm 1)}={\cal P}_{\rm LLL}\prod_{j<k=1}^N(z_j-z_k)^{2}\Phi_{\pm n}
\label{PsiCF}
\ee
where ${\cal P}_{\rm LLL}$ denotes LLL projection and $\Phi_n$ is the wave function of $n$ filled LLs, with $\Phi_{-n}=[\Phi_n]^*$.  At first it may appear that the above mentioned dichotomy is present also in the microscopic theory of composite fermion, as we illustrate by considering the fraction $\nue={n+1\over 2n+1}$. (Similar considerations apply to $\nu=n/(2n+1)$.) According to the CF theory, there are two ways of constructing a FQHE state at this fraction: (i) As the electron partner of the $\nuhs=n$ integer QHE (IQHE) of \hcf s in positive $B^*$, with wave function given by $C_{\rm p-h}{\cal P}_{\rm LLL}\prod_{j<k}(z_j-z_k)^2\Phi_n$ where $C_{\rm p-h}$ represents p-h transformation.  (ii) as the $\nues=n+1$ IQHE of \ecf s in a negative $B^*$, with wave function given by ${\cal P}_{\rm LLL}\prod_{j<k}(z_j-z_k)^2[\Phi_{n+1}]^*$. A priori, these appear to represent two distinct FQHE states, and one may ask which one applies to the real system. However, explicit evaluations\cite{Wu93,Davenport12} have demonstrated the nontrivial result that these two descriptions represent the same state. They predict identical quantum numbers for the ground state and the excitations (see Supplemental Materials (SM) \cite{SM}), and there is an almost prefect overlap between the two wave functions wherever it has been evaluated (e.g. for the 10 particle 2/3 state, the two wave functions have overlaps of 0.996 and 0.994 with the exact Coulomb state\cite{Wu93}). 
We have evaluated the $g(r)$'s of $\Psi_{(n+1)/(2n+1)}$ and $\Psi_{n/(2n+1)}$ and found them to be related by p-h symmetry to a very high accuracy (See SM \cite{SM}). 
The wave functions $\Psi_{n/(2n\pm 1)}$ produce, in the limit of $n\rightarrow \infty$, the same CFFS from either side, because $\Phi(B^*=0)$ is real\cite{Rezayi94}. Further, Rezayi and Haldane\cite{Rezayi00} have directly constructed the wave function for the CFFS on a torus and found that, for $N=16$ particles, it has has an overlap of 
0.9994 with its hole conjugate, and 0.9925 with the exact p-h symmetric Coulomb ground state\cite{Rezayi00}.  The degree to which the microscopic wave functions of the CF theory satisfy the p-h symmetry may seem surprising, 
but is a byproduct of the fact that these are excellent approximations of the exact Coulomb states in the LLL which satisfy p-h symmetry exactly.

The understanding of FQHE at $\nu=(n+1)/(2n+1)$ as $\nu^*=n+1$ IQHE of \ecf s in a negative $B^*$ becomes essential when we consider the spin degree of freedom, because it is the only known way to explain the non-fully spin-polarized FQHE states here, e.g. the spin singlet state at $\nu=2/3$. (Recall that for spinful states, p-h symmetry relates $\nu$ to $2-\nu$.) An extensive experimental\cite{Eisenstein89,Du95,Yeh99,Kukushkin99,Kukushkin00,Freytag01,Melinte00,Bishop07,Padmanabhan09,Tiemann12,Feldman12,Feldman13,Liu14a} and theoretical\cite{Wu93,Park98,Park99,Park01,Davenport12,Archer13b,Sodemann14,Balram15,Balram15a} literature on spin phase transitions has validated the explanation of the $\nue=(n+1)/(2n+1)$ as $\nues=n+1$ IQHE of \ecf s.

\begin{figure}[t]
\begin{center}
\includegraphics[width=7cm,height=3.9cm]{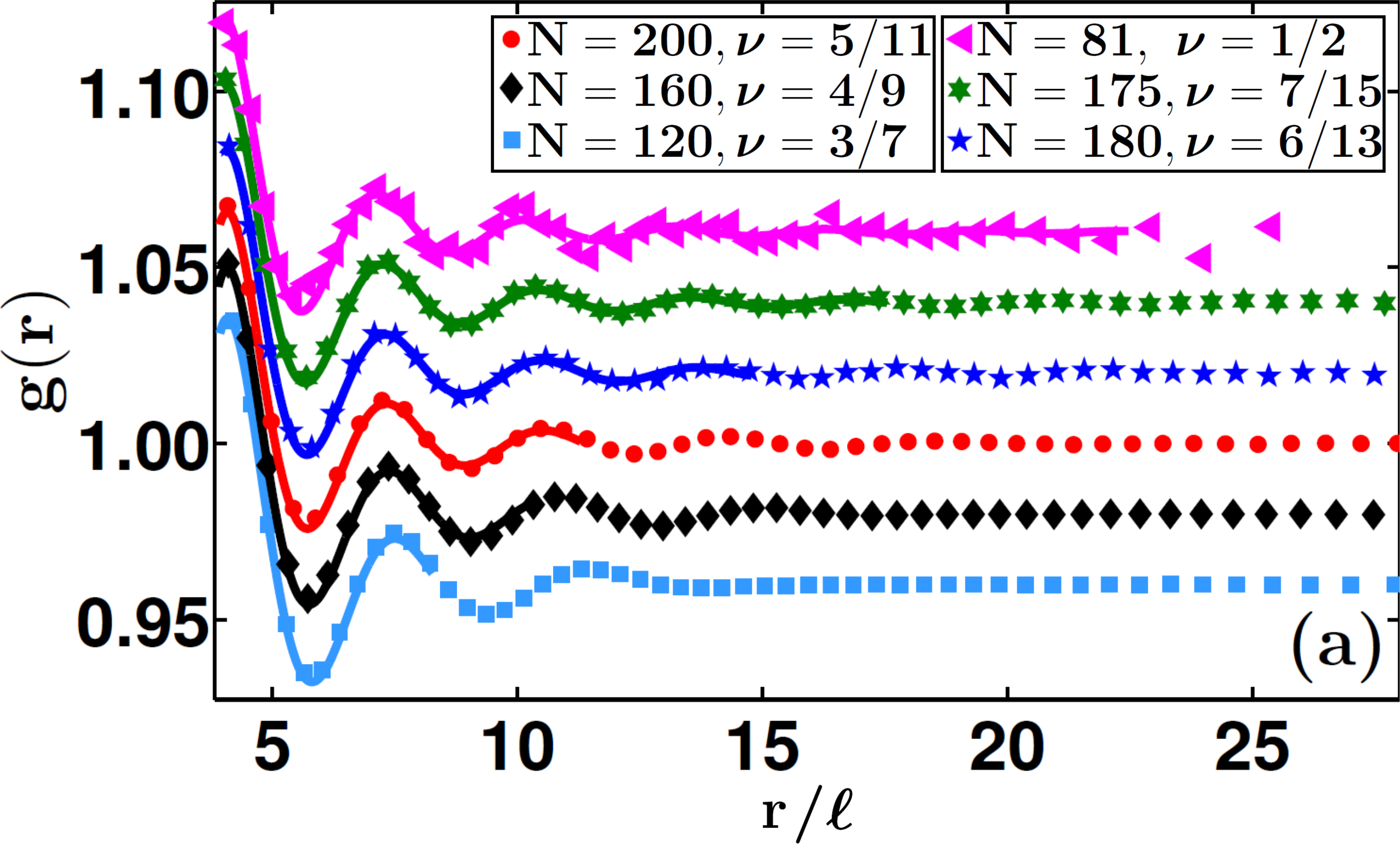}
\includegraphics[width=7cm,height=3.9cm]{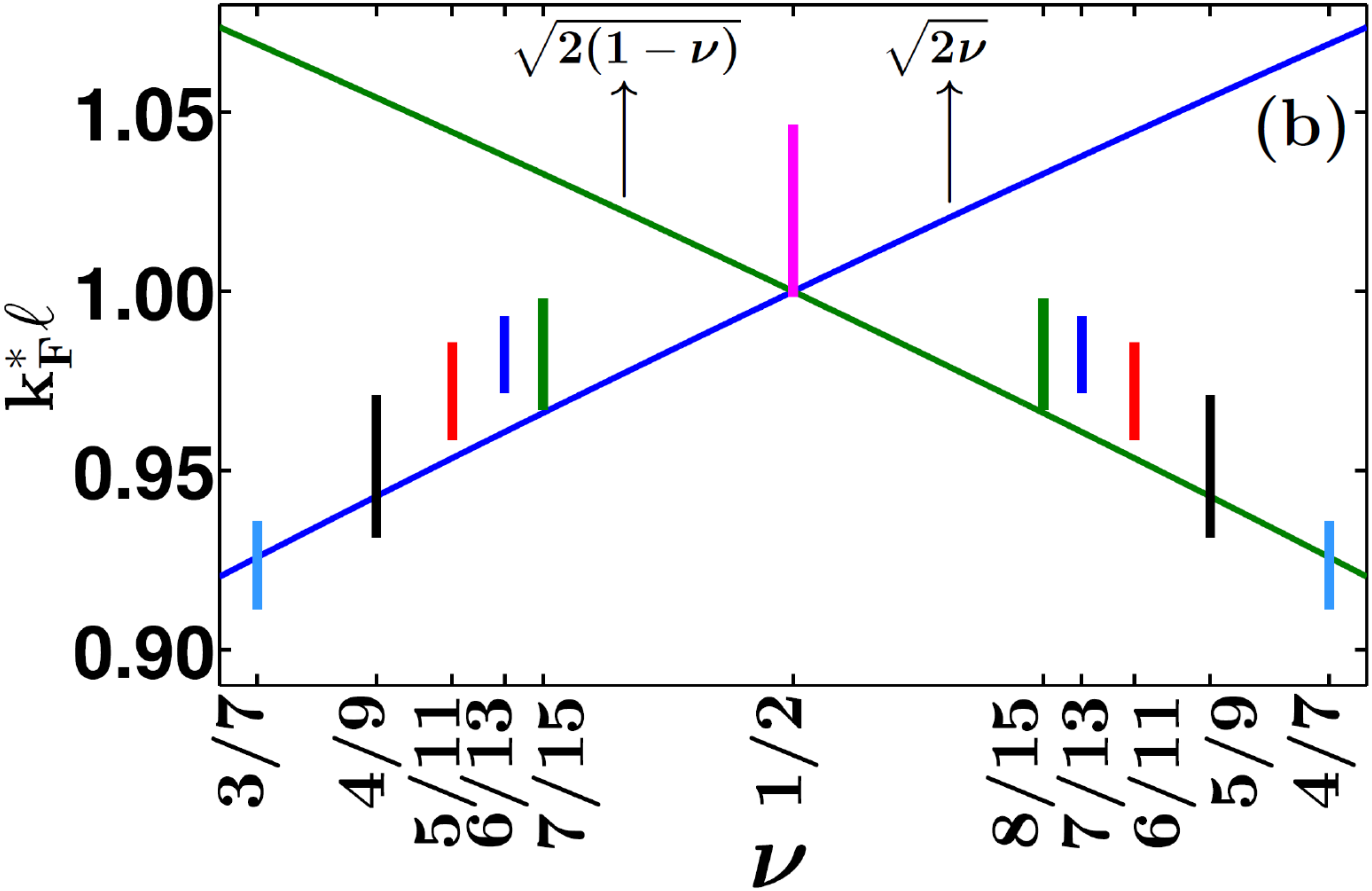}
\end{center}
\caption{(a) Pair correlation function $g(r)$ as a function of $r/\ell$, where $r$ is the arc distance on the sphere.  The projected wave functions in Eq.~\ref{PsiCF} have been used for its evaluation. The solid lines are fits using Eq.~\ref{pair_correlation_fitting_form} for the initial oscillations. For clarity, the curves (except for 5/11) have been shifted up or down by multiples of 0.02. (b) The calculated thermodynamic values of $k_{\rm F}^*\ell$ as a function of $\nu$. The mean-field values $\sqrt{2\nu}$ and $\sqrt{2(1-\nu)}$ are also shown for reference.}
\label{kF}
\end{figure}

The validity of $\Psi_{n/(2n\pm 1)}$ for the {\em incompressible} states has been established by extensive numerical studies\cite{Dev92,Dev92a,Wu93,Rezayi94,Balram13,Jain07}. We will  
make the {\em assumption} that $\Psi_{n/(2n\pm 1)}$ remain valid to arbitrarily high $n$, i.e., that the compressible region around $\nu=1/2$ consists of unresolved IQHE states of composite fermions. 
We stress that we cannot rule out the possibility that the $^{\rm e}$CFFS and $^{\rm h}$CFFS are in reality topologically distinct, as proposed in Ref.~\onlinecite{Barkeshli15}, and a spontaneous breaking of the p-h symmetry selects one of them. This would happen, for example, if the half filled ground state were unstable to CF pairing \cite{Greiter91,Cipri14,Wang14, Metlitski15} and $^{\rm e}$CFFS and $^{\rm h}$CFFS are the normal states of the topologically distinct Moore-Read Pfaffian and anti-Pfaffian paired-CF states \cite{Moore91,Read00,Levin07,Lee07}. Nonetheless, the presently known facts do admit the possibility of a p-h symmetric CFFS, and our aim here is to deduce its properties, so experiments may distinguish between the different proposals. 

\begin{figure}[t]
\begin{center}
\includegraphics[width=7cm,height=3.9cm]{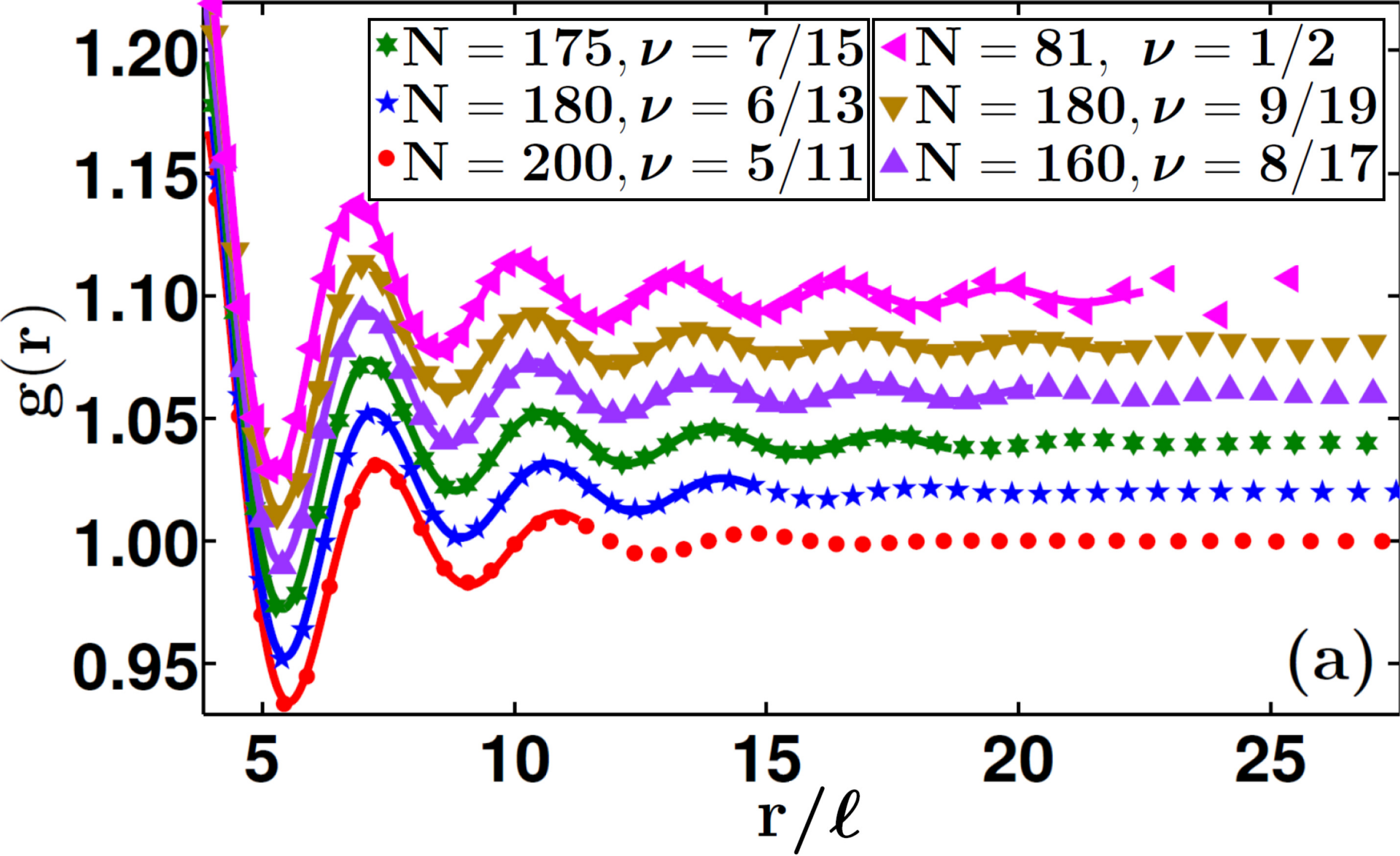}
\includegraphics[width=7cm,height=3.9cm]{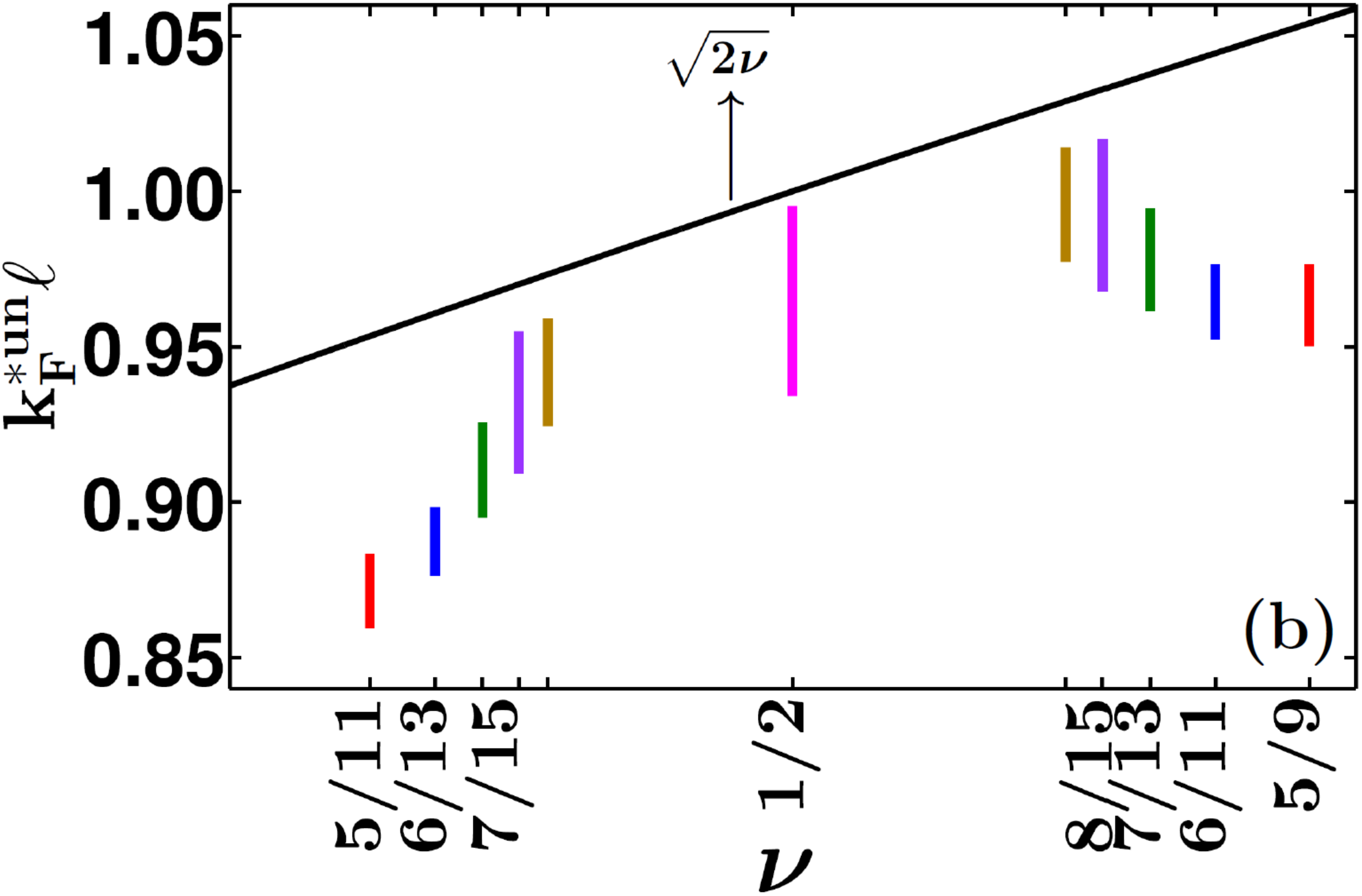}
\end{center}
\caption{Same as in Fig. \ref{kF} but for the unprojected wave functions $\Psi^{\rm un}_{n/(2n\pm 1)}$. Also shown for reference is $k_{\rm F}^{\rm *MF} \ell=\sqrt{2\nu}$ corresponding to the Chern-Simons mean field theory for $^{\rm e}$CFs.}
\label{kFun}
\end{figure}

We have calculated the pair correlation function for $\nu=n/(2n+1)$ up to 7/15 using the Metropolis Monte Carlo method. For technical reasons, we find it convenient to use the standard spherical geometry\cite{Haldane83}; see SM for details\cite{SM}.  The results extrapolated to the thermodynamic limit apply to the planar geometry of the experiments. The spherical analogs of the above wave functions, as well as the details of LLL projection can be found in the literature\cite{Jain97,Jain97b,Jain07}. All wave functions considered below have uniform density and are translationally invariant (i.e. have orbital angular momentum $L=0$ on the sphere). The $g(r)$'s for the largest systems in our study are shown in Fig.~\ref{kF}(a). For incompressible systems the pair correlation function is expected to decay in a gaussian manner in the limit of $r\rightarrow \infty$, but there is a range of intermediate $r$ where it exhibits well defined oscillations from which a Fermi wave vector can be extracted. In fitting $g(r)$ to Eq.~\ref{pair_correlation_fitting_form}, we avoid very small $r$ (where short distance correlations are important) and very large $r$ (where curvature effects become non-negligible). From the results for finite systems we obtain the thermodynamic limits for the $k^*_{\rm F}\ell$ (see \cite{SM}). We find that very large systems ($N>100$) are needed for a satisfactory thermodynamic extrapolation of $k^*_{\rm F}\ell$. The thermodynamic limits are shown in Fig.~\ref{kF}(b). [We have assumed exact p-h symmetry, which implies that the $k^*_{\rm F}\ell$ at $\nu=(n+1)/(2n+1)$ is the same as that at $\nu=n/(2n+1)$.]
The range of $k^*_{\rm F}\ell$ includes uncertainly in the fits (estimated by linear and quadratic fitting in $1/N$ for $g(r)$) as well as uncertainty due to the curvature of the spherical geometry (estimated by considering fits with $r$ chosen as the chord or the arc distance). For reference, Fig.~\ref{kF}(b) also shows the values $k^{\rm *MF}_{\rm F}\ell=\sqrt{2\nu}$ and  $k^{\rm *MF}_{\rm F}\ell=\sqrt{2(1-\nu)}$ as expected from a MF picture for the $^{\rm e}$CFFS and $^{\rm h}$CFFS.

For $\nu=1/2$, we have estimated $k^*_{\rm F} \ell$ by an extrapolation, in the spherical geometry, of filled shell CF systems at zero effective flux\cite{Rezayi94} occurring at $N=n^2$. For technical reasons, we are not able to go to systems with $N>81$ (which requires filling the 10th Landau-like level of composite fermions, where the numerics become unstable). We have therefore also studied the CFFS in the torus geometry\cite{Haldane85,Rezayi00,Shao15} where we can go up to $N=153$, and find that the results are consistent with our spherical results. Results in the torus geometry are presented in \cite{SM}.

For $\nu$ away from 1/2, our calculated $k^*_{\rm F} \ell$ is close, but not equal, to the smaller of $\sqrt{2\nu}$ and $\sqrt{2(1-\nu)}$.  Both from extrapolation of the results from the sequences $n/(2n\pm 1)$ and from calculations directly at $\nu=1/2$, our calculations suggest that the CFFS area at $\nu=1/2$ slightly deviates from the value expected from the Luttinger rule.  

The physics of the CFFS at $\nu=3/2$ is analogous to that at $\nu=1/2$ once the $B$ dependence of the density of either \ecf s or \hcf s in the spin-reversed LL is accounted for\cite{Kamburov12,Kamburov14}. Near $\nu=1/4$, both $n/(4n+1)$ and $n/(4n-1)$ are understood only in terms of \ecf s, and thus one expects $k^*_{\rm F} \ell\approx \sqrt{2\nu}$ (as opposed to $k_F^*\ell\approx \sqrt{2(1-\nu)}$), as observed experimentally \cite{Kamburov14b,Kamburov14} and also in our calculations\cite{SM}. Analogous consideration for the CFFS at $\nu=3/4$ gives $k^*_{\rm F} \ell\approx\sqrt{2(1-\nu)}$. 

We next investigate how robust the CFFS area is to LL mixing. LL mixing requires a formulation in terms of electrons (rather than holes of the LLL), and the LLL electronic wave functions in Eq.~\ref{PsiCF} can be used as a starting point to address this issue\cite{Melik-Alaverdian95}.  A realistic treatment of LL mixing is outside the scope of the current study, but we have considered the ``unprojected" Jain wave functions $\Psi^{\rm un}_{n/(2n\pm 1)}=\prod_{j<k=1}^N (z_j-z_k)^{2}\Phi_{\pm n}$, which contain some amplitude outside of the LLL \cite{Trivedi91,Kamilla97b}. Even though they do not give a realistic account of LL mixing, it is likely that they are adiabatically connected to the projected wave functions (as explicitly demonstrated for $\nu=2/5$ \cite{Rezayi91}), and hence to the actual Coulomb ground states. For these wave functions, the $g(r)$'s of $\nu=n/(2n-1)$ and $\nu=n/(2n+1)$ are identical for a given $N$ when plotted in units of the sphere radius, which in the thermodynamic limit implies the relation
\be
(k^{\rm * un}_{\rm F} \ell)_{n\over 2n-1} = \left({2n +1 \over 2n-1}\right)^{1/2} (k^{\rm * un}_{\rm F} \ell)_{n \over 2n+1}
\label{kFrelation}
\ee
The calculated values of $k_{\rm F}^{\rm * un}\ell$ using the $^{\rm e}$CF description (see SM for details) are shown in Fig.~\ref{kFun}. Our calculations thus provide evidence that $k_{\rm F}^*\ell$ depends on LL mixing. 
Another approximate wave function with LL mixing is the CS MF state $\Psi^{\rm MF}_{n/(2n\pm 1)}=\prod_{j<k=1}^N [(z_j-z_k)/|z_j-z_k|]^2\Phi_{\pm n}(B^*)$. Given that its $g(r)$ coincides with that for $\Phi_{\pm n}(B^*)$, we get $k_{\rm F}^{\rm *MF}=\sqrt{4\pi \rho_{e}}$, i.e., $k_{\rm F}^{\rm *MF} \ell=\sqrt{2\nu}$ for all $\nu=n/(2n\pm 1)$. For the unprojected or the CS-MF wave functions, $k_{\rm F}^{*}\ell$ does not obey particle-hole symmetry, as expected in the presence of LL mixing. 

In summary, we have shown that, within the microscopic theory of composite fermions, it is valid to consider electron (or hole) based composite fermions for $\nu<1/2$ as well as $\nu>1/2$. We have calculated the CF Fermi wave vector in the vicinity of $\nu=1/2$ and 
find that it is closer, but not equal, to the smaller of $\sqrt{4\pi\rho_{\rm e}}$ and $\sqrt{4\pi\rho_{\rm h}}$. In terms of electron based composite fermions, this implies that the Luttinger theorem is slightly (substantially) violated for $\nu<1/2$ ($\nu>1/2$). At $\nu=1/2$, our results suggest, but do not conclusively demonstrate, that the $k^*_{\rm F}$ differs slightly (by a few percent) from the value $\sqrt{4\pi\rho_{\rm e}}$ predicted by the Luttinger area rule. We also provide evidence that $k^*_{\rm F}$ varies as a function of LL mixing.

{\rm Note Added:} Since the completion of this work, several other articles have appeared addressing the nature of the CFFS and the role of p-h symmetry \cite{Kachru15,Wang15,Geraedts15,Murthy15,Liu15}.

\begin{acknowledgments}
We are grateful to Maissam Barkeshli, Matthew Fisher, Steve Kivelson, Subir Sachdev, Vijay Shenoy and especially Mansour Shayegan for insightful discussions. This work was supported by the U.S. Department of Energy, Office of Science, Basic Energy Sciences, under Award No. DE-SC0005042 and Hungarian Scientific Research Funds No. K105149 (C.T.).  We acknowledge the Research Computing and Cyberinfrastructure at Pennsylvania State University which is in part funded by the National Science Foundation Grant No. OCI-0821527 and the HPC facility at the Budapest University of Technology and Economics. We thank A. W\'ojs for providing exact diagonalization results. C.T. was supported by the Hungarian Academy of Sciences.
\end{acknowledgments}

\bibliography{../../Latex-Revtex-etc./biblio_fqhe}
\bibliographystyle{apsrev}

\pagebreak

\setcounter{figure}{0}
\setcounter{equation}{0}
\renewcommand\thefigure{S\arabic{figure}}
\renewcommand\thetable{S\arabic{table}}
\renewcommand\theequation{S\arabic{equation}}

\centerline{\bf Supplemental material}
\hspace{1cm}

The Supplemental Material (SM) sections contains (i) a primer on the spherical geometry; (ii) a discussion of particle-hole (p-h) symmetry within the composite fermion (CF) theory; (iii) the relation between pair correlation functions of two states related by the p-h symmetry for finite $N$; (iv) extrapolations of the CF Fermi wave vector and the power law exponent $\alpha$; (v) a discussion of the CF Fermi Sea (CFFS) at $\nu=1/4$; and (vi) our calculations for the CFFS at $\nu=1/2$ and $\nu=1/4$ in the torus geometry. 

A note on the notation: Quantities defined with the subscript or superscript `e' are for electrons and those with superscript or subscript `h' are for holes. 

\section{Haldane's Spherical geometry}
\begin{figure}[htpb]
\begin{center}
\includegraphics[width=4cm,height=4cm]{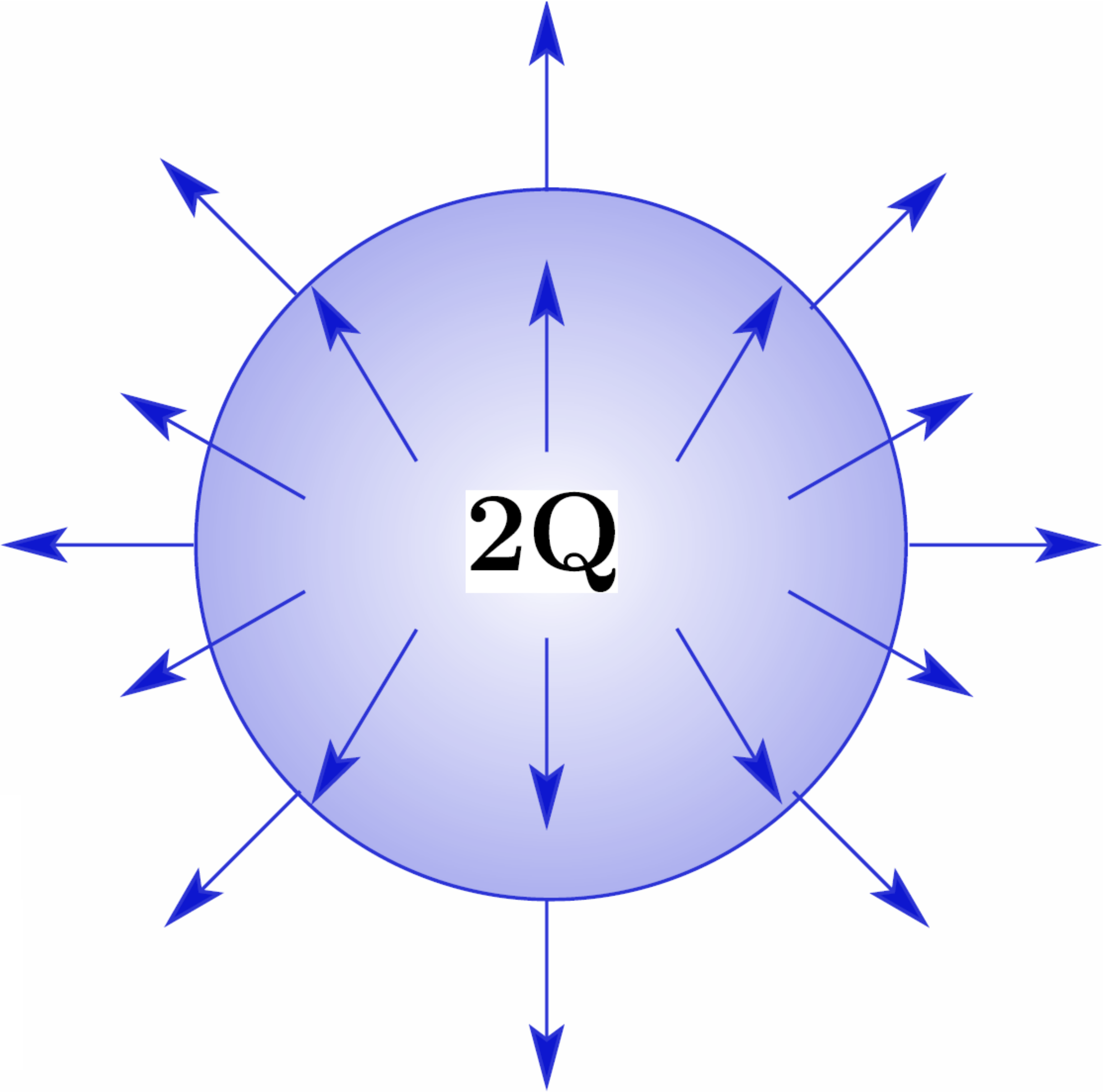}
\end{center}
\caption{A schematic of the Haldane sphere, in which $N$ electrons reside on the surface of sphere. A Dirac monopole at the center produces a flux of $2Q$ (an integer) in units of the flux quantum $\phi_0=hc/e$.}
\label{fig:sphere}
\end{figure}

For many of our calculations we employ Haldane's spherical geometry\cite{Haldane83} shown in Fig. \ref{fig:sphere}. In this geometry, $N$ electrons move on the surface of the sphere and the magnetic field perpendicular to the surface is generated by a monopole of strength $Q$ placed at the center of the sphere. The total flux through the surface of the sphere is $2Qhc/e$ and the radius of the sphere is $\sqrt{Q}\ell$. The electron coordinates are denoted by ${\bf \Omega}\equiv (\theta,\phi)$, and 
the single particle eigenstates, called monopole harmonics, are given by (see, e.g. the review in Ref.~\onlinecite{Jain07}):
\begin{eqnarray}
Y_{Q,l,m}=&N_{Q,l,m}(-1)^{l-m}v^{Q-m}u^{Q+m} \\
&\times \sum_{s=0}^{l-m}(-1)^s\binom{l-Q}{s}\binom{l+Q}{l-m-s}(v^{*}v)^{l-Q-s}(u^{*}u)^{s} \nonumber
\end{eqnarray}
\begin{eqnarray}
N_{Q,l,m}&=&\Bigg[\frac{(2l+1)}{4\pi}\frac{(l-m)!(l+m)!}{(l-Q)!(l+Q)!}\Bigg]^{\frac{1}{2}} \\
u&=&\cos(\theta/2)e^{i\frac{\phi}{2}} \\
v&=&\sin(\theta/2)e^{-i\frac{\phi}{2}} 
\end{eqnarray}
where the 
orbital angular momentum $l$ takes values $|Q|,|Q|+1,|Q|+2,...$ corresponding to the $n=0$ (lowest)$,~1$ (second)$,~2$ (third)$,...$ Landau Levels (LLs) and $m=-l,-l+1,...,l-1,l$. Consequently the degeneracy of each LL
is $2l+1$. Positive and negative values of $Q$ respectively stand for a flux extending radially outward and inward through the surface of the sphere. The wave functions in the lowest Landau Level are obtained by setting $l=Q$ in the above equation which gives:
\begin{equation}
Y_{Q,Q,m}=\Big[\frac{2Q+1}{4\pi}\binom{2Q}{2Q-m}\Big]^{\frac{1}{2}}(-1)^{Q-m}v^{Q-m}u^{Q+m} 
\end{equation}
The wave function for a state with an integer number of completely filled LLs is a Slater determinant formed from the filled single particle states. Using the above equation, one can show that the form of the wave function for one filled LL (apart from a normalization) has the following nice form:
\begin{eqnarray}
\Phi_1=\prod_{j<k}(u_jv_k-u_kv_j)
\end{eqnarray}

In the main article, the wave functions of composite fermions have been written in coordinates appropriate for the planar geometry. They can be expressed in the spherical geometry in the standard fashion explained in the literature\cite{Jain07}. We have also used the torus geometry (see below) for an additional substantiation of some of our results. The choice of which geometry is used is a matter of technical convenience. For example, the spherical geometry is more suitable for the FQHE states at $n/(2n\pm 1)$ than the other two geometries. We carefully extrapolate all of our results to the thermodynamic limit to eliminate any effects of the geometry employed. 

\section{Particle hole symmetry in the CF theory}
In the CF theory, the states at $\nu=n/(2n+1)$ maps into $\nu^*=n$ of $^{\rm e}$CFs, whereas its hole partner at $\nu=1-n/(2n+1)=(n+1)/(2n+1)$ maps into $\nu^*=n+1$ of $^{\rm h}$CFs. This may seem to be inconsistent with p-h symmetry, but we show that is not the case. 

From the standard CF theory on sphere, the state at $\nu=n/(2n+1)$
occurs at flux value 
\be
2Q={2n+1\over n}N_{\rm e} -(2+n)
\ee
Using the relation 
$2Q_{\rm e}^*=2Q-2N_{\rm e}+2$ 
we have 
$2Q_{\rm e}^*=N_{\rm e}/n-n$,
which corresponds to precisely $n$ filled $\Lambda$ levels of $^{\rm e}$CFs (CF Landau-like levels are termed $\Lambda$ levels). The single particle orbital angular momentum of the topmost occupied $\Lambda$L is 
\be
l^*_{\rm e}=Q_{\rm e}^*+n-1
\ee
Now let us ask what happens if we model this state in terms of $^{\rm h}$CFs. In terms of $N_{\rm h}=2Q+1-N_{\rm e}$ holes, the  
$\nu=(n+1)/(2n+1)$ state occurs at 
\be
2Q={2n+1\over n+1}N_{\rm h} -(1-n)
\ee
This maps into a negative effective flux of 
$
2Q^*_h=-N_{\rm h}/(n+1)+n+1
$
for $^{\rm h}$CFs, which corresponds to $n+1$ filled $\Lambda$Ls of $^{\rm h}$CFs. The single particle orbital angular momentum of the topmost occupied $\Lambda$L is 
\be
l^*_{\rm h}=|Q_{\rm h}^*|+n.
\ee
Using the above relations, it follows that $l^*_{\rm e}=l^*_{\rm h}$, i.e., the topmost fully occupied $\Lambda$L has the same single particle orbital angular momentum, and thus the same number of particles, independent of whether we model the problem in terms of $^{\rm e}$CFs or $^{\rm h}$CFs. This implies that the $^{\rm e}$CFs and $^{\rm h}$CFs descriptions produce identical structures, at the mean field level, for the excited bands for energies up to $n\hbar\omega_c^*$, where $\hbar\omega_c^*$ is the effective cyclotron energy of composite fermions. We further note that the cyclotron energies for both are given by $\hbar\omega_c^*=\hbar eB^*/m^*c=\hbar eB/(2n+1)m^*c$, where $B^*$ is the effective magnetic field for composite fermions and $m^*$ is their mass. This implies that that for a given magnetic field, the two descriptions produce the same cyclotron energy, and thus the same excitation gaps at the mean field level. As noted in the text, the microscopic wave functions of composite fermions also obey the particle-hole (p-h) symmetry to a very good approximation.

\section{Relation between $g_{\nu}$ and $g_{1-\nu}$ in the spherical geometry}
In this section we shall derive the exact formula relating the pair correlation functions of two states related by particle-hole symmetry, in the spherical geometry for finite $N$. (The relation derived in the main text assumes disk geometry and translational invariance, and thus is valid only in the thermodynamic limit.) We then test how well this relation is satisfied by the pair correlation functions of ${\cal P}_{\rm LLL}\prod_{j<k}(z_j-z_k)^2\Phi_n$ and ${\cal P}_{\rm LLL}\prod_{j<k}(z_j-z_k)^2[\Phi_{n+1}]^*$. That would be a test of the degree to which these wave functions are related by p-h conjugation.

The lowest Landau level projected electron field annihilation operator in the spherical geometry is written as:
\begin{equation}
\hat{\Psi}(\vec{r})=\sum_{m=-Q}^{m=+Q} c_{m} Y_{Q,Q,m}(\vec{r})
\label{sphere_LLL_field_op}
\end{equation}
where $\vec{r}\equiv {\bf \Omega} \equiv(u,v)\equiv(\theta,\phi)$ is the coordinate on the sphere [$\theta$ and $\phi$ are the polar and azimuthal angles on the sphere, $u$ and $v$ are spinorial coodinates defined as: $u=\cos(\theta/2)e^{i\phi/2}$ and $v=\sin(\theta/2)e^{-i\phi/2}$] and $Y_{Q,Q,m}$ are the so-called spherical monopole harmonics\cite{Jain07}. The following identities are satisfied by the electron field operator:
\begin{eqnarray}
\{\hat{\Psi}^{\dagger}(\vec{r}),\hat{\Psi}(\vec{r}')\}&=&\bar{\delta}(\vec{r}',\vec{r}) \nonumber \\
\langle \phi_{\nu} | \hat{\Psi}(\vec{r})\hat{\Psi}^{\dagger}(\vec{r}') | \phi_{\nu} \rangle &=& (1-\lambda)\bar{\delta}(\vec{r},\vec{r}'),\nonumber \\
\langle \phi_{\nu} | \hat{\Psi}^{\dagger}(\vec{r})\hat{\Psi}(\vec{r}') | \phi_{\nu} \rangle &=& \lambda\bar{\delta}(\vec{r}',\vec{r}) 
\label{matrix_elements_anti_comm}
\end{eqnarray}
where\cite{Jain07}:
\begin{equation}
\bar{\delta}(\vec{r},\vec{r}')=\frac{2Q+1}{4\pi Q\ell^2}(u^{*}u'+v^{*}v')^{2Q},~\lambda=\frac{N}{2Q+1}
\label{identities_sphere}
\end{equation}

To derive the relation between the pair correlation functions at $\nu$ and $1-\nu$ we start with Eq. 5 of the main text. Using the relations in Eq. \ref{matrix_elements_anti_comm} we can show that:
\begin{eqnarray}
g_{1-\nu}(\vec{r},\vec{r}')&=&\frac{(1-2\lambda)}{\rho_{1-\nu}^2}[\bar{\delta}(\vec{r},\vec{r})^2-\bar{\delta}(\vec{r},\vec{r}')^2]\nonumber \\
&&+\frac{\rho_{\nu}^2}{\rho_{1-\nu}^2}g_{\nu}(\vec{r},\vec{r}')
\end{eqnarray}
Next we make use of the fact that incompressible states have both rotational and translational invariance, and choose $\vec{r}'$ to be the north pole. Substituting the values from Eq.~\ref{identities_sphere}, we find that
\begin{widetext}
\begin{equation}
g_{1-\nu}(r_{c})=\frac{(2Q+1)(2Q+1-2N)}{(2Q+1-N)^2}\Bigg(1-\Bigg[1-\Big(\frac{r_c}{2\sqrt{Q}\ell}\Big)^2\Bigg]^{2Q}\Bigg) +\Big(\frac{N}{2Q+1-N}\Big)^{2} g_{\nu}(r_{c}) 
\label{gr_sphere_ph}
\end{equation} 
\end{widetext}
where $r_{c}$ is the chord distance of the point $\vec{r}$ from the north pole. It can be verified that this reduces to Eq.~7 of the main text in the thermodynamic limit $N, 2Q\rightarrow\infty$ with $N/2Q\rightarrow\nu$. 

\begin{figure*}[htpb]
\begin{center}
\includegraphics[width=8cm,height=4.5cm]{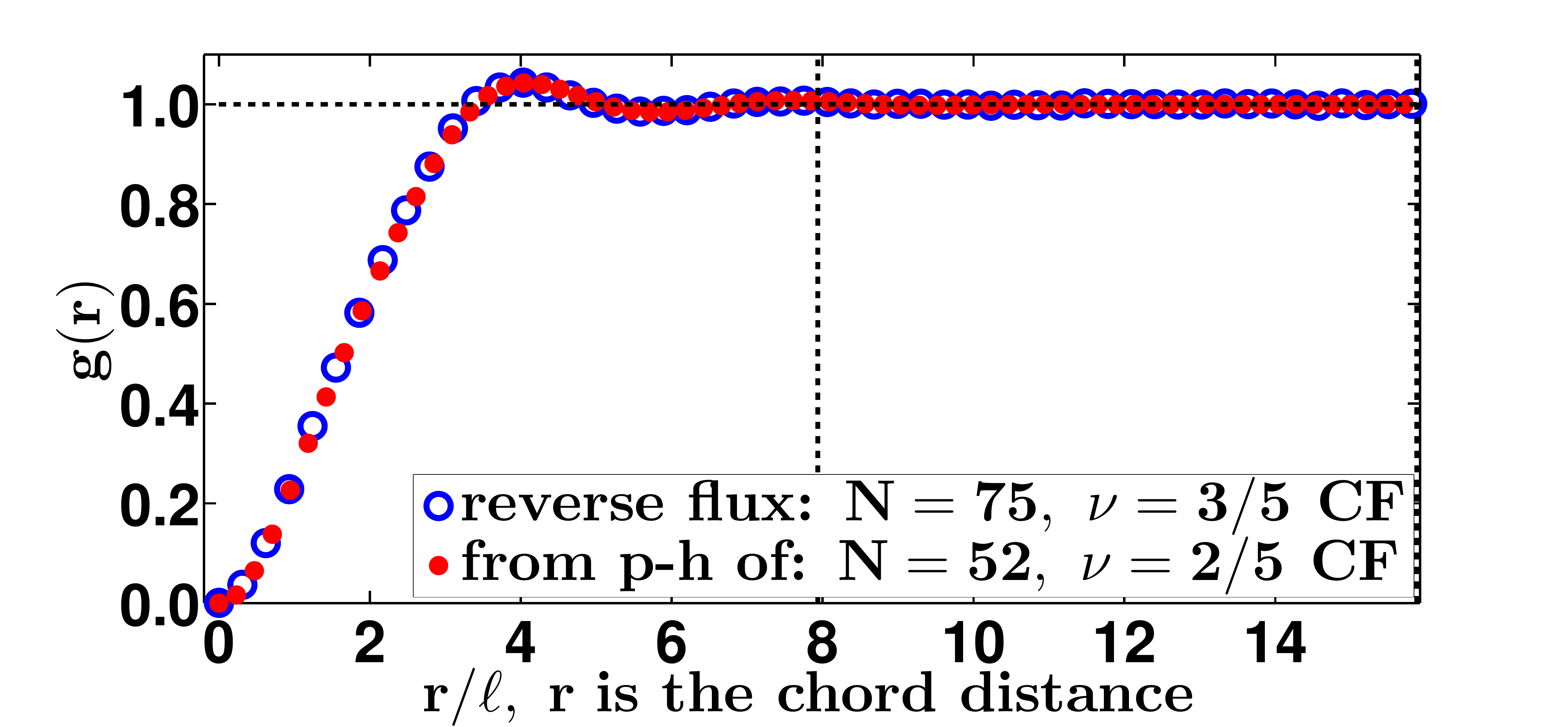}
\includegraphics[width=8cm,height=4.5cm]{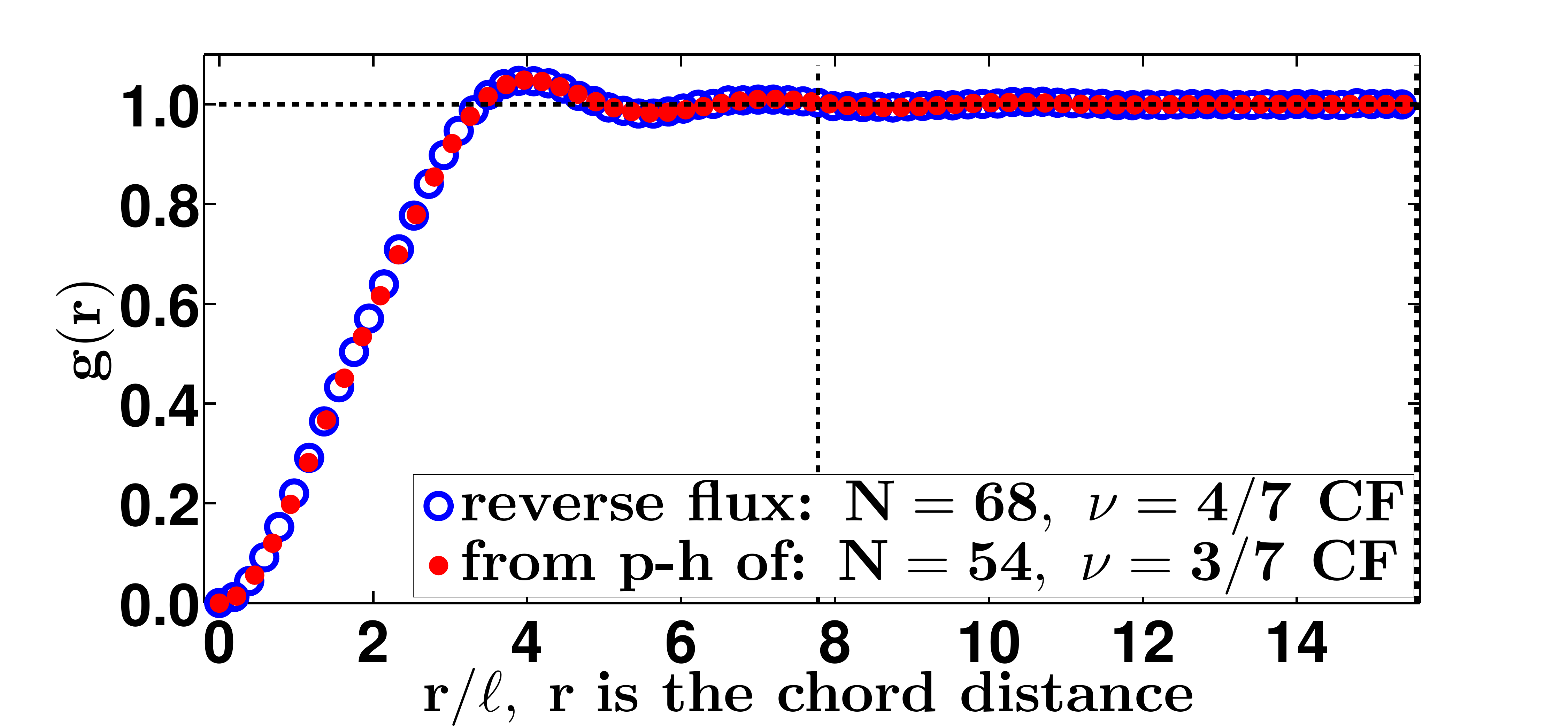}
\includegraphics[width=8cm,height=4.5cm]{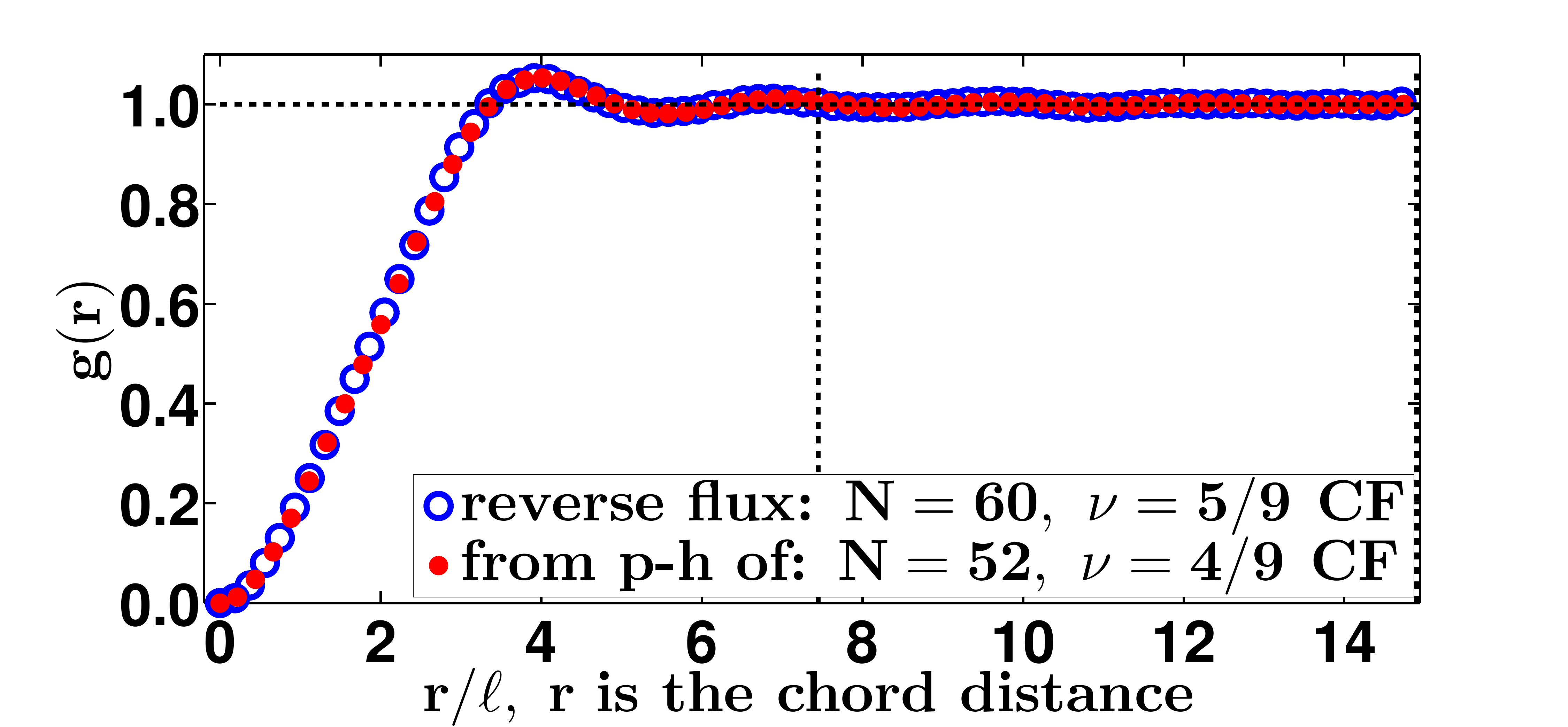}
\includegraphics[width=8cm,height=4.5cm]{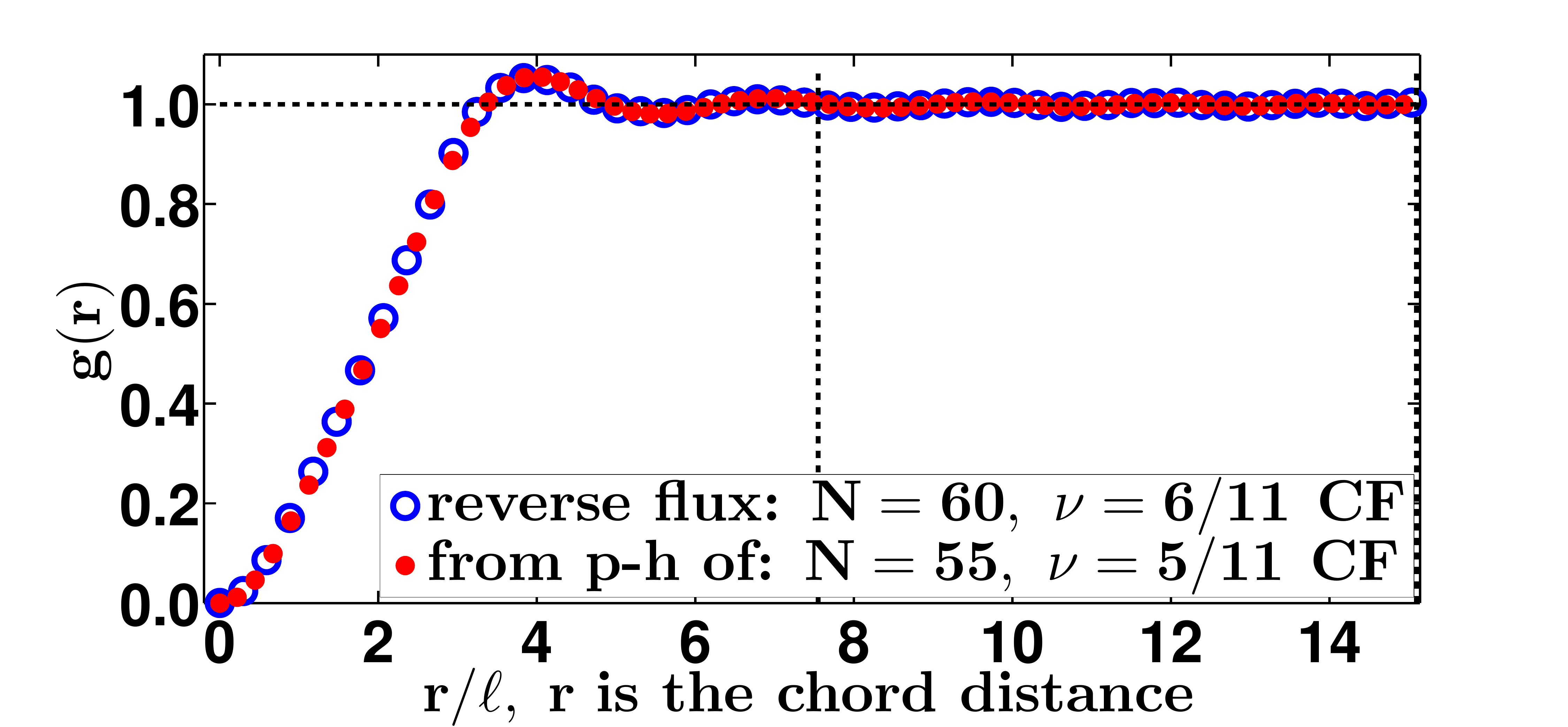}
\end{center}
\caption{Pair correlation function $g(r)$ as a function of $r/\ell$, where $r$ is the chord distance. The blue circles give the $g(r)$ obtained directly from the lowest Landau level projected wave functions $\Psi_{(n+1)/(2n+1)}$ for 3/5 (top left panel), 4/7 (top right panel), 5/9 (bottom left panel) and 6/11 (bottom right panel). The red dots give the $g(r)$ for $C_{\rm p-h}\Psi_{n/(2n+1)}$, namely the hole conjugates of projected  wave functions $\Psi_{n/(2n+1)}$.}
\label{gr_ph}
\end{figure*}

In Fig. \ref{gr_ph} we show a comparison of the pair correlation functions of $\Psi_{(n+1)/(2n+1)}$ and $C_{\rm p-h}\Psi_{n/(2n+1)}$. (The $g(r)$ of the latter is obtained from the $g(r)$ of $\Psi_{n/(2n+1)}$ following Eq.~\ref{gr_sphere_ph}). The excellent agreement between the two pair correlation functions confirms that the wave functions $\Psi_{n/(2n+1)}$ and $\Psi_{(n+1)/(2n+1)}$ are related by p-h symmetry to an extremely good approximation.

\section{Extrapolations in the spherical geometry}
\begin{figure*}[htpb]
\begin{center}
\includegraphics[width=8cm,height=5cm]{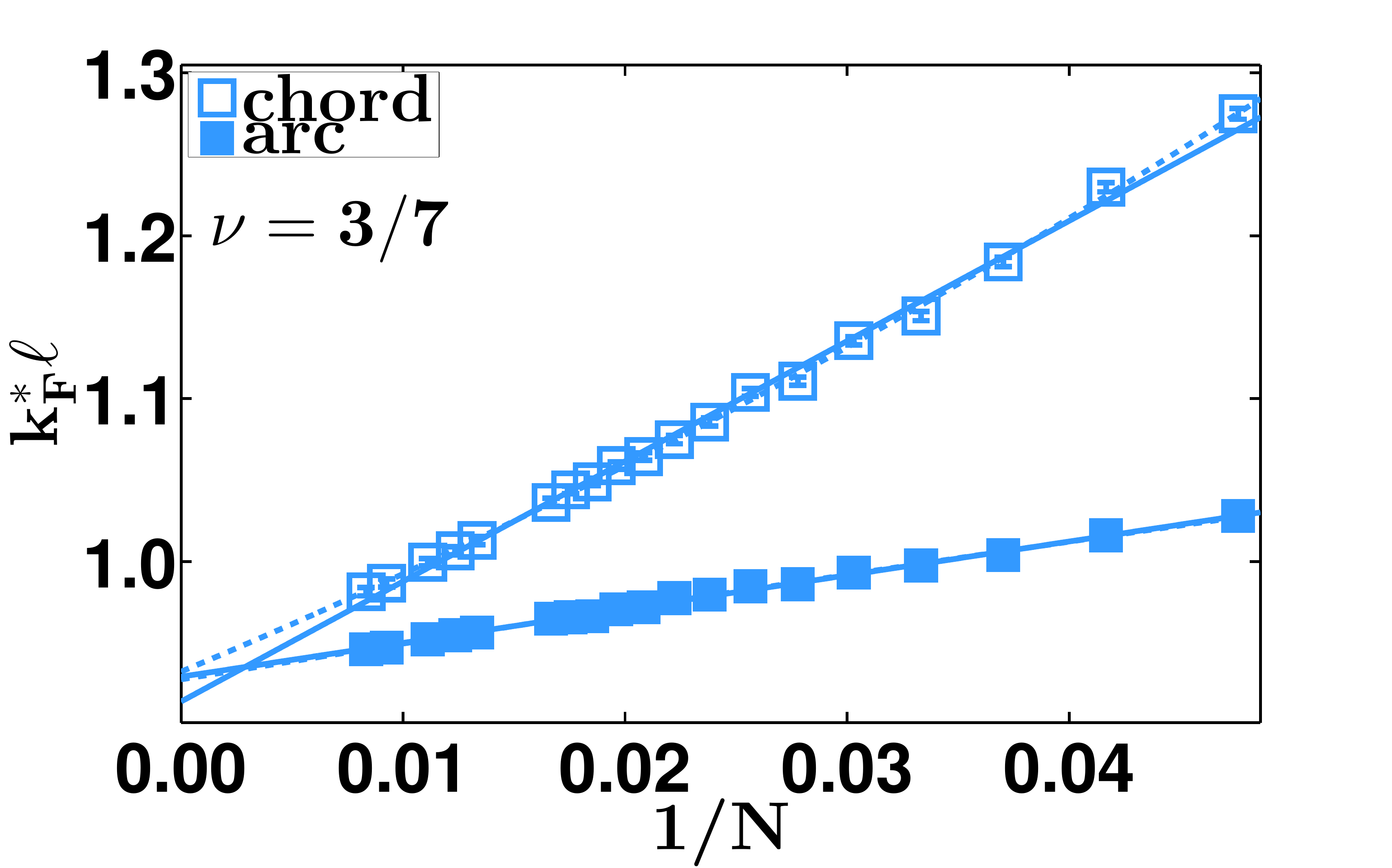}
\includegraphics[width=8cm,height=5cm]{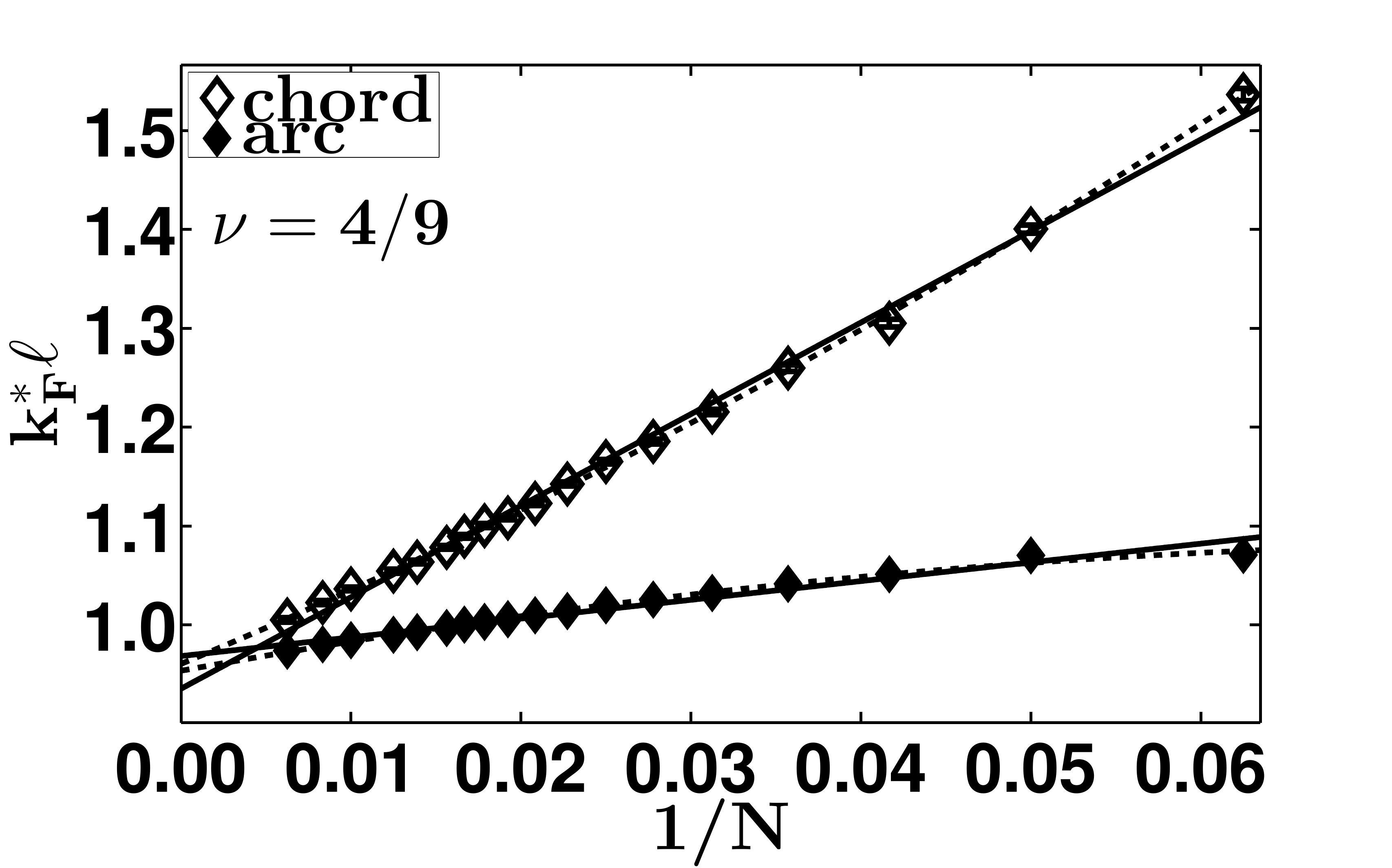}
\includegraphics[width=8cm,height=5cm]{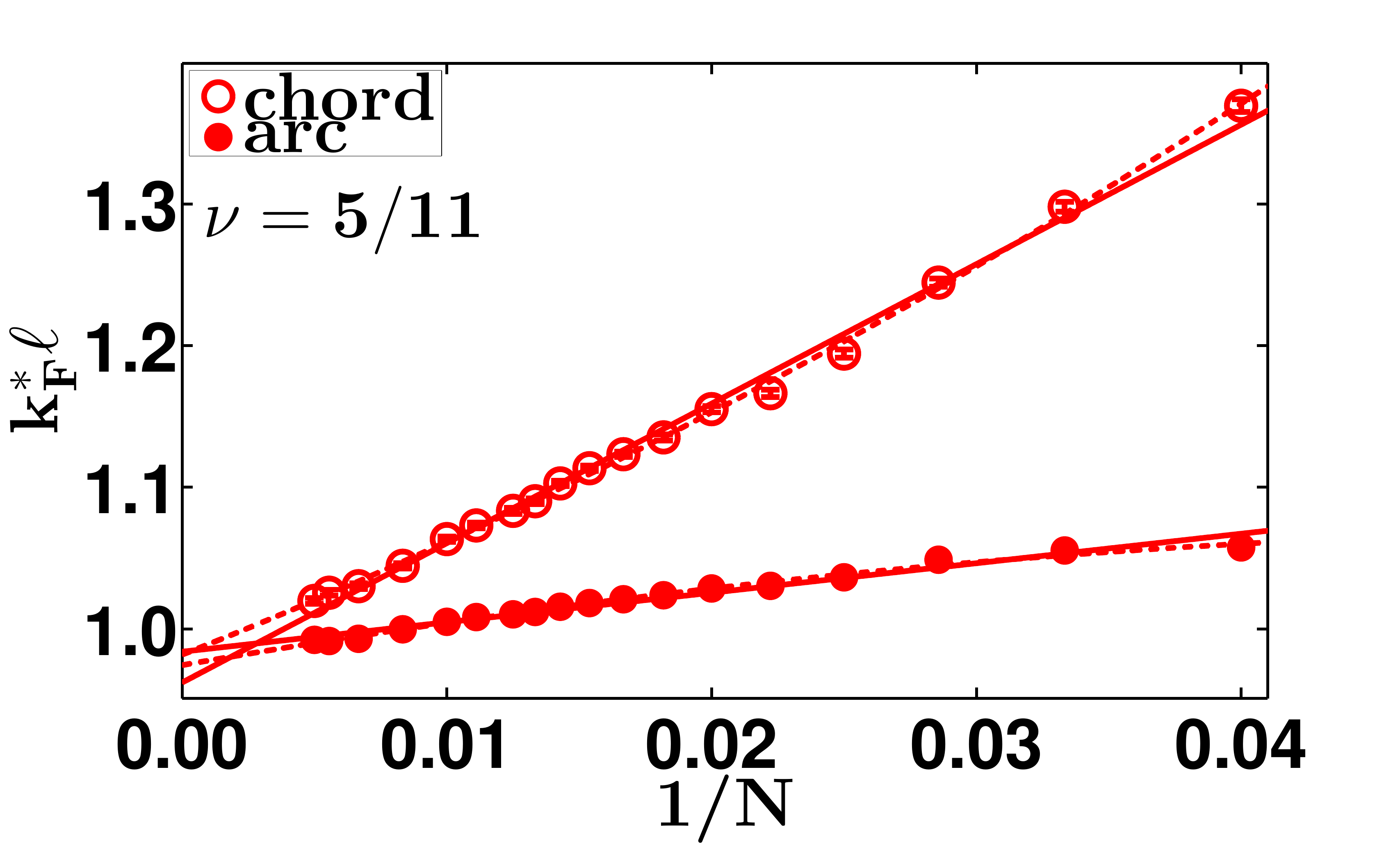}
\includegraphics[width=8cm,height=5cm]{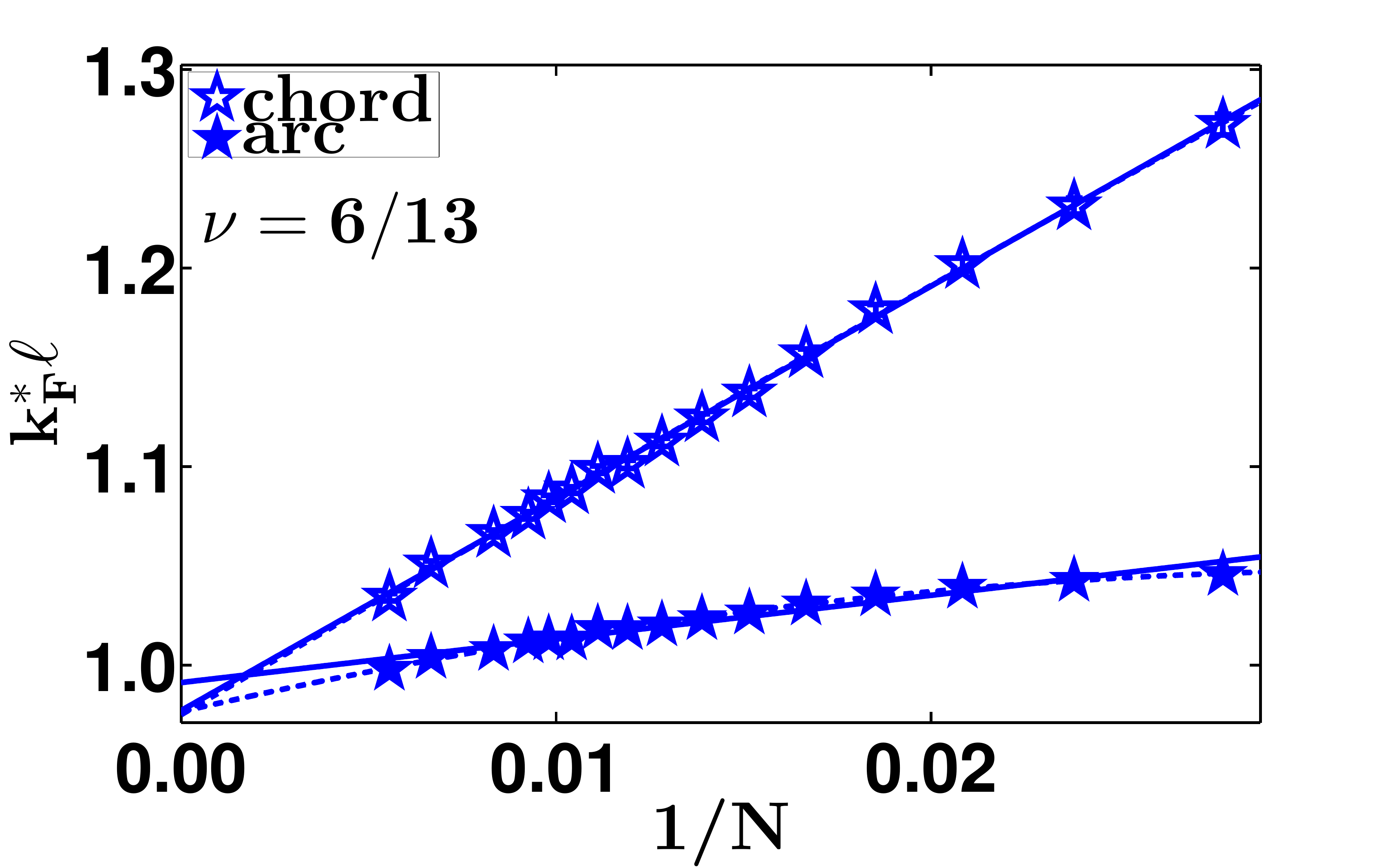}
\includegraphics[width=8cm,height=5cm]{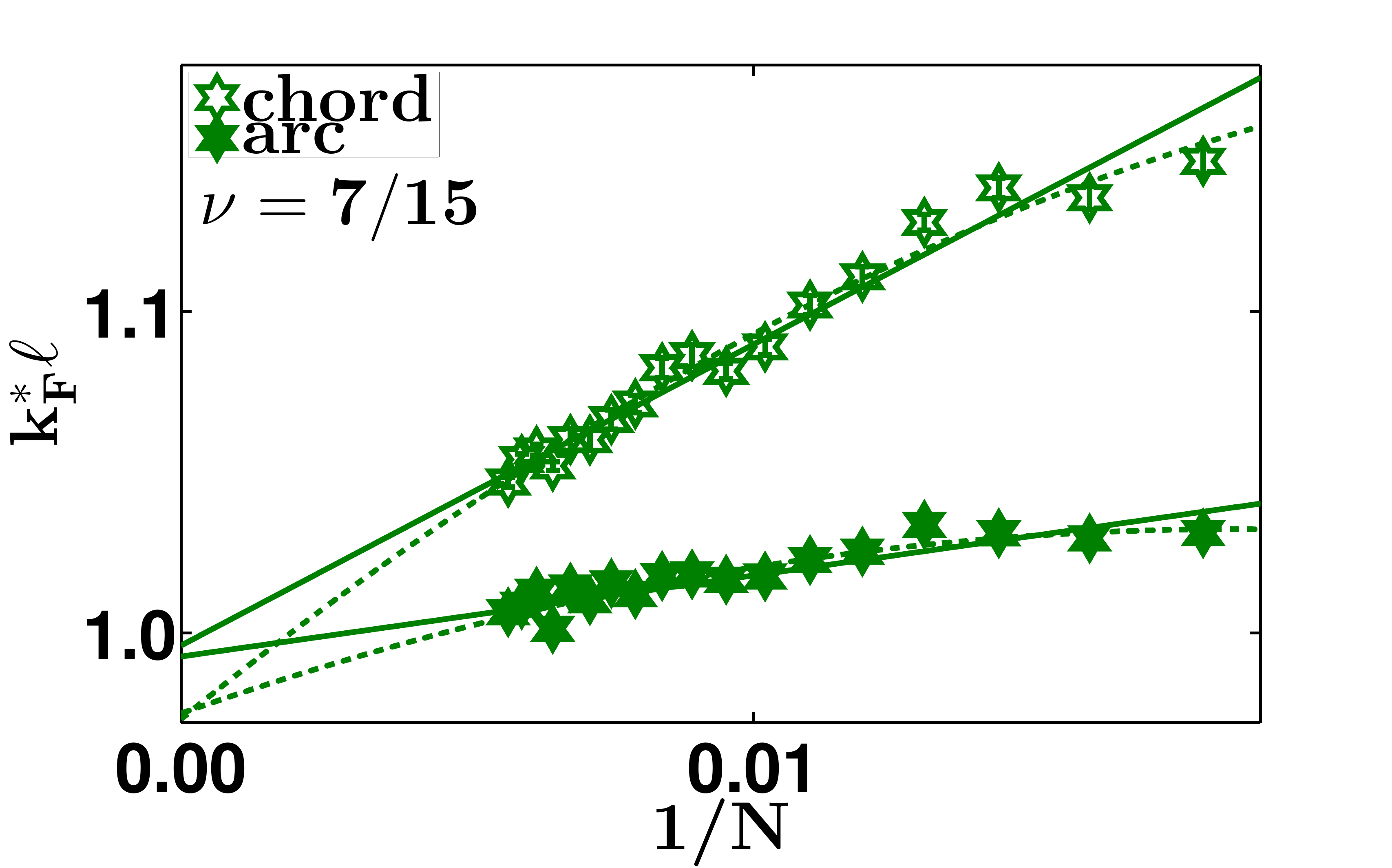}
\includegraphics[width=8cm,height=5cm]{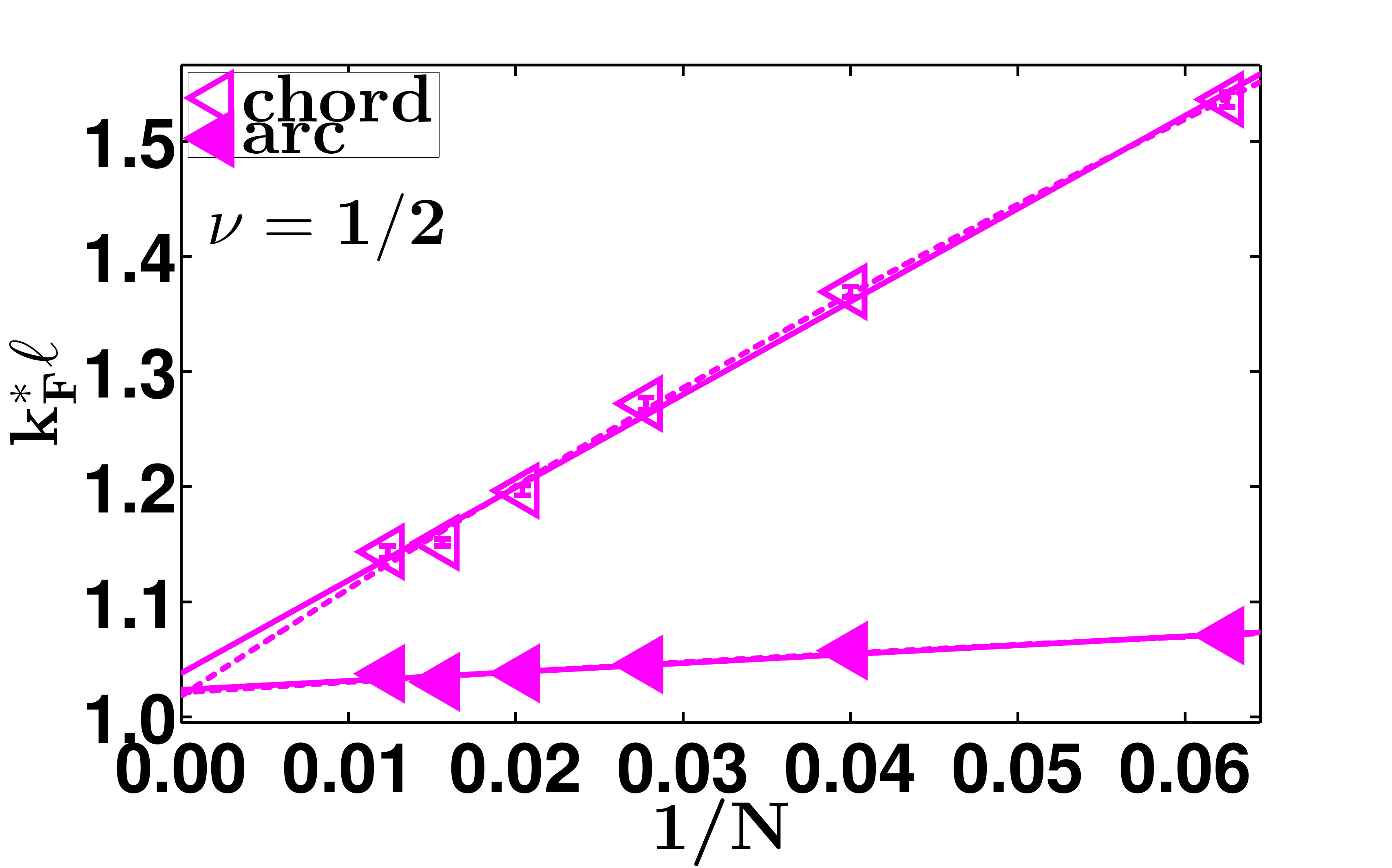}
\end{center}
\caption{Thermodynamic extrapolation of the Fermi wave vector $k_{\rm F}^{*}\ell$ for the projected Jain wave function at various filling factors. In this, as well as the following seven figures, the empty (filled) symbols correspond to the values obtained from the chord (arc) distance on the sphere and the thick (dashed) lines show linear (quadratic) fits to these values as a function of $1/N$.}
\label{kFsm}
\end{figure*}

\begin{figure*}[htpb]
\begin{center}
\includegraphics[width=8cm,height=5cm]{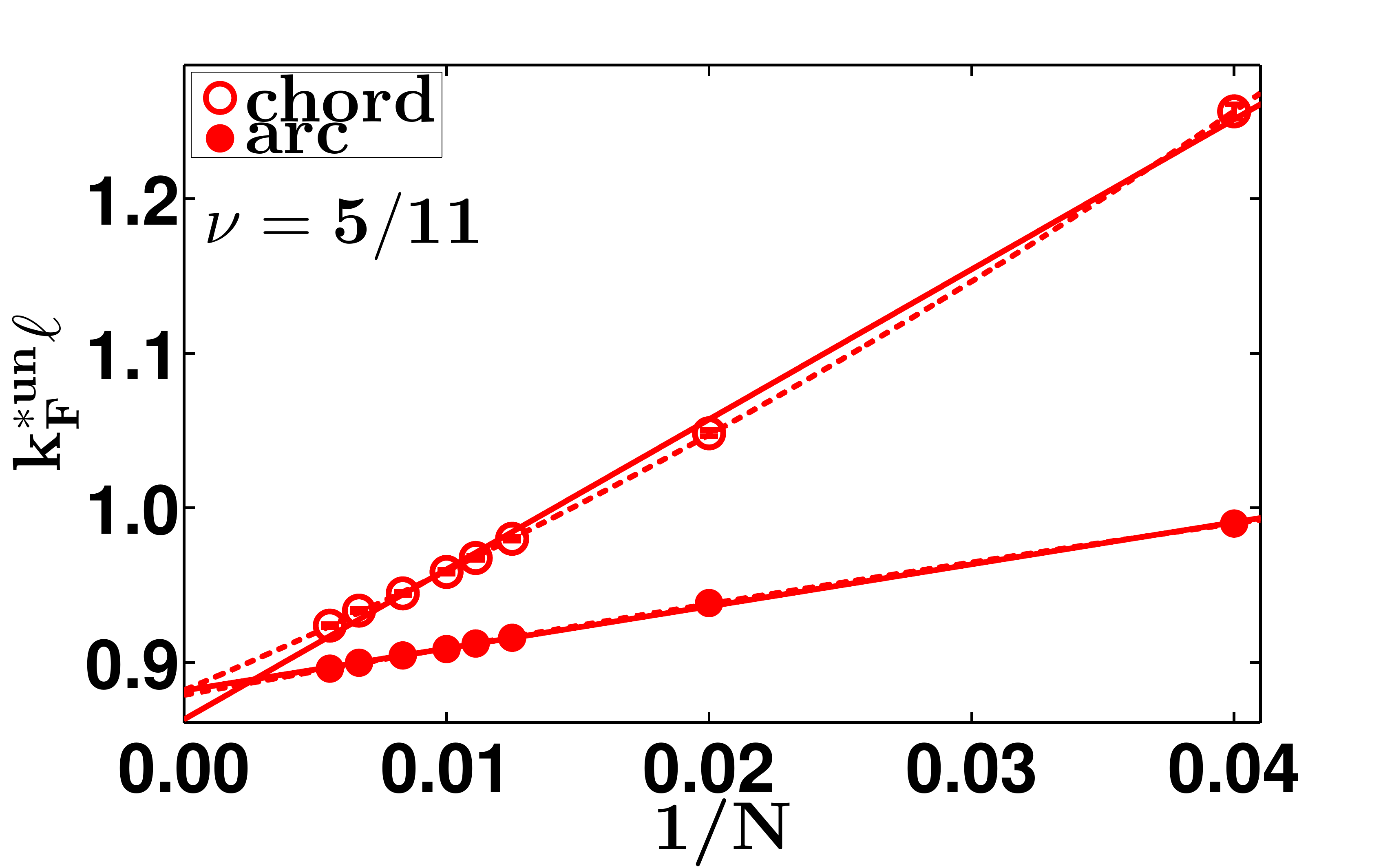}
\includegraphics[width=8cm,height=5cm]{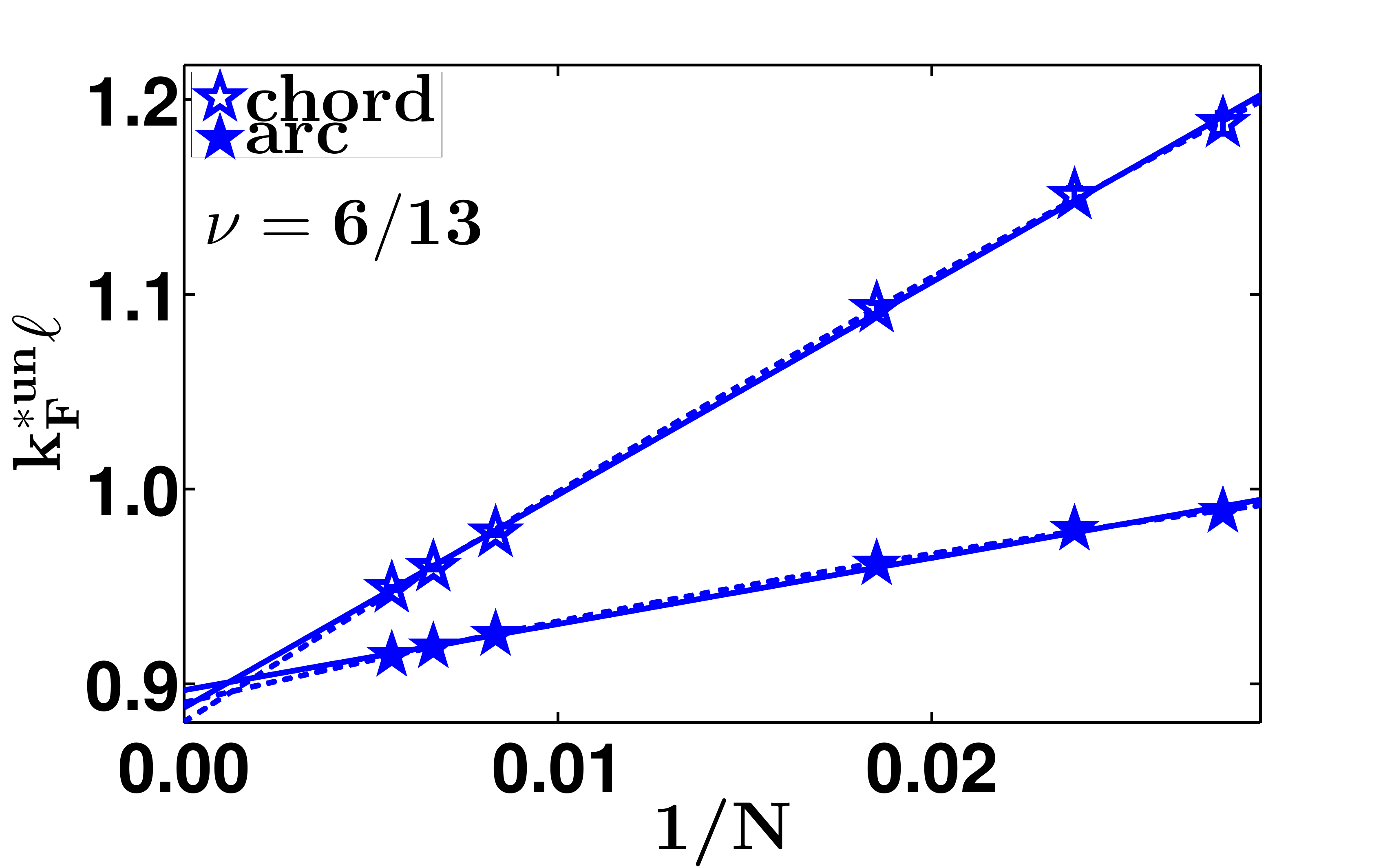}
\includegraphics[width=8cm,height=5cm]{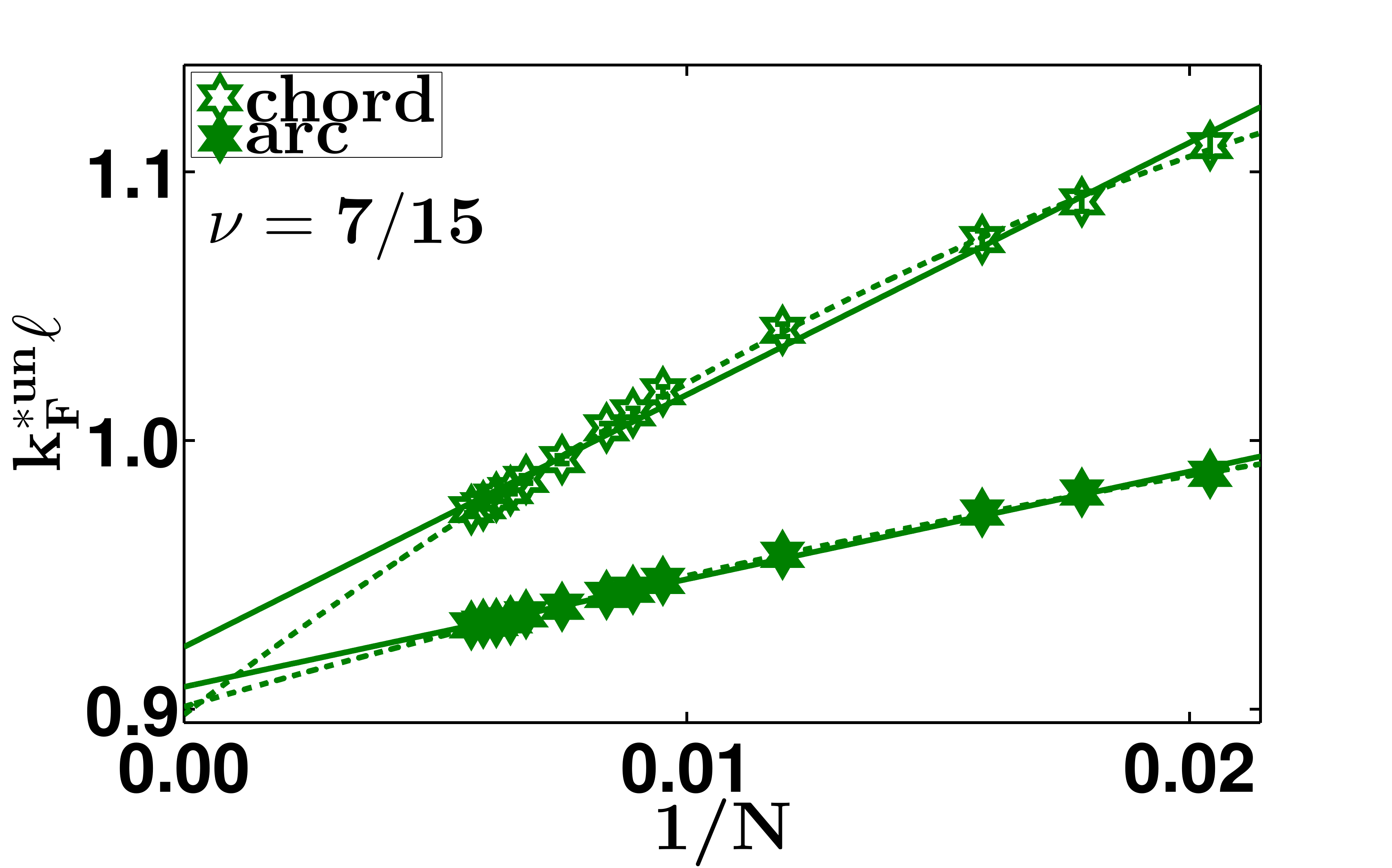}
\includegraphics[width=8cm,height=5cm]{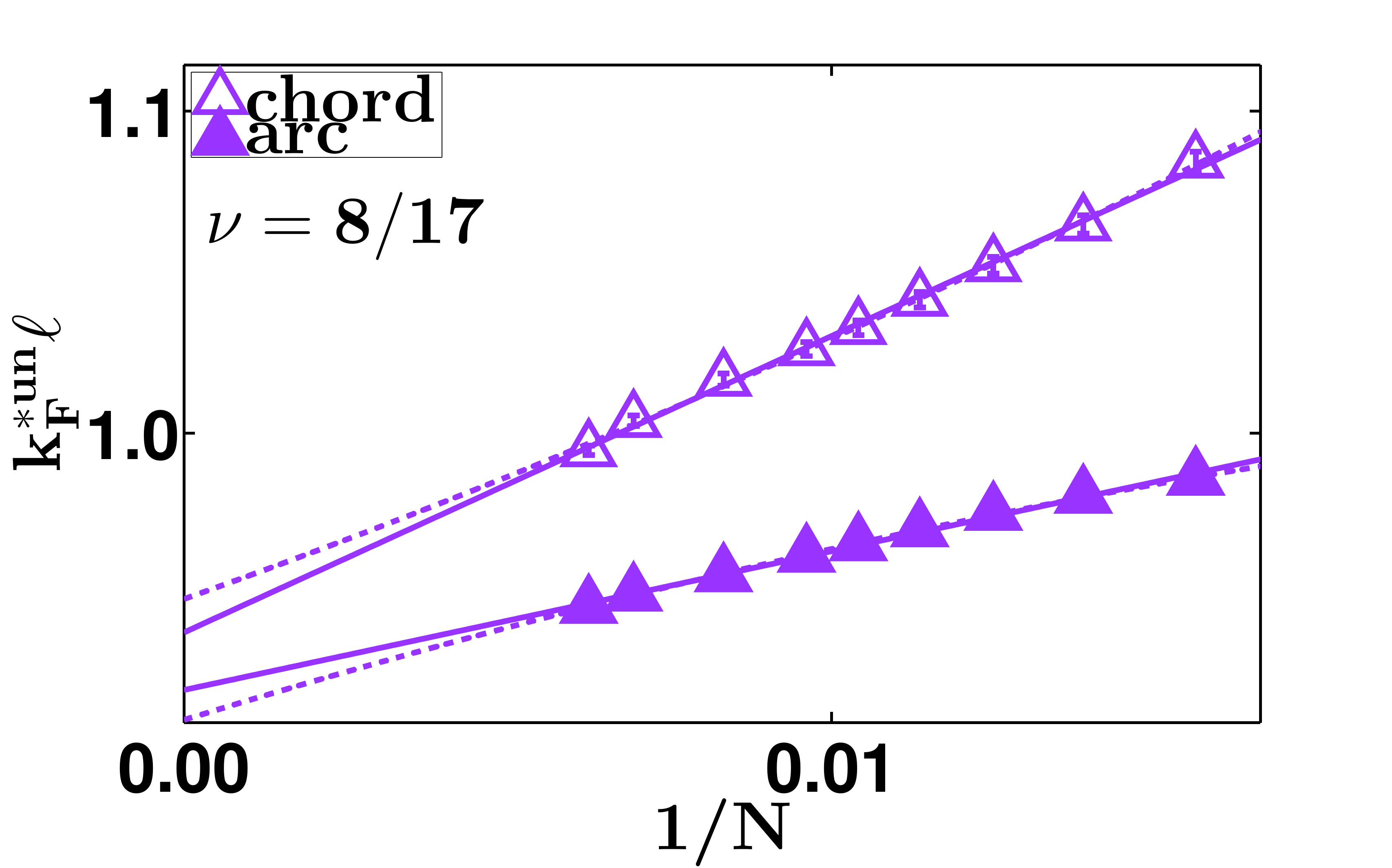}
\includegraphics[width=8cm,height=5cm]{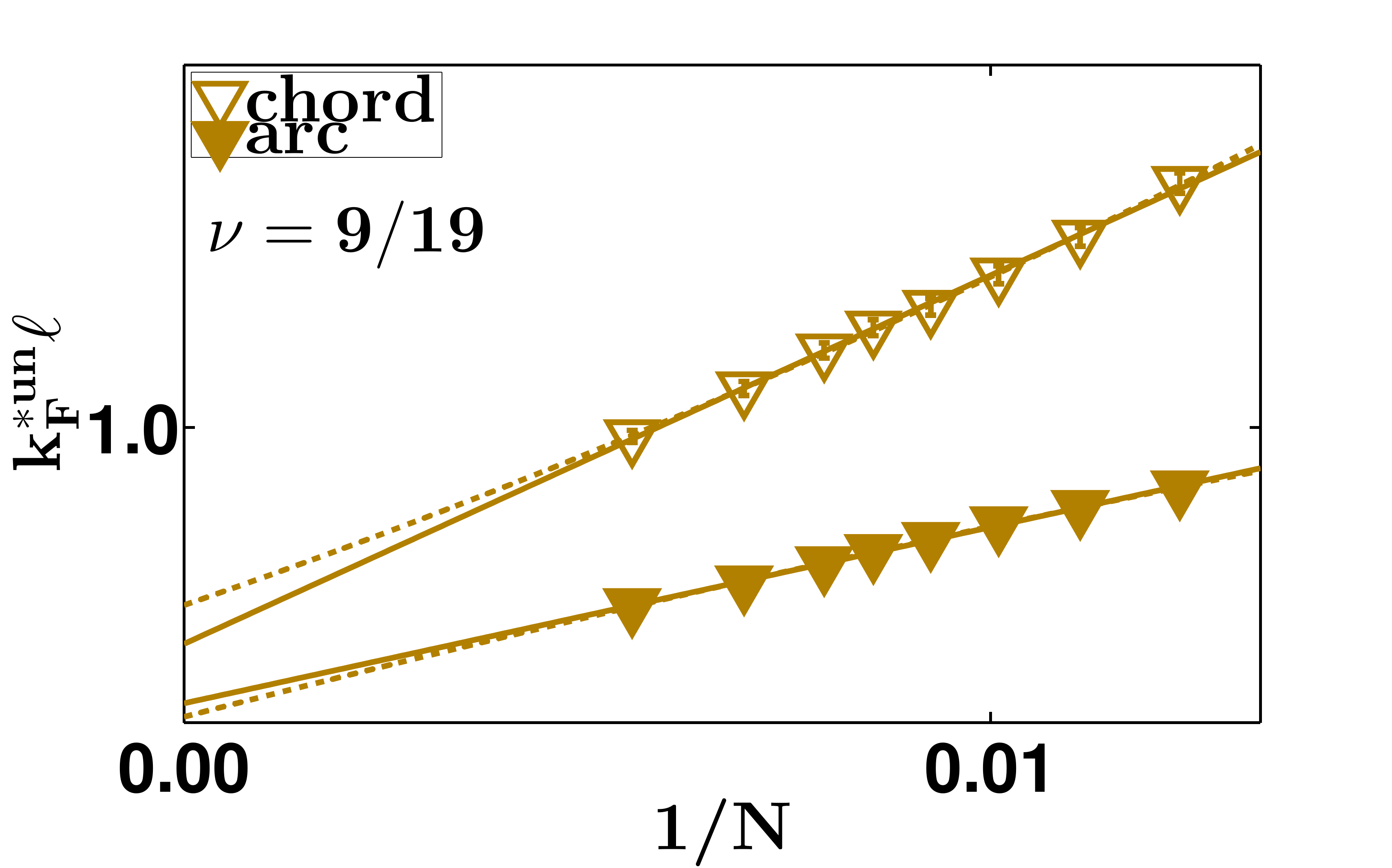}
\includegraphics[width=8cm,height=5cm]{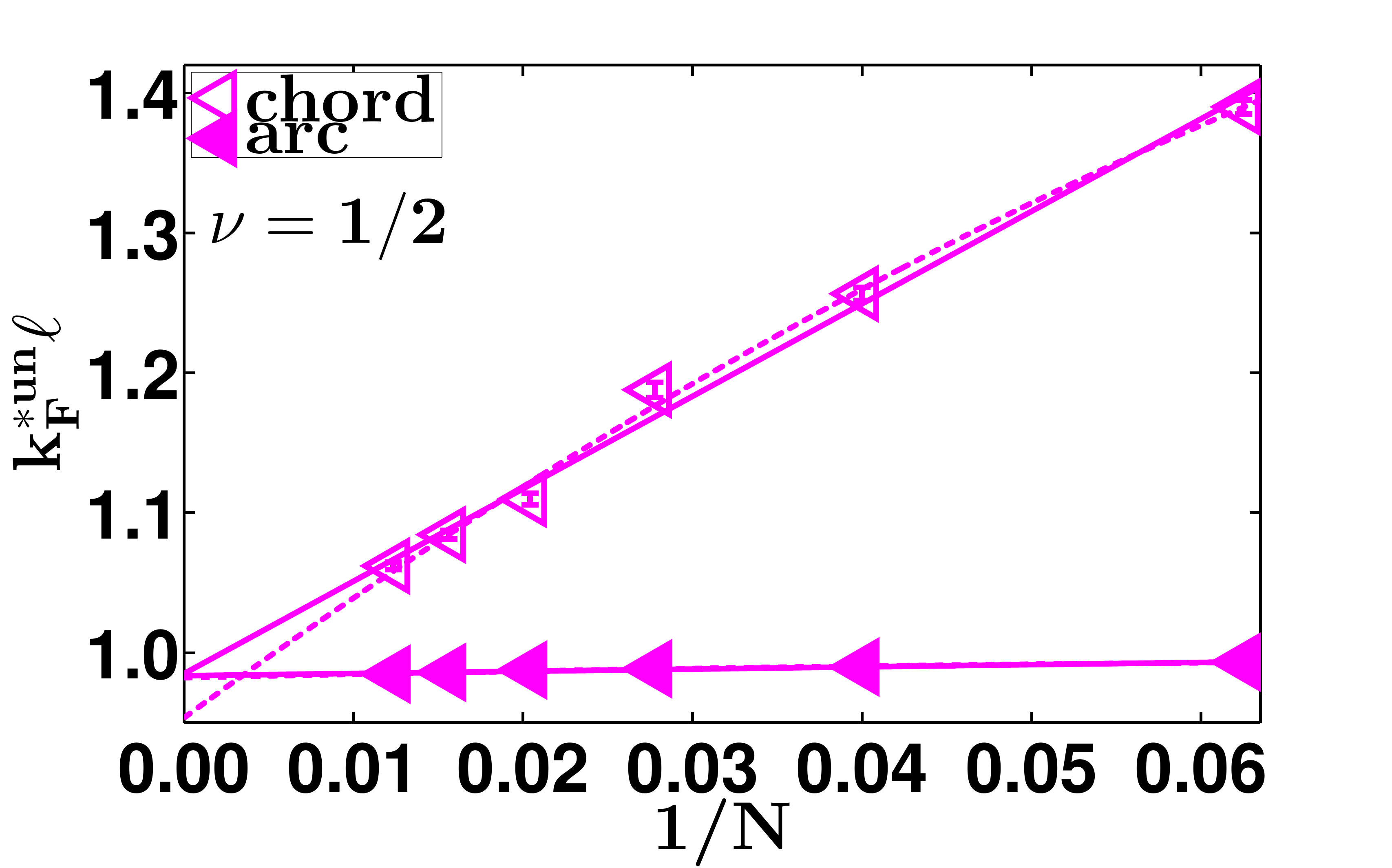}
\end{center}
\caption{Thermodynamic extrapolation of the Fermi wave vector $k_{\rm F}^{\rm *un}\ell$ for the {\em unprojected}  wave functions $\Psi^{\rm un}_{n/(2n\pm 1)}$ defined in the main text  at various filling factors of the form $\nu=n/(2n+1)$.}
\label{kFunsm}
\end{figure*}

\begin{figure*}[htpb]
\begin{center}
\includegraphics[width=8cm,height=5cm]{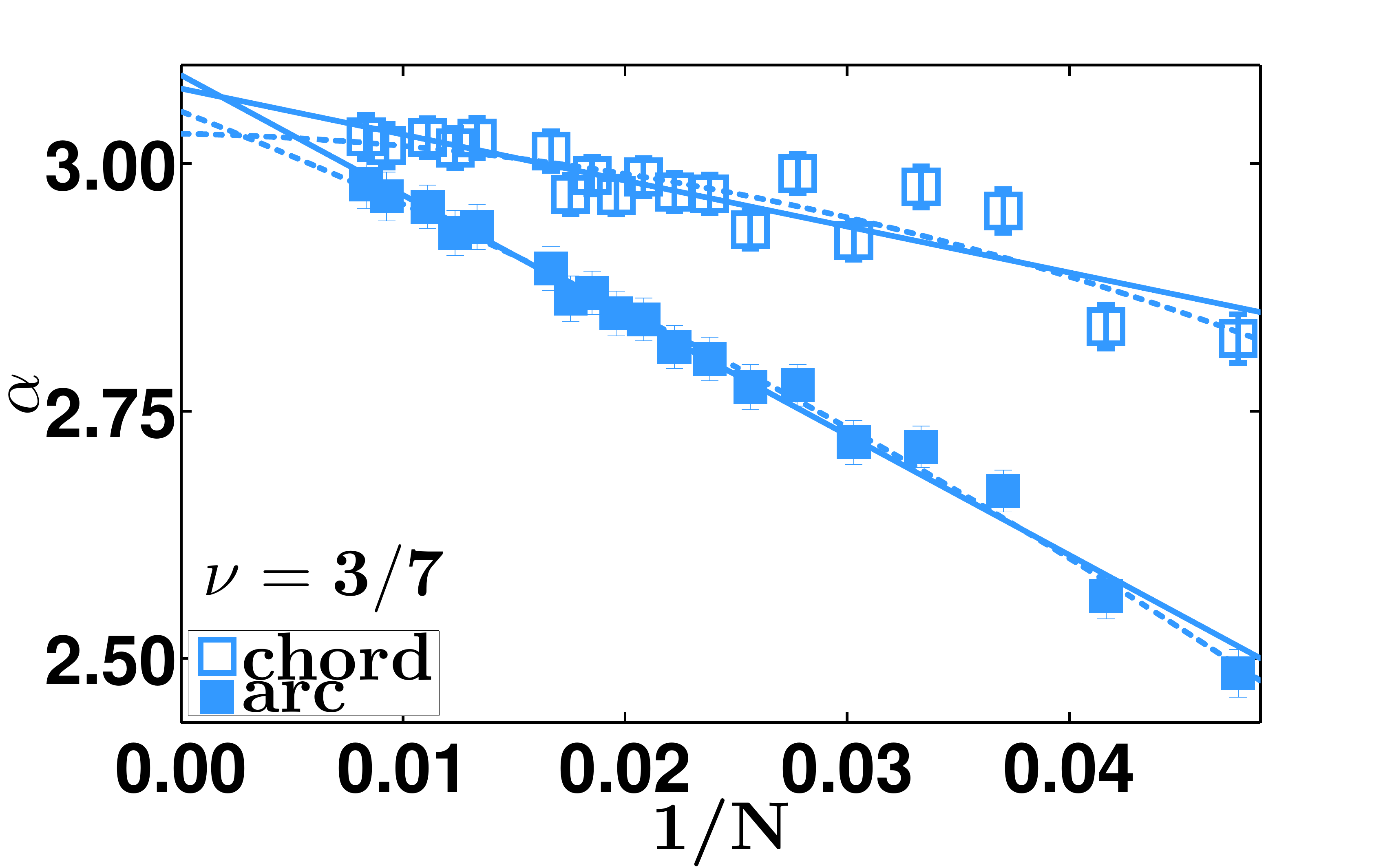}
\includegraphics[width=8cm,height=5cm]{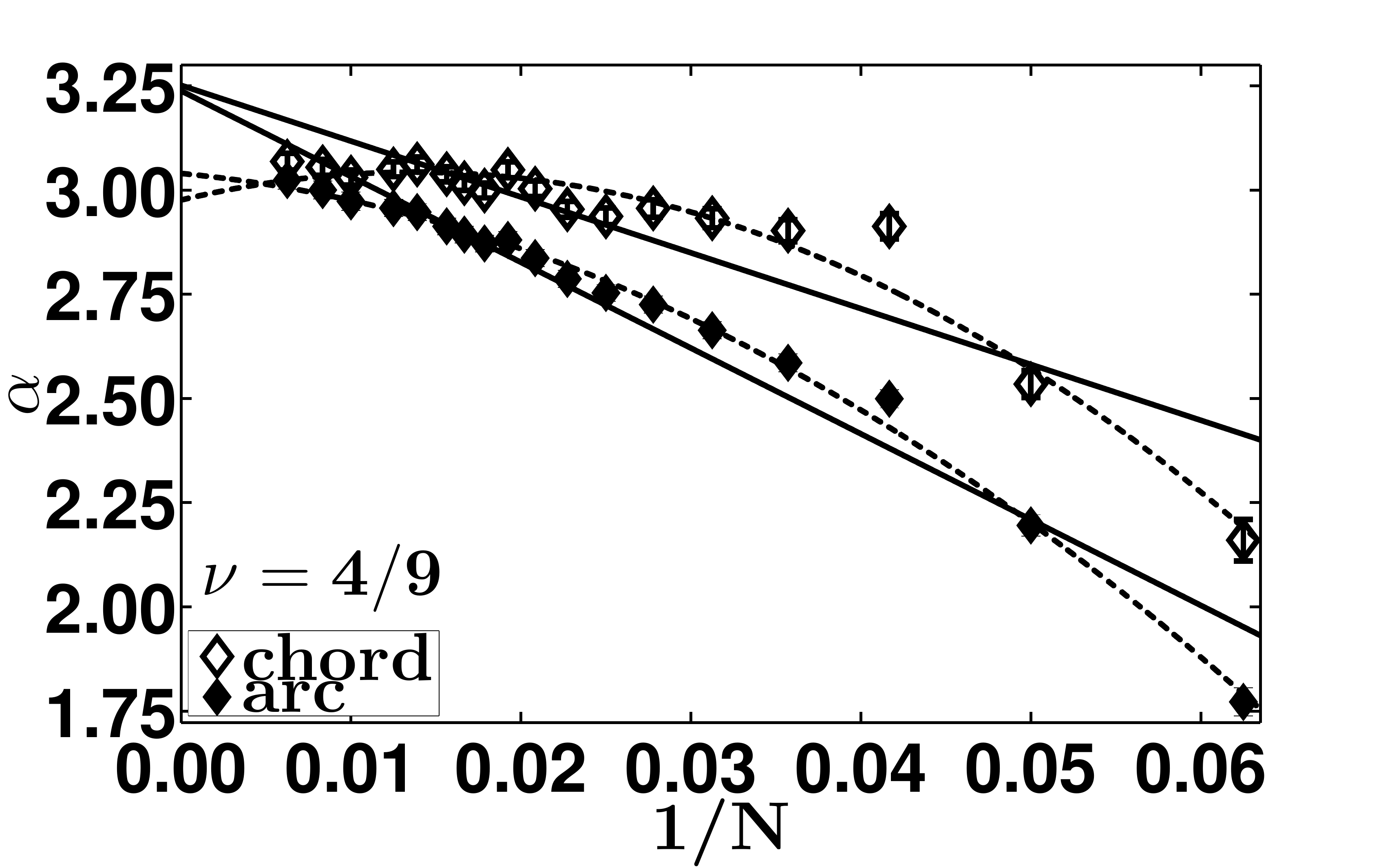}
\includegraphics[width=8cm,height=5cm]{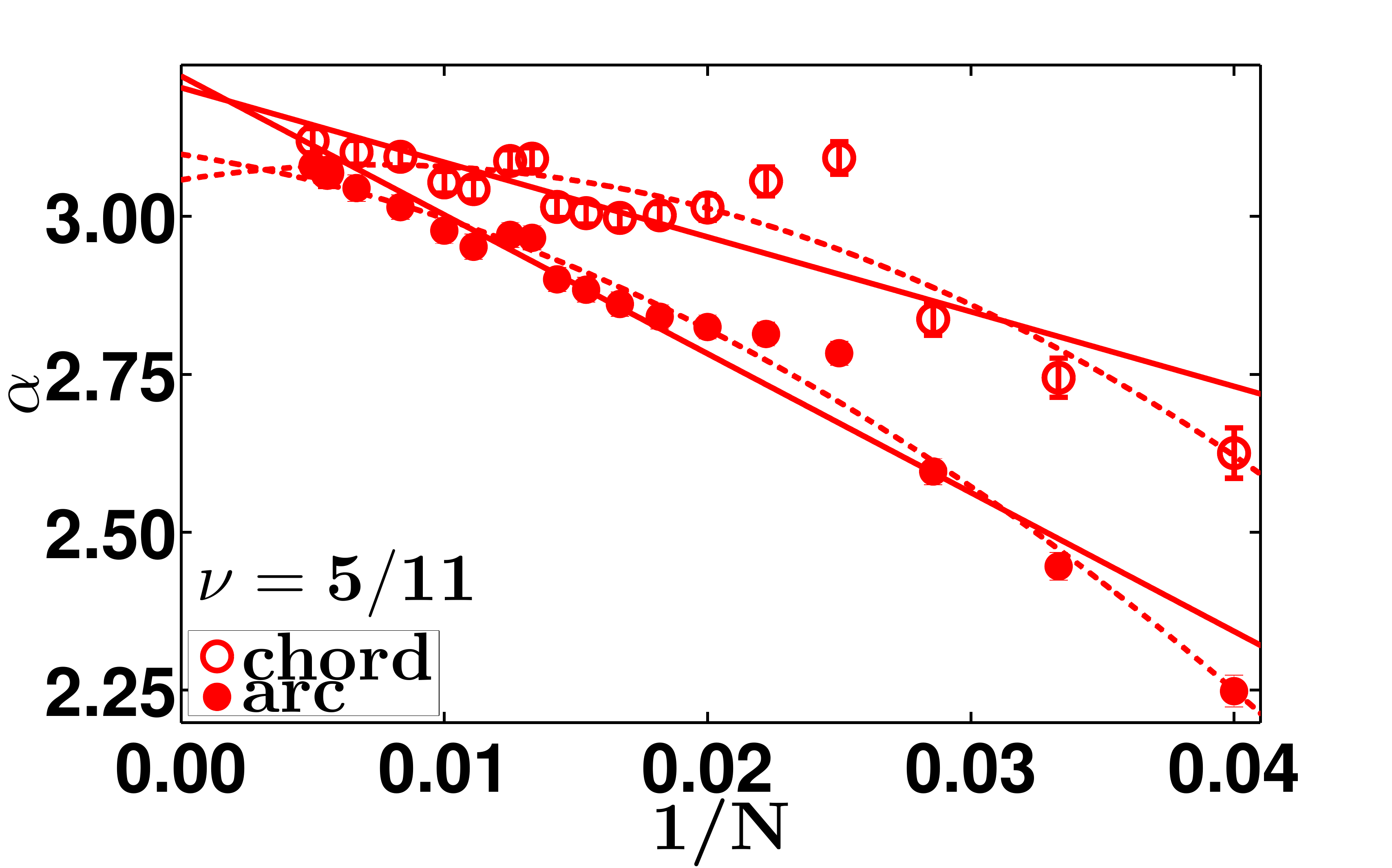}
\includegraphics[width=8cm,height=5cm]{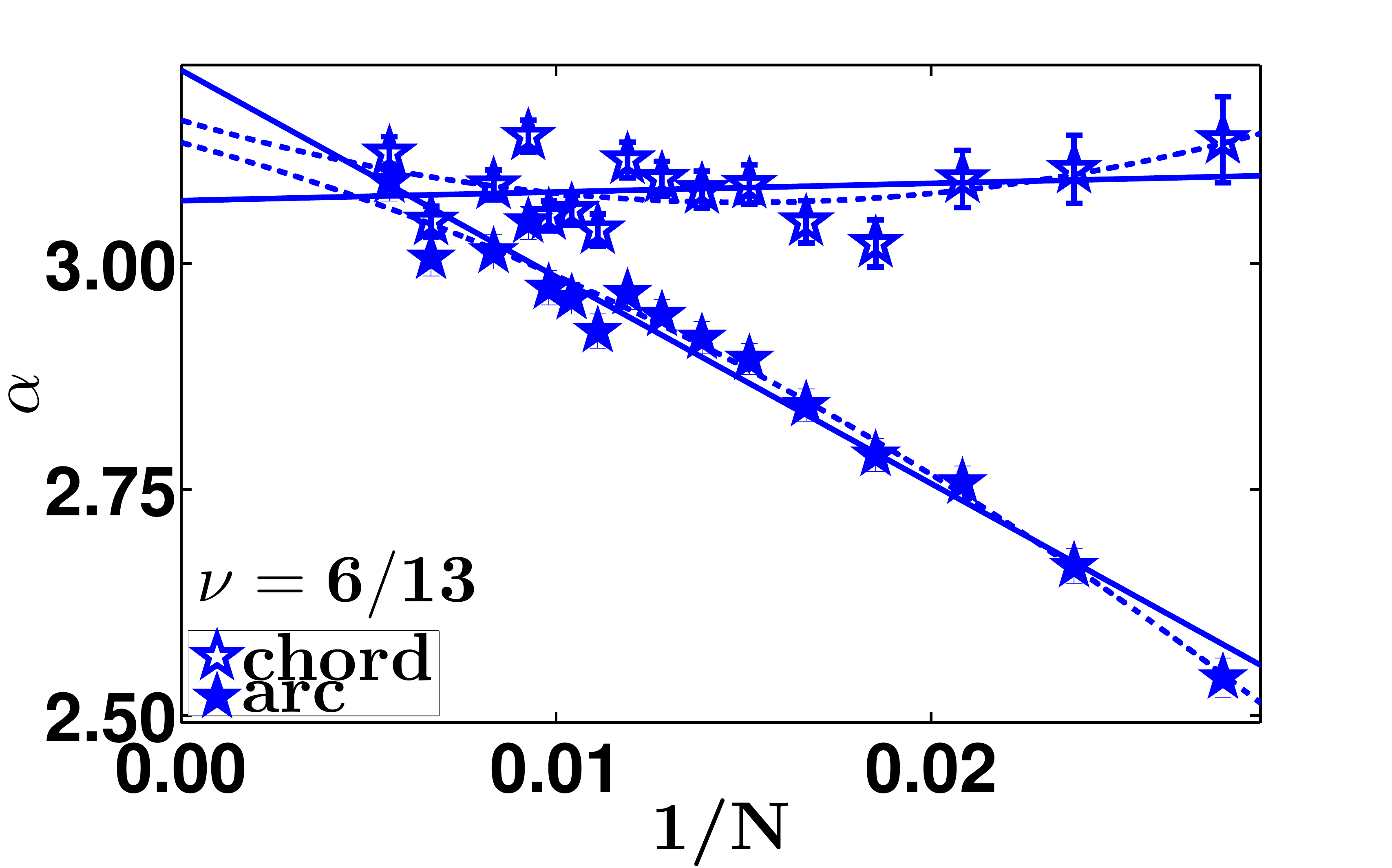}
\includegraphics[width=8cm,height=5cm]{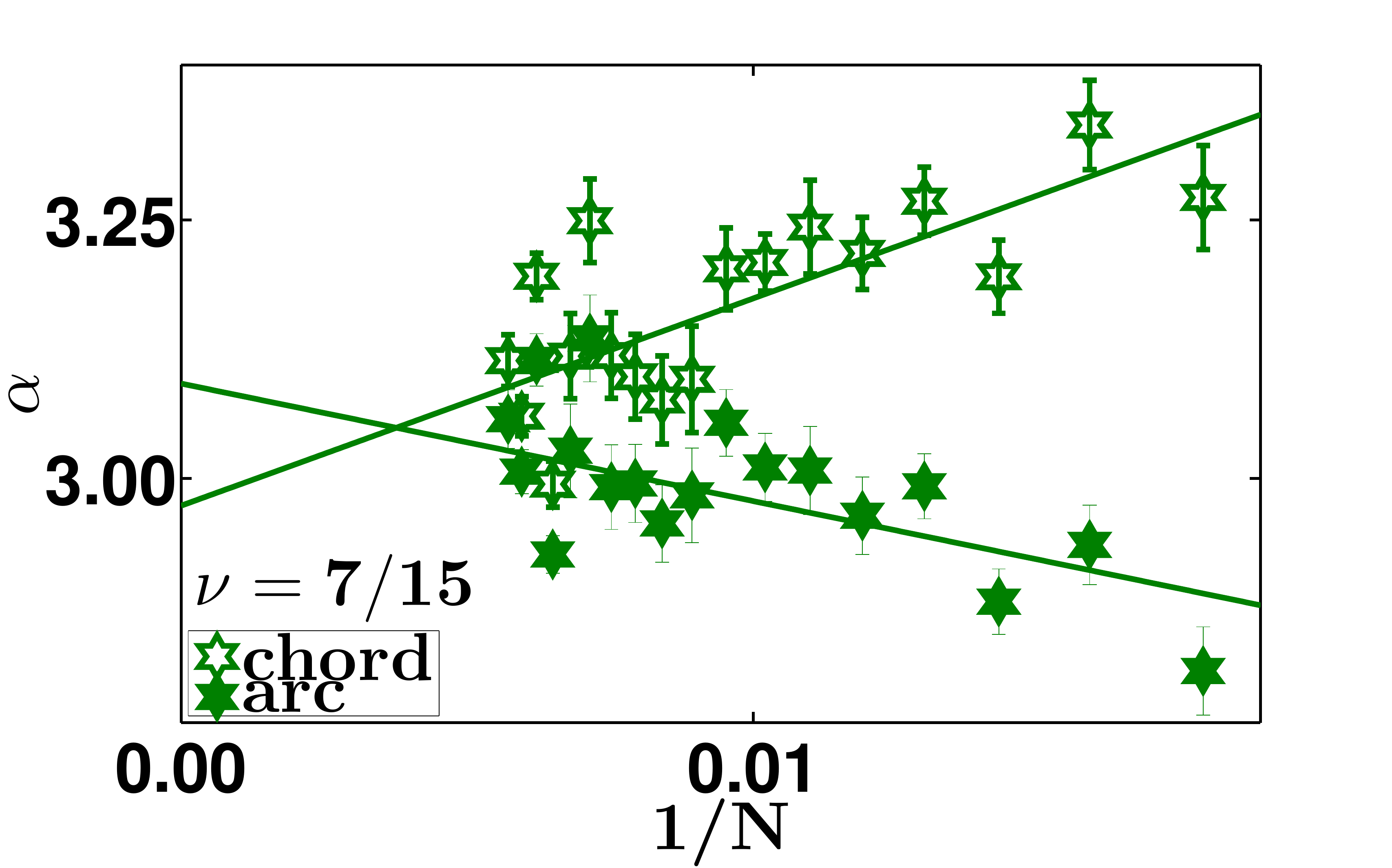}
\includegraphics[width=8cm,height=5cm]{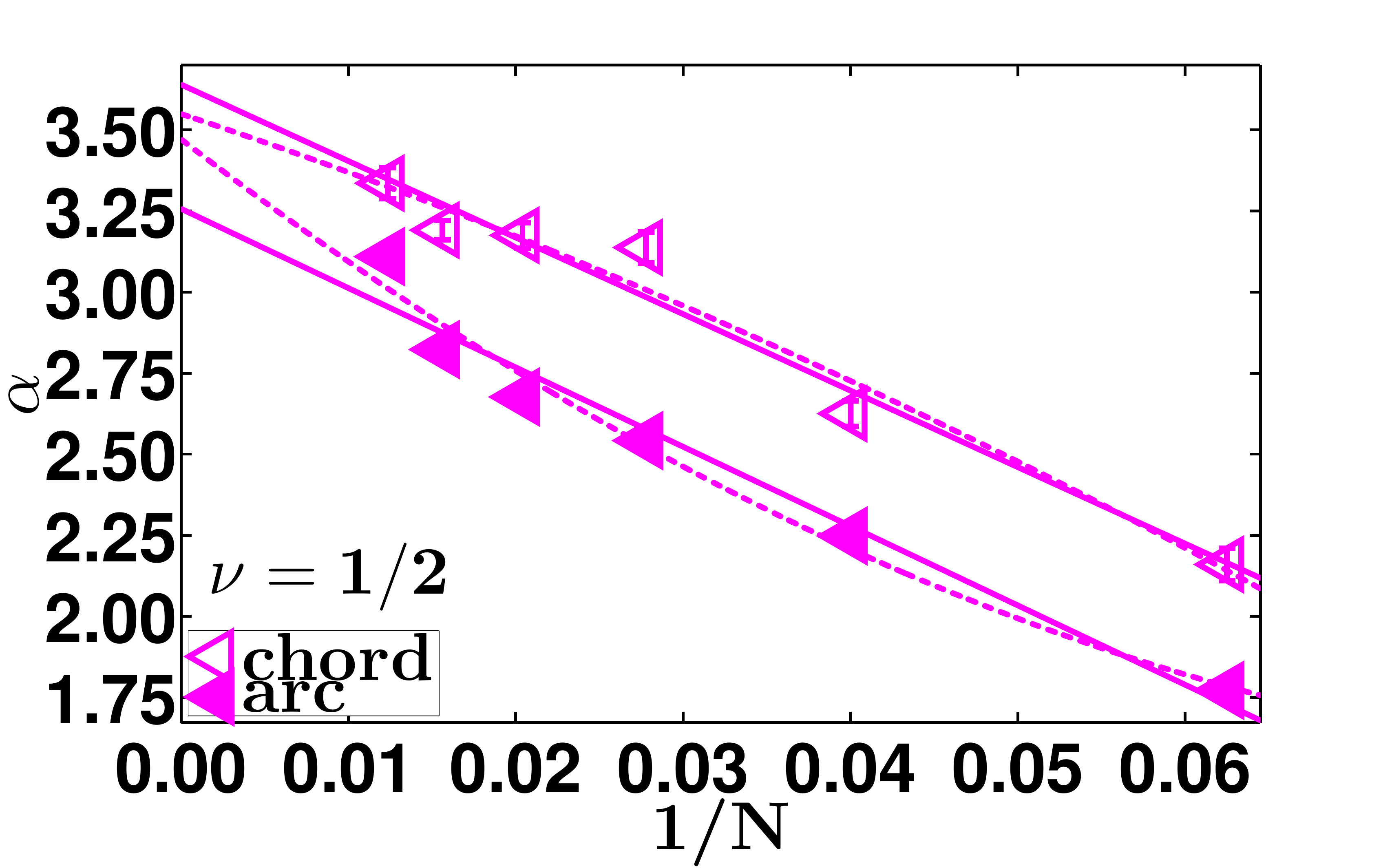}
\includegraphics[width=8cm,height=5cm]{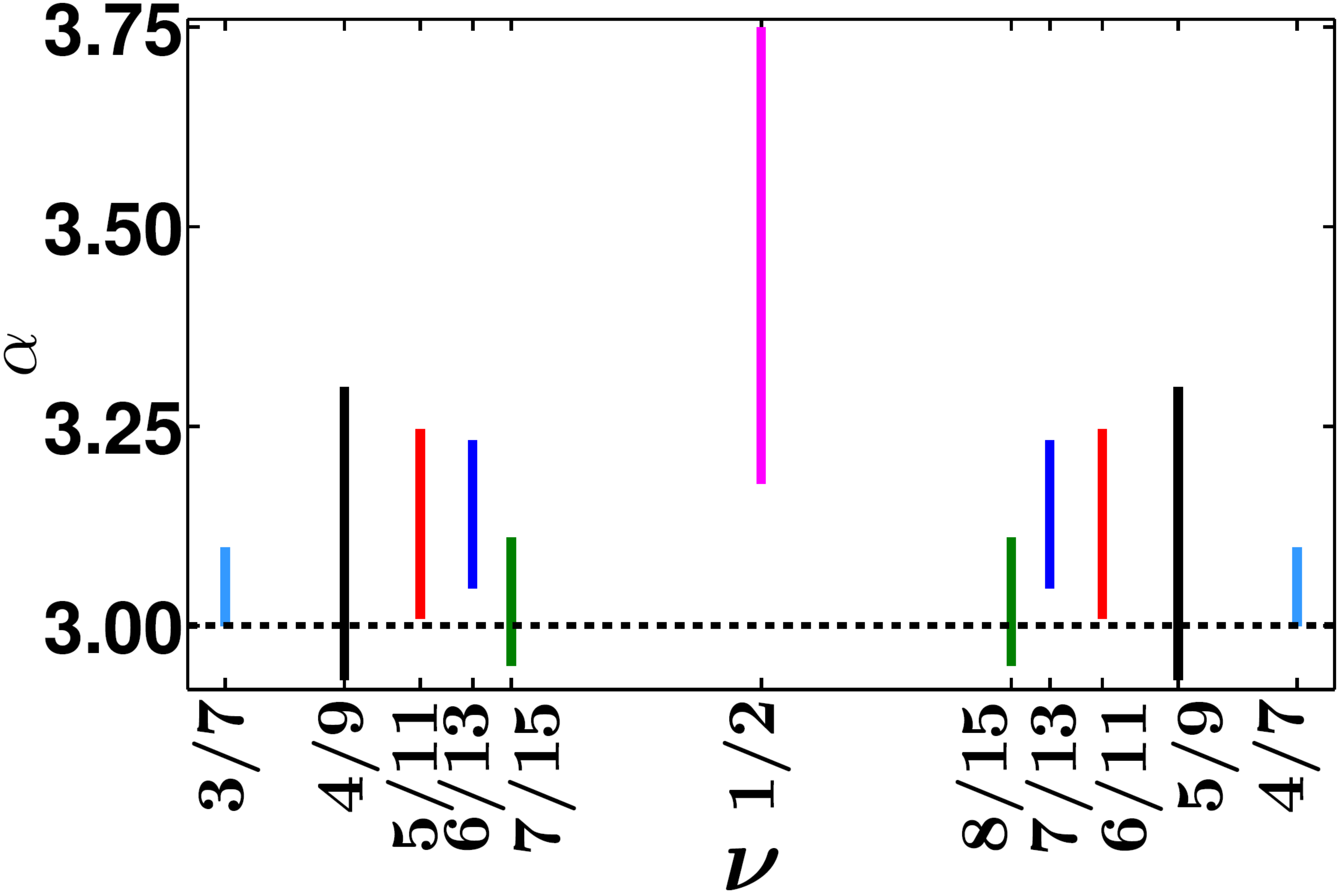}
\end{center}
\caption{Thermodynamic extrapolation of the power law exponent $\alpha$ (defined in Eq. 1 of the main text) evaluated using the projected Jain wave functions at $n/(2n\pm 1)$. The bottom-most panel shows thermodynamic values of $\alpha$ as a function of the filling factor $\nu$.}
\label{alpha_sm}
\end{figure*}

\begin{figure*}[htpb]
\begin{center}
\includegraphics[width=8cm,height=5cm]{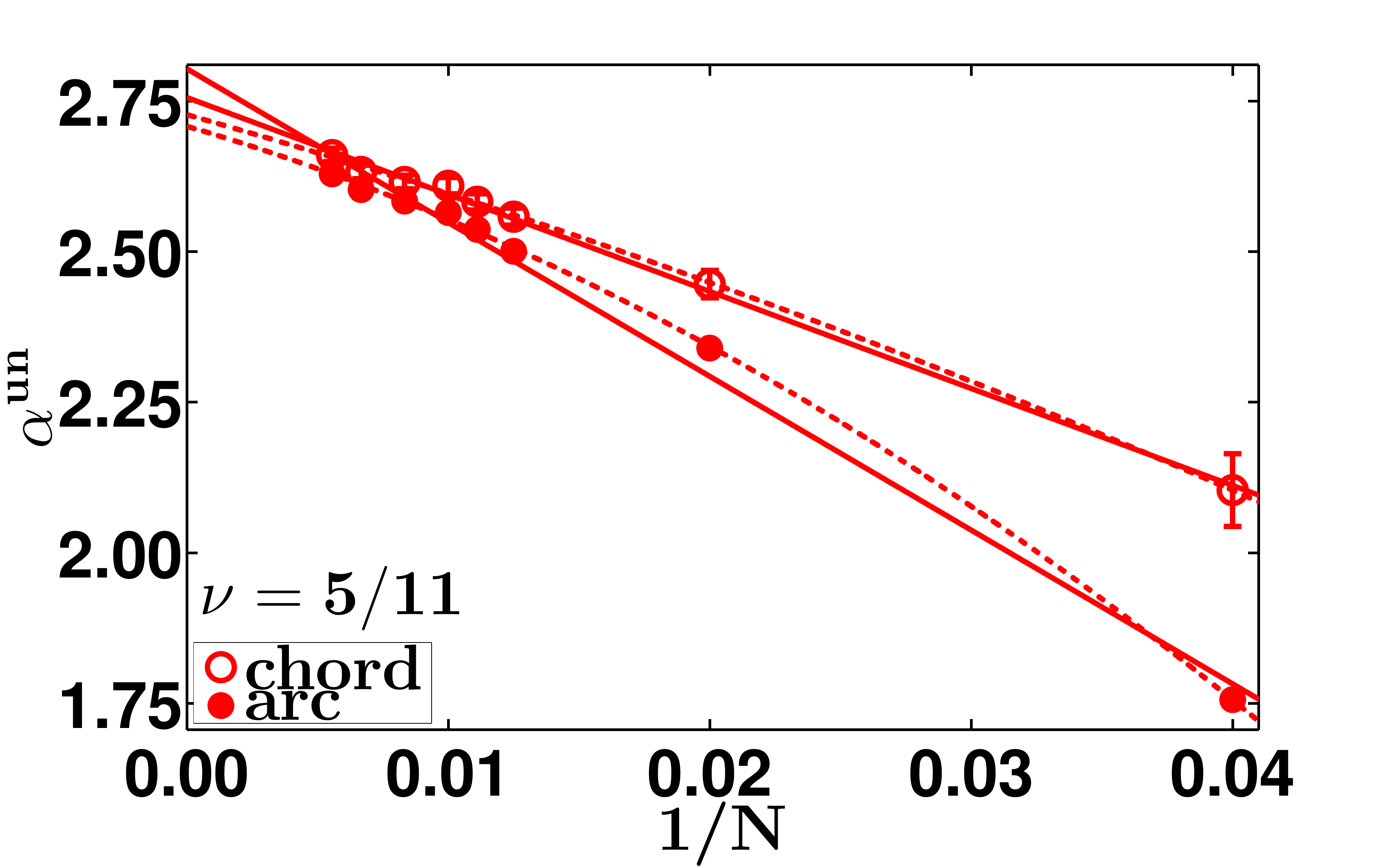}
\includegraphics[width=8cm,height=5cm]{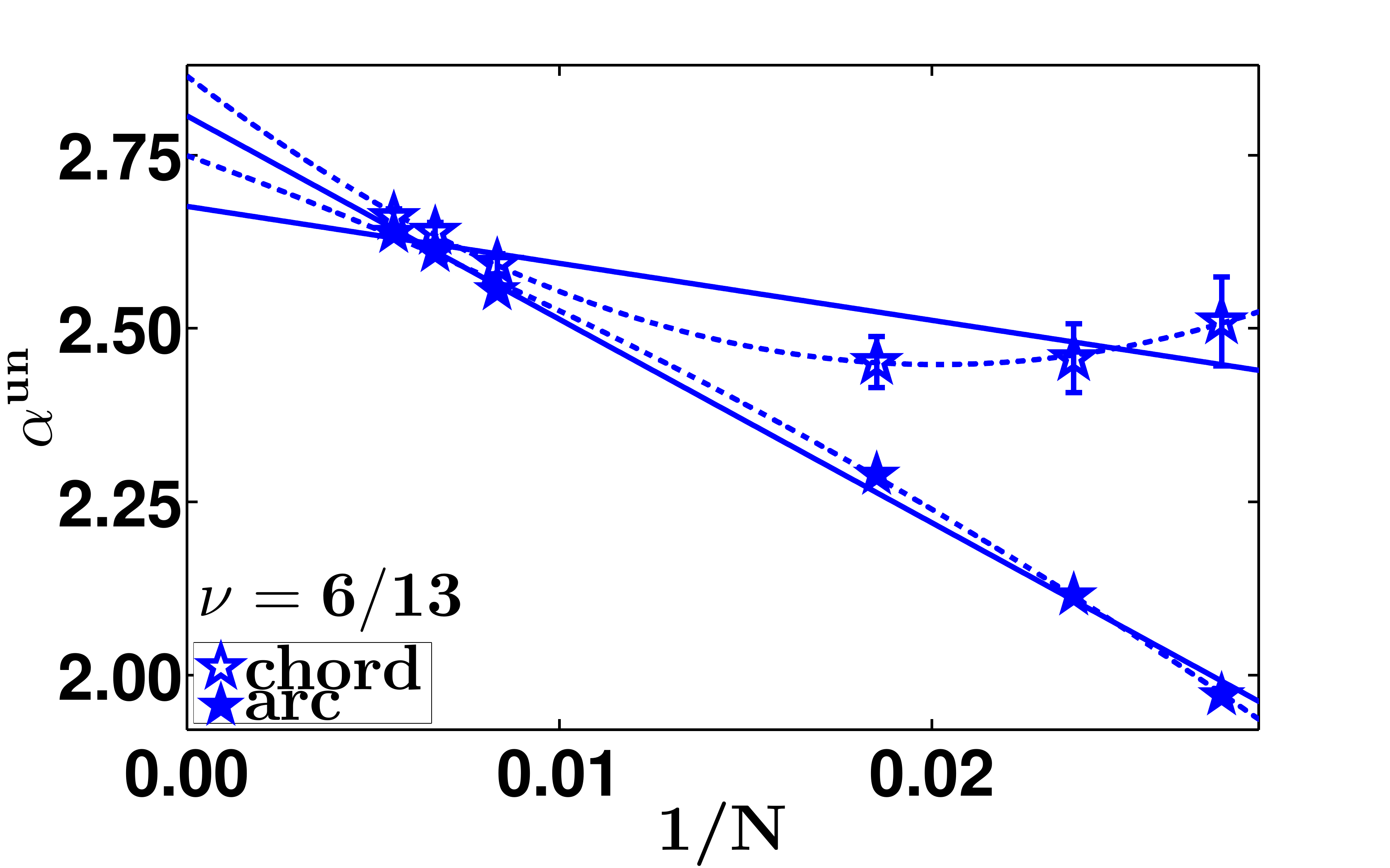}
\includegraphics[width=8cm,height=5cm]{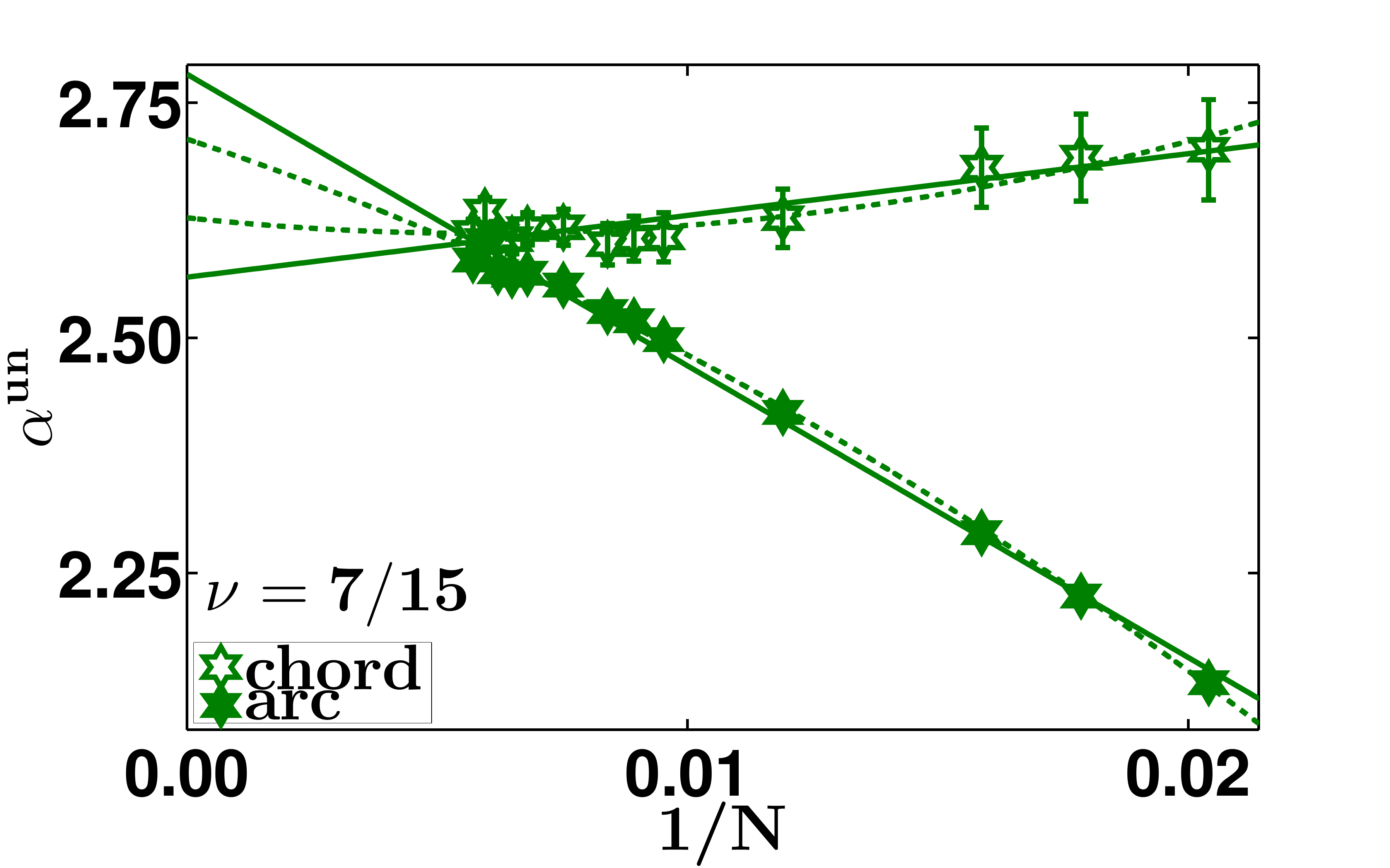}
\includegraphics[width=8cm,height=5cm]{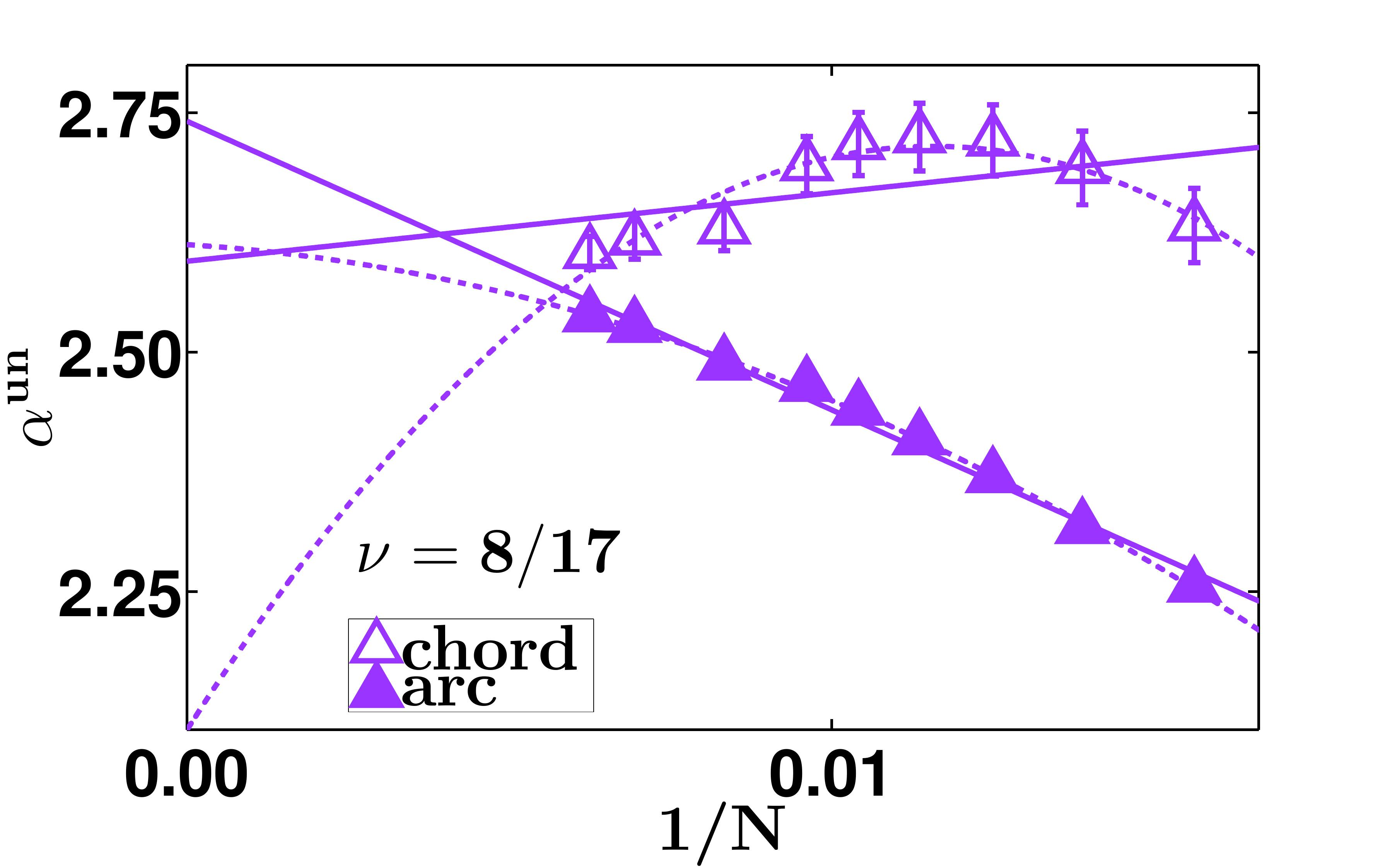}
\includegraphics[width=8cm,height=5cm]{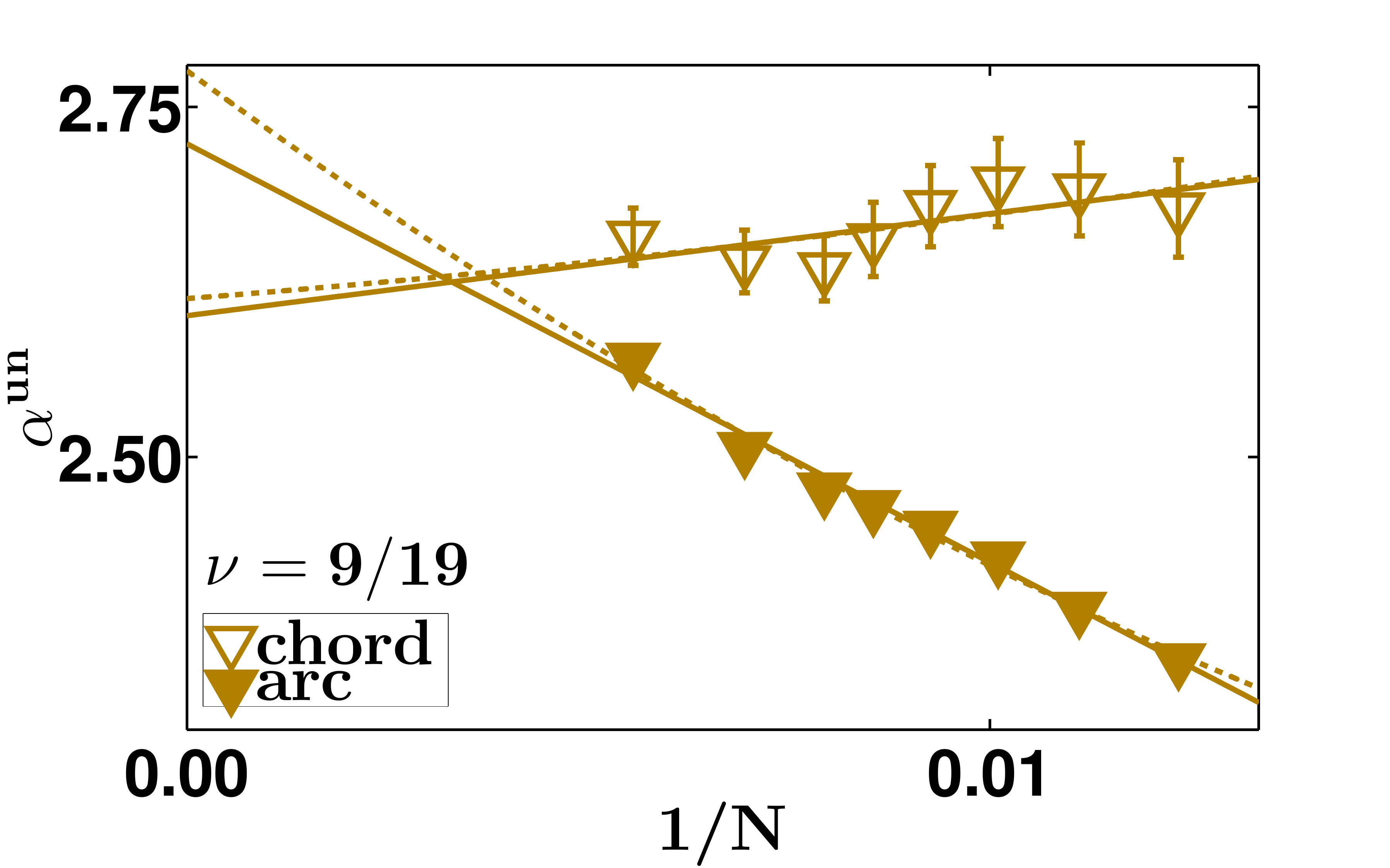}
\includegraphics[width=8cm,height=5cm]{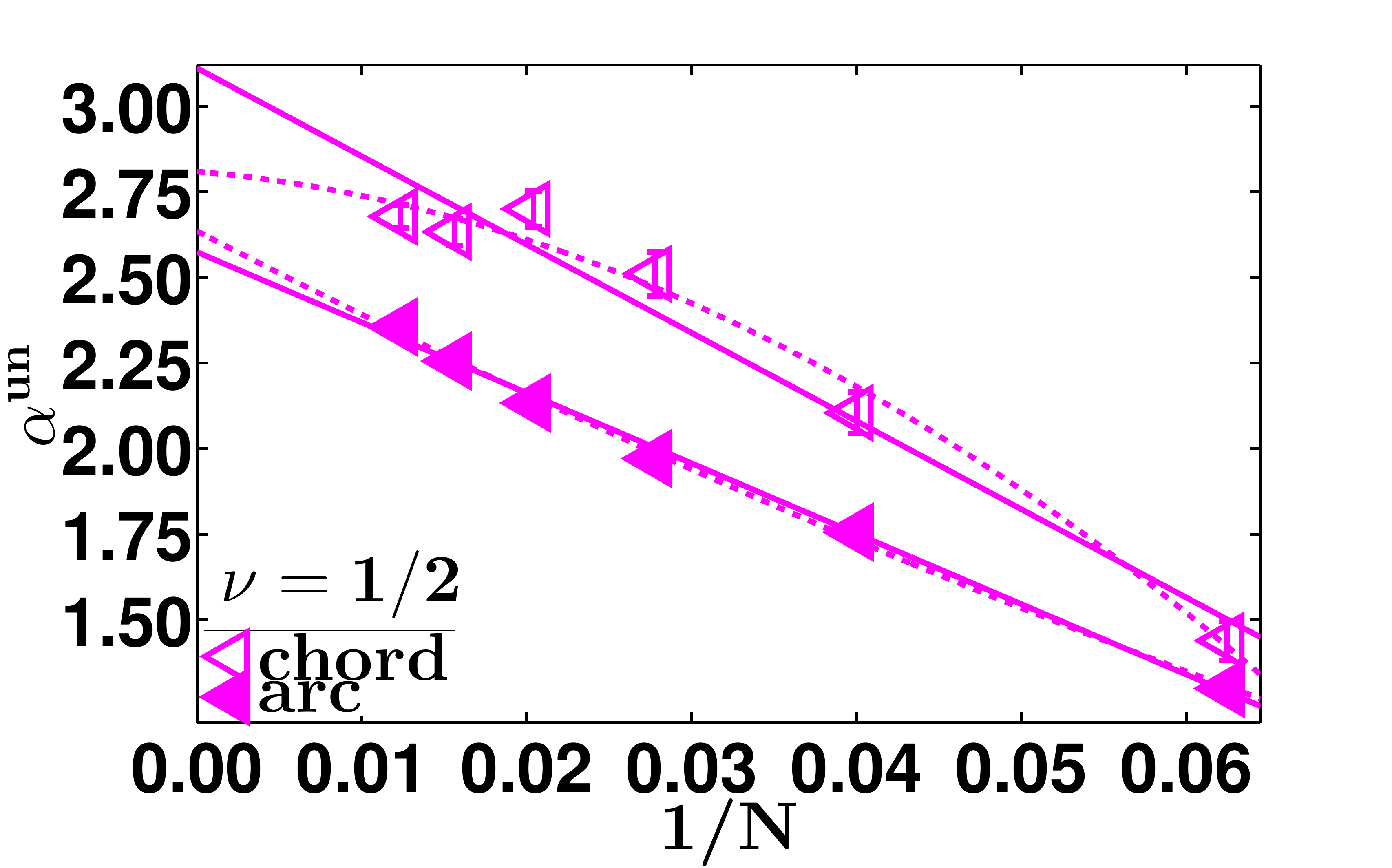}
\includegraphics[width=8cm,height=5cm]{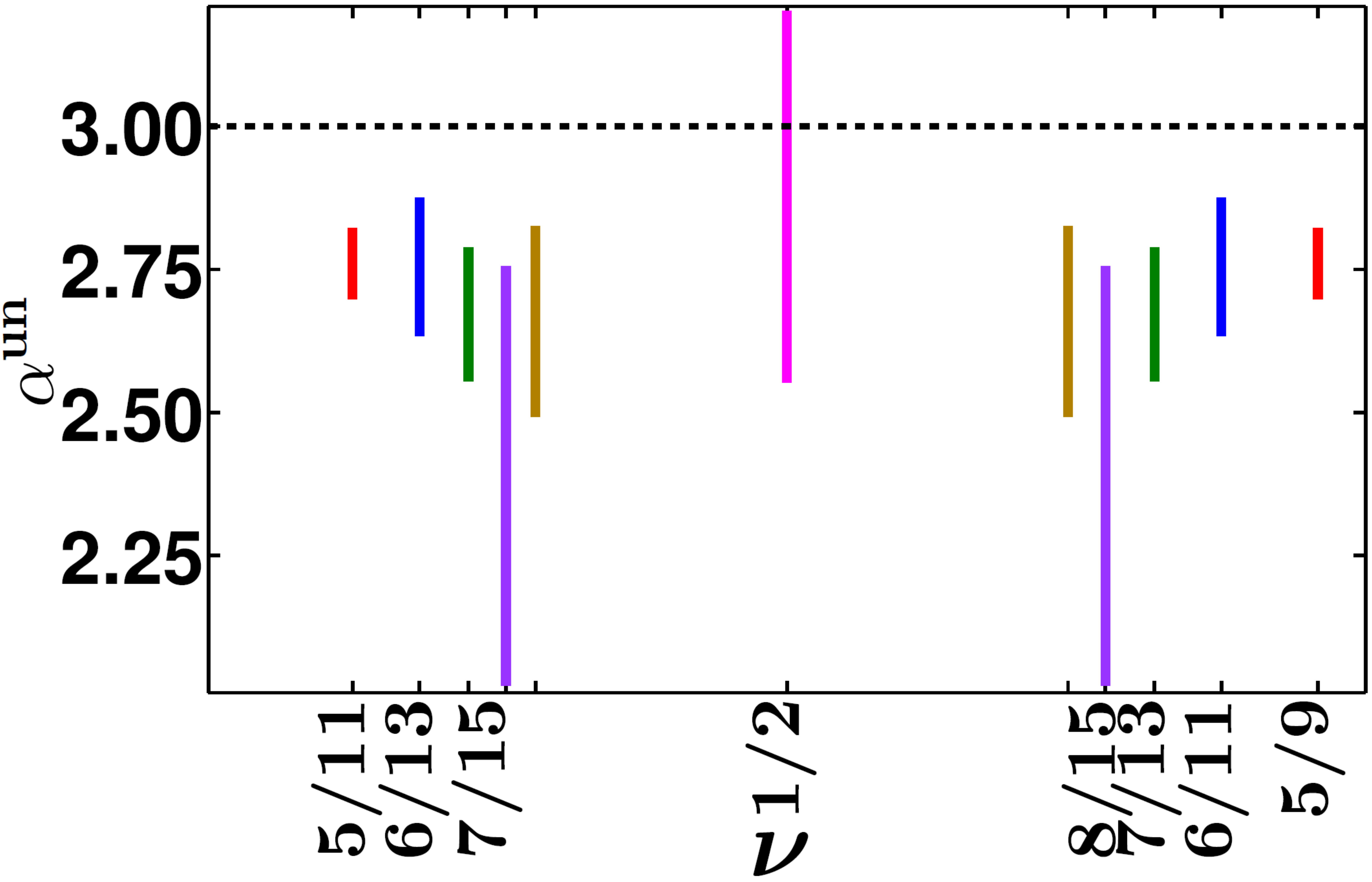}
\end{center}
\caption{Thermodynamic extrapolations for the power law exponent $\alpha^{\rm un}$ (as defined in Eq. 1 of the main text) evaluated using the {\em unprojected} wave functions $\Psi^{\rm un}_{n/(2n+ 1)}$ for filling factors of the form $n/(2n+1)$. The bottom-most panel shows the thermodynamic value of $\alpha^{\rm un}$ as a function of the filling factor $\nu$.}
\label{alphaun_sm}
\end{figure*}

In this section, we provide some further details associated with the results shown in the main article. 
Fig.~\ref{kFsm} (\ref{kFunsm}) shows the extrapolations of $k_{\rm F}^*\ell$ ($k_{\rm F}^{*un}\ell$) to the thermodynamic limit for the projected (unprojected) wave functions of Eq.~8 of the main article. The extrapolated values are shown in Fig.~1 and Fig.~2 of the main article. In each case, we ascertain the range of $k_{\rm F}^*\ell$ by fitting it as a function of $1/N$ with a linear and a quadratic form, using both the arc and the chord distances; we believe the error estimated in this fashion adequately takes into account finite size as well as curvature effects.  We also show the extrapolation of the power law exponent $\alpha$ ($\alpha^{un}$) as defined in Eq. 1 of the main text using the projected (unprojected) Jain wave functions at various filling factors in Fig. \ref{alpha_sm} (\ref{alphaun_sm}).

\section{CF Fermi sea at $\nu=1/4$}
The FQHE states near $\nu=1/2$ can be obtained in two seemingly distinct fashions. For example, the state at $n/(2n-1)$ can be thought of as either filling factor $\nu^*=n$ of \ecf s in a negative effective magnetic field, or as the electron partner of $\nu^*=n-1$ of \hcf s in a positive effective magnetic field. For the FQHE states near $\nu=1/4$ the situation is different. The only way to obtain states at $\nu=n/(4n+1)$ ($\nu=n/(4n-1)$) is in terms of \ecf s at $\nu^*=n$ in positive (negative) effective magnetic field. As a result, we have a unique CFFS at $\nu=1/4$ and the apparent dichotomy encountered above near $\nu=1/2$ does not present itself near $\nu=1/4$. Viewing this state in terms of holes at $\nu_{\rm h}=3/4$, and then composite fermionizing the holes does not produce a CFFS of \hcf s at zero effective magnetic field. From this perspective, it is to be expected that the the Fermi wave vector will be close to $k_{\rm F}^{*}\ell=\sqrt{4\pi\rho_{\rm e}}$ near $\nu=1/4$. 

We have performed extensive calculations to confirm this expectation. Our calculations are carried out at filling factors $\nu=n/(4n\pm 1)$ which arise from IQHE of \ecf s carrying four vortices. The projected Jain wave functions for the ground states at these filling factors are given by:
\be 
\Psi_{n/(4n\pm 1)}={\cal P}_{\rm LLL}\prod_{j<k=1}^N(z_j-z_k)^4\Phi_{\pm n}
\label{Psi_4CFs}
\ee
where ${\cal P}_{\rm LLL}$ denotes LLL projection and $\Phi_n$ is the wave function of $n$ filled LLs, with $\Phi_{-n}=[\Phi_n]^*$. The {\em unprojected} Jain wave functions for the ground states are defined as:
\be 
\Psi_{n/(4n\pm 1)}=\prod_{j<k=1}^N(z_j-z_k)^4\Phi_{\pm n}
\label{Psiun_4CFs}
\ee
For $\nu= 1/4$, akin to the case of $\nu=1/2$, we have calculated $k_{\rm F}^{*}$ by considering filled shell CF systems occurring at zero effective flux.

\begin{figure*}[htpb]
\begin{center}
\includegraphics[width=0.4\textwidth,height=0.25\textwidth]{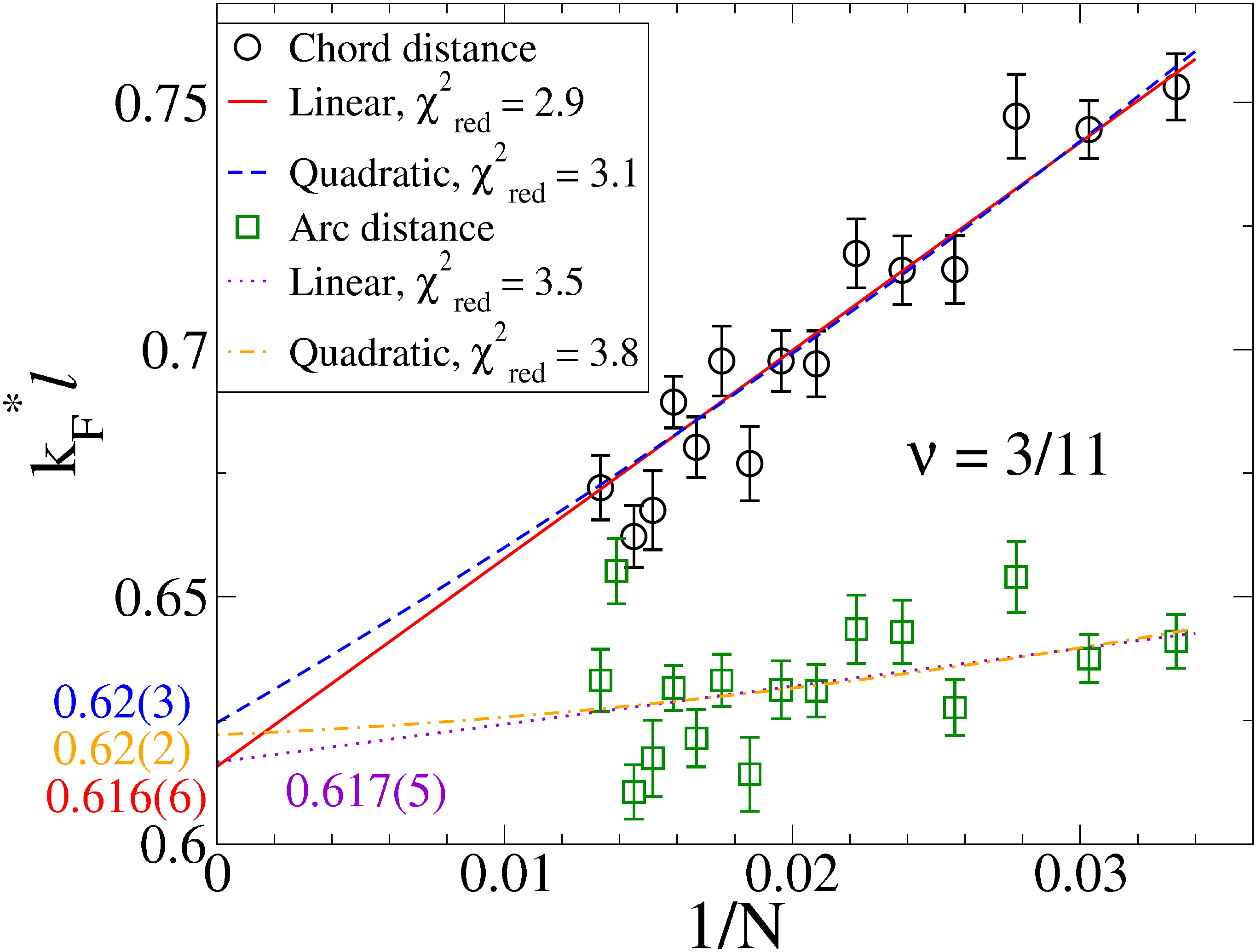}
\includegraphics[width=0.4\textwidth,height=0.25\textwidth]{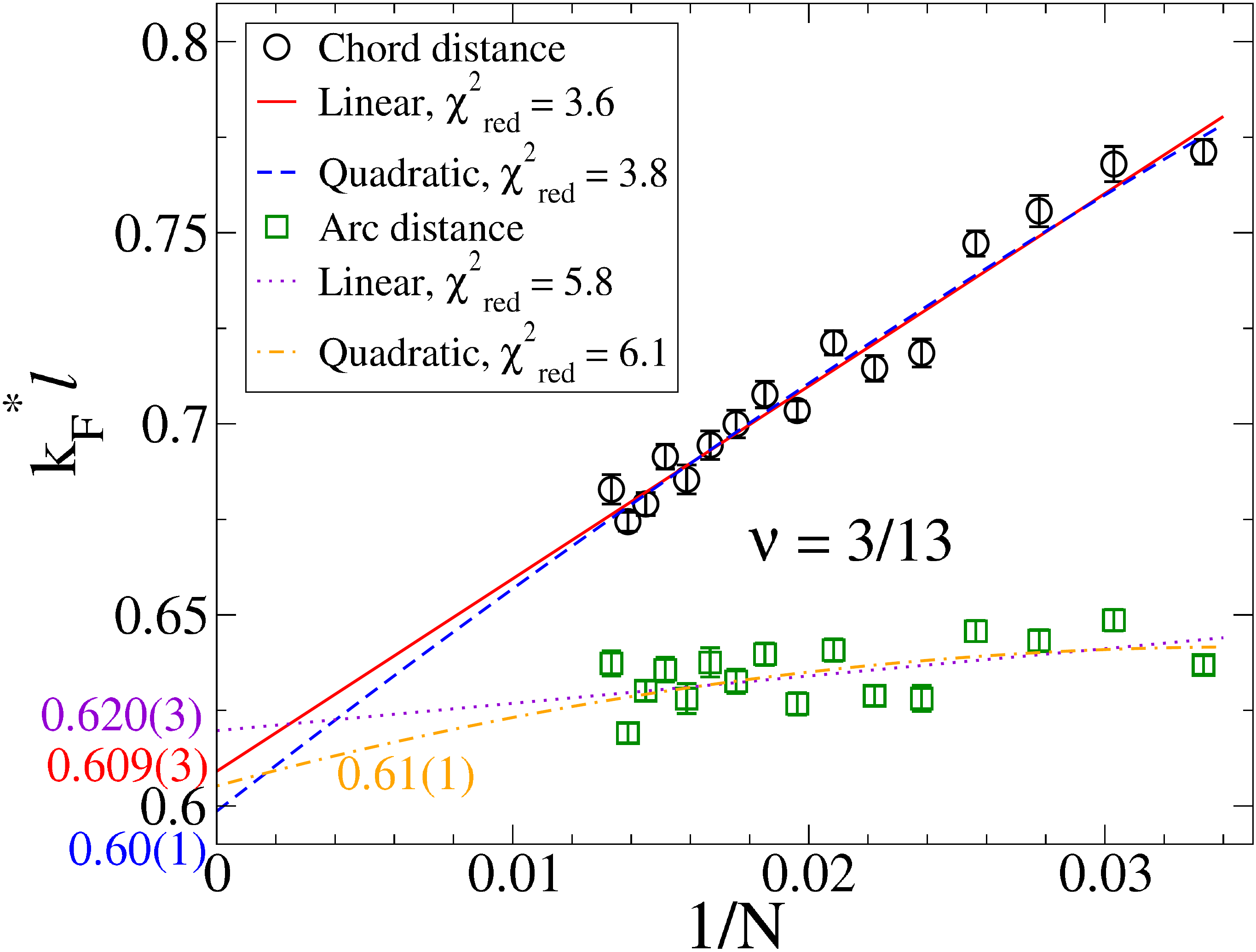}

\includegraphics[width=0.4\textwidth,height=0.25\textwidth]{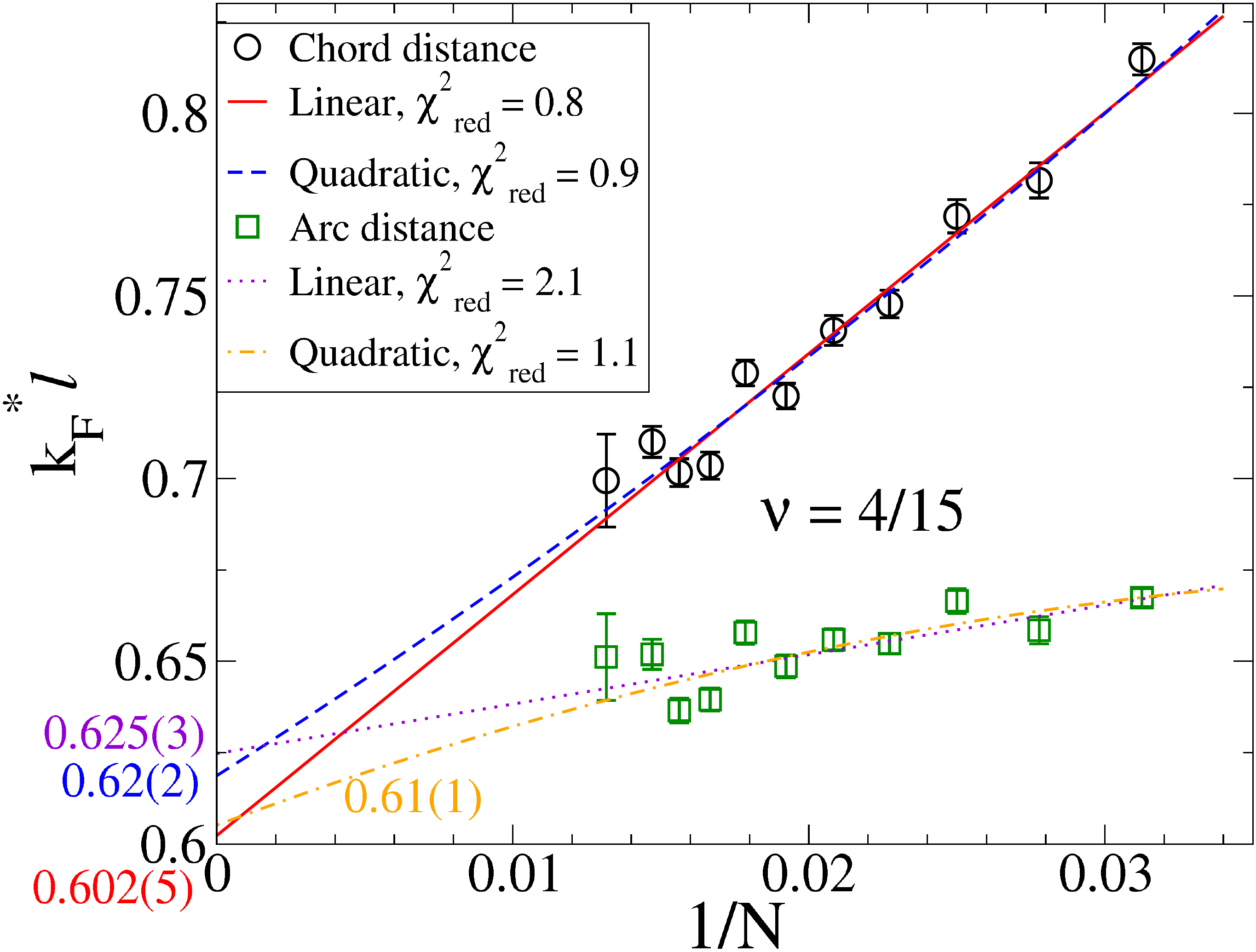}
\includegraphics[width=0.4\textwidth,height=0.25\textwidth]{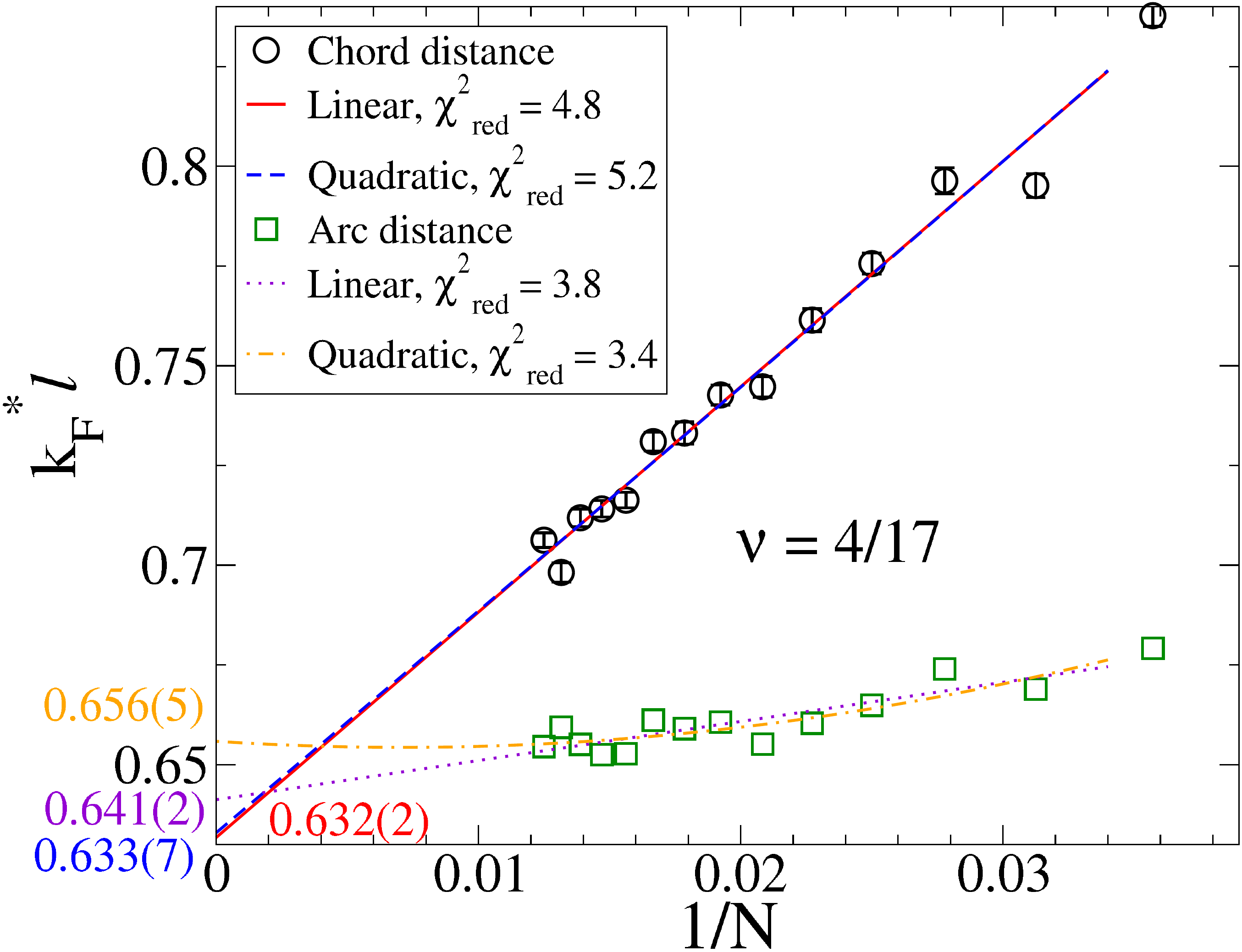}

\includegraphics[width=0.4\textwidth,height=0.25\textwidth]{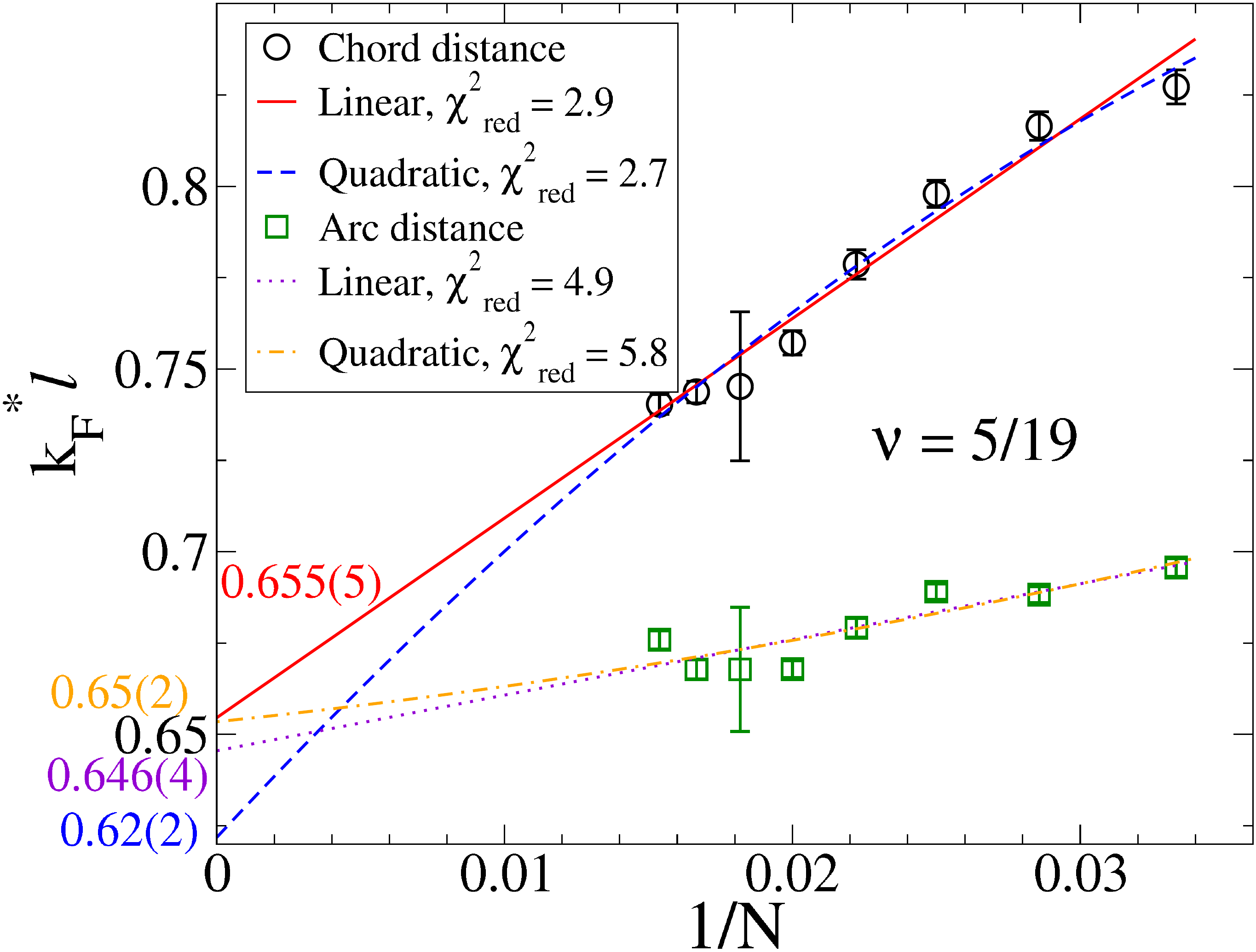}
\includegraphics[width=0.4\textwidth,height=0.25\textwidth]{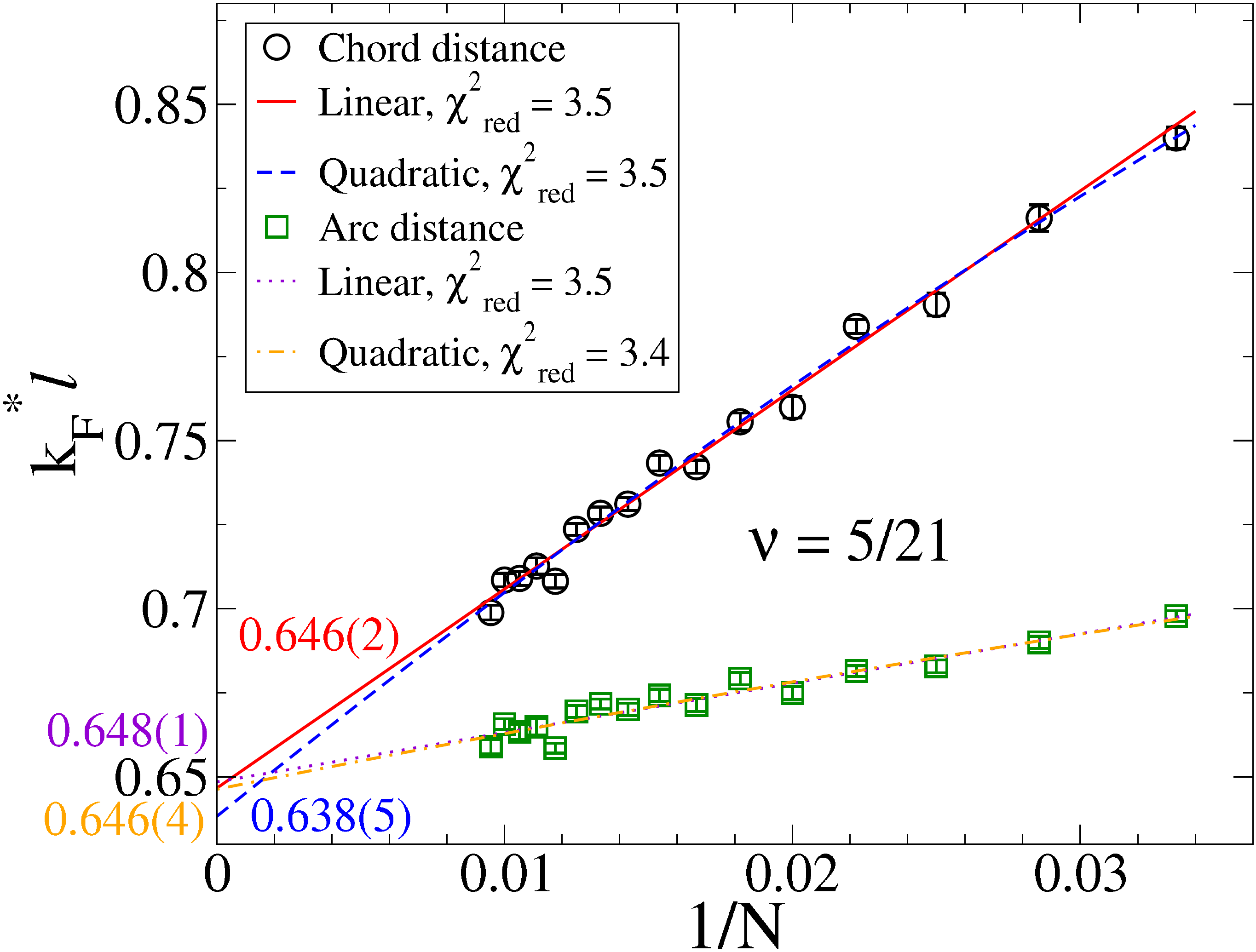}

\includegraphics[width=0.4\textwidth,height=0.25\textwidth]{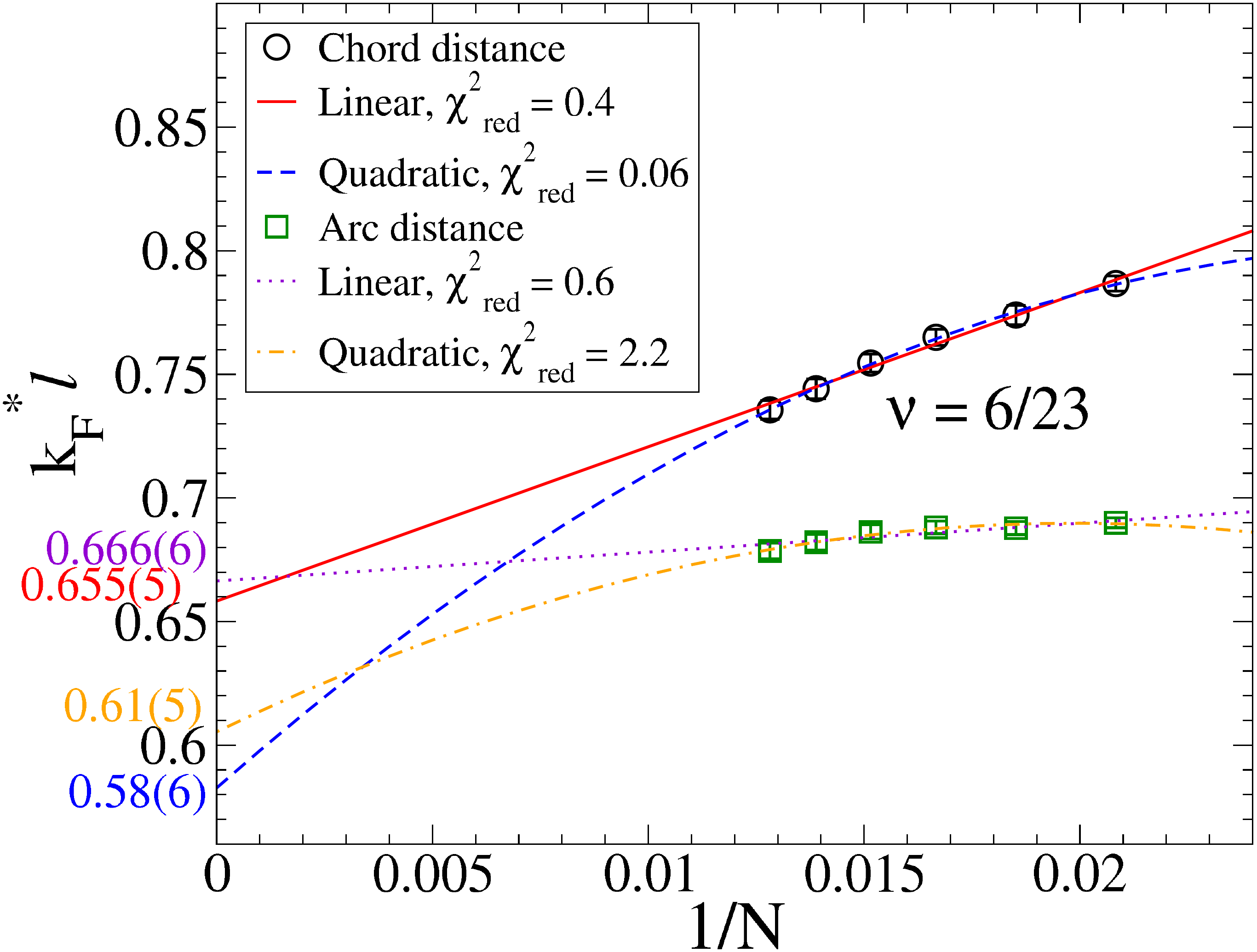}
\includegraphics[width=0.4\textwidth,height=0.25\textwidth]{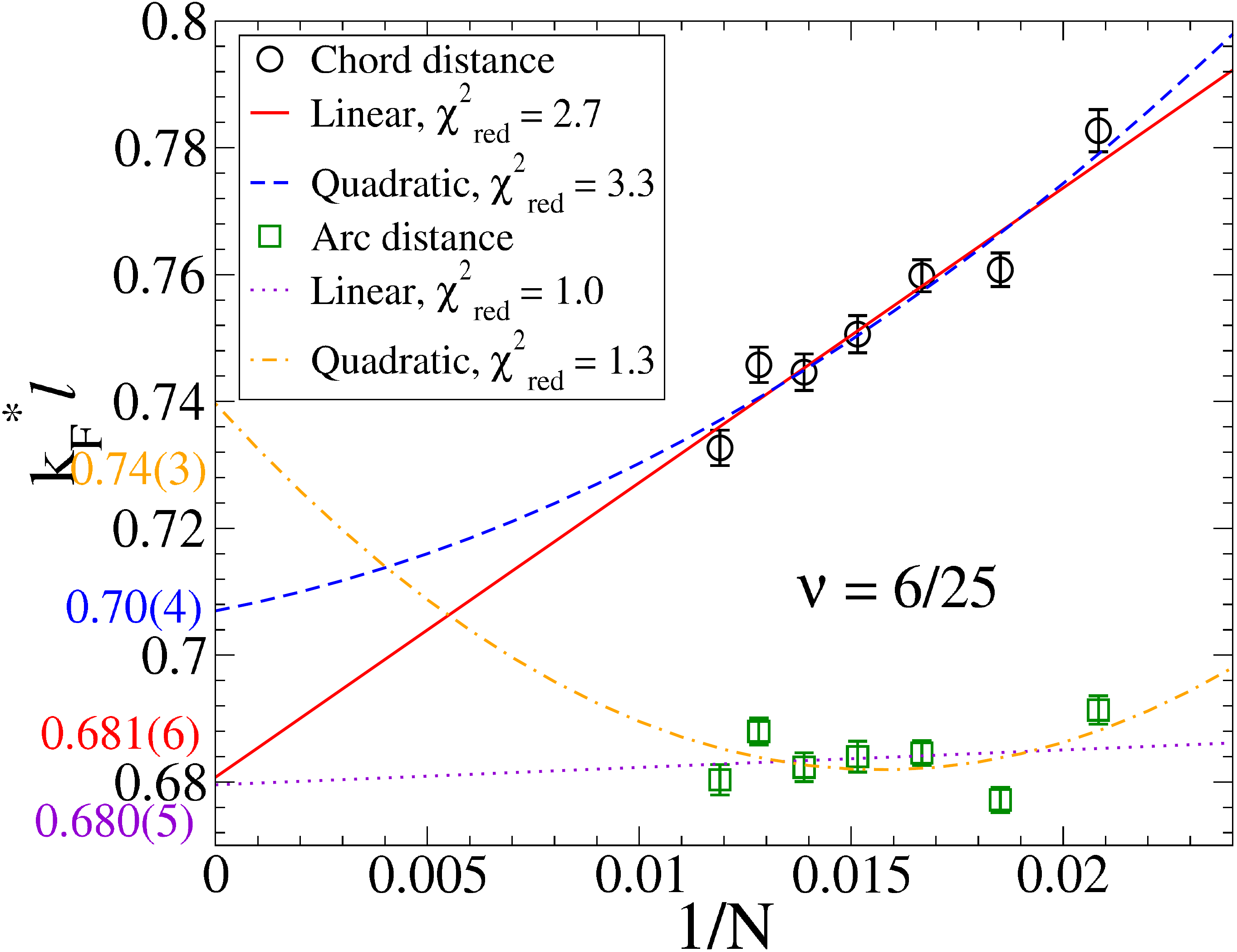}

\includegraphics[width=0.4\textwidth,height=0.25\textwidth]{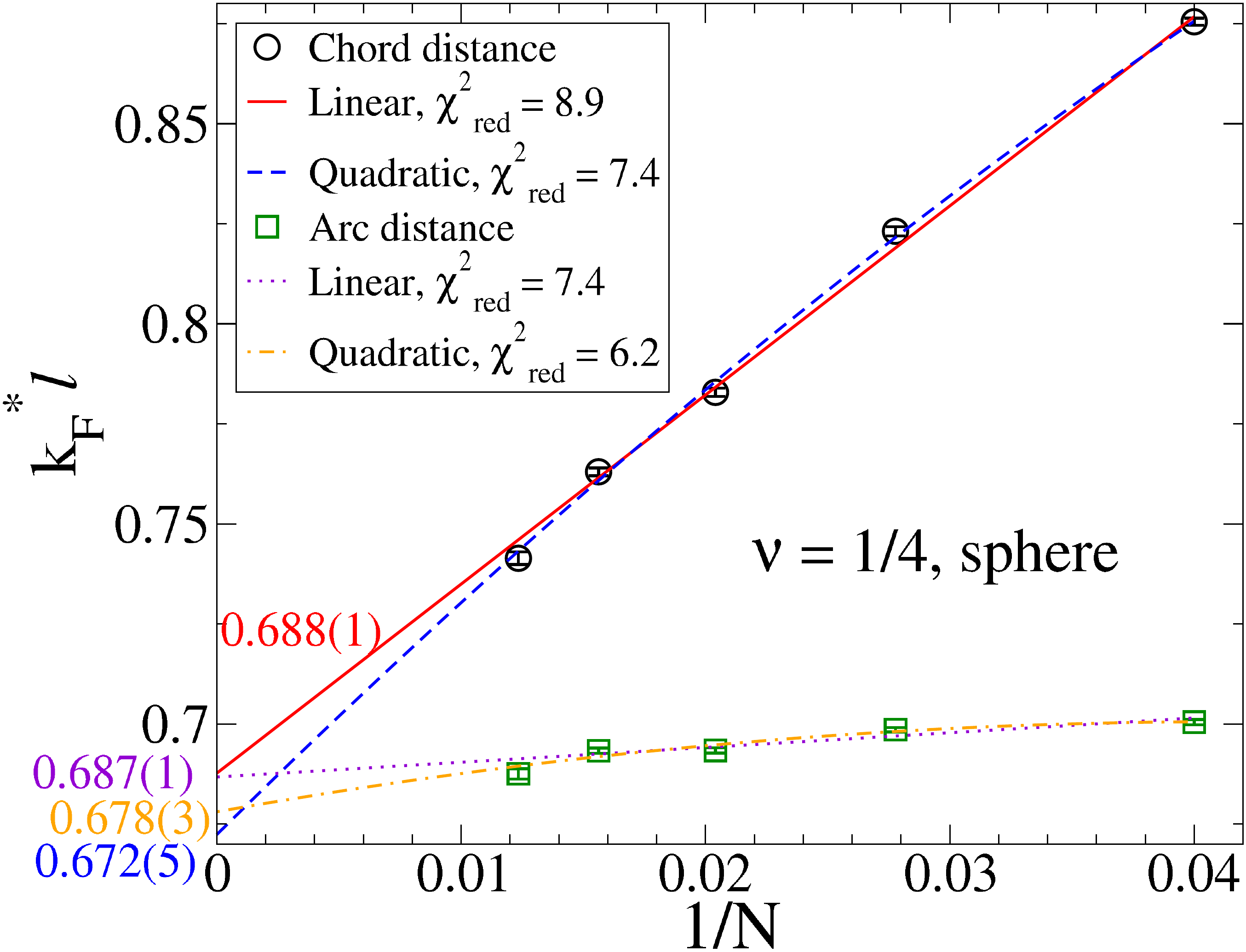}
\includegraphics[width=0.4\textwidth,height=0.25\textwidth]{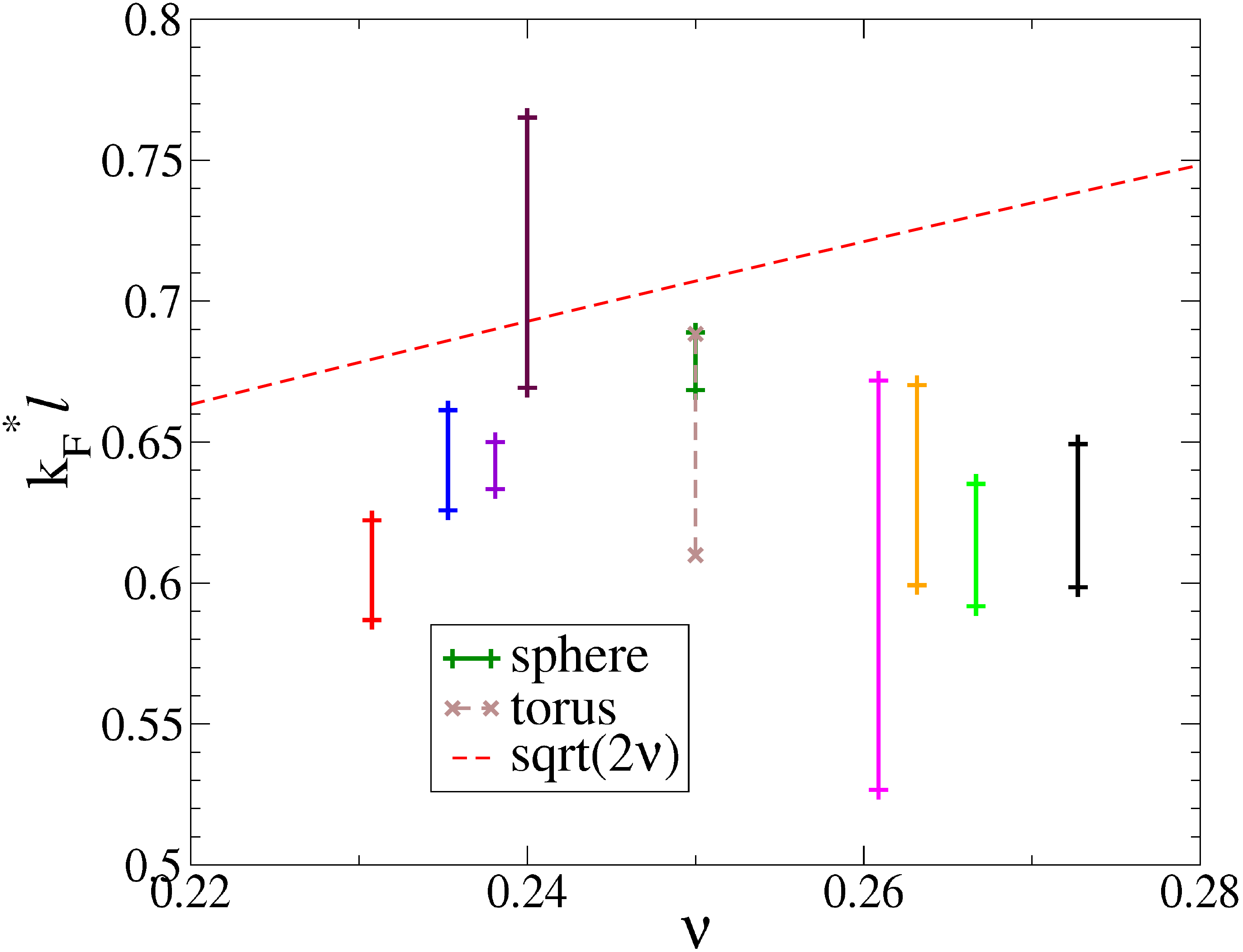}
\end{center}
\caption{Thermodynamic extrapolation of the Fermi wave vector $k_{\rm F}^{*}\ell$ obtained using the projected wave functions $\Psi_{n/(4n\pm 1)}$ at $n/(4n\pm1)$. The bottom-right most panel shows thermodynamic values of $k_{\rm F}^{*}\ell$ as a function of the filling factor $\nu$.}
\label{kFsm_4CFs}
\end{figure*}

\begin{figure*}[htpb]
\begin{center}
\includegraphics[width=0.4\textwidth,height=0.25\textwidth]{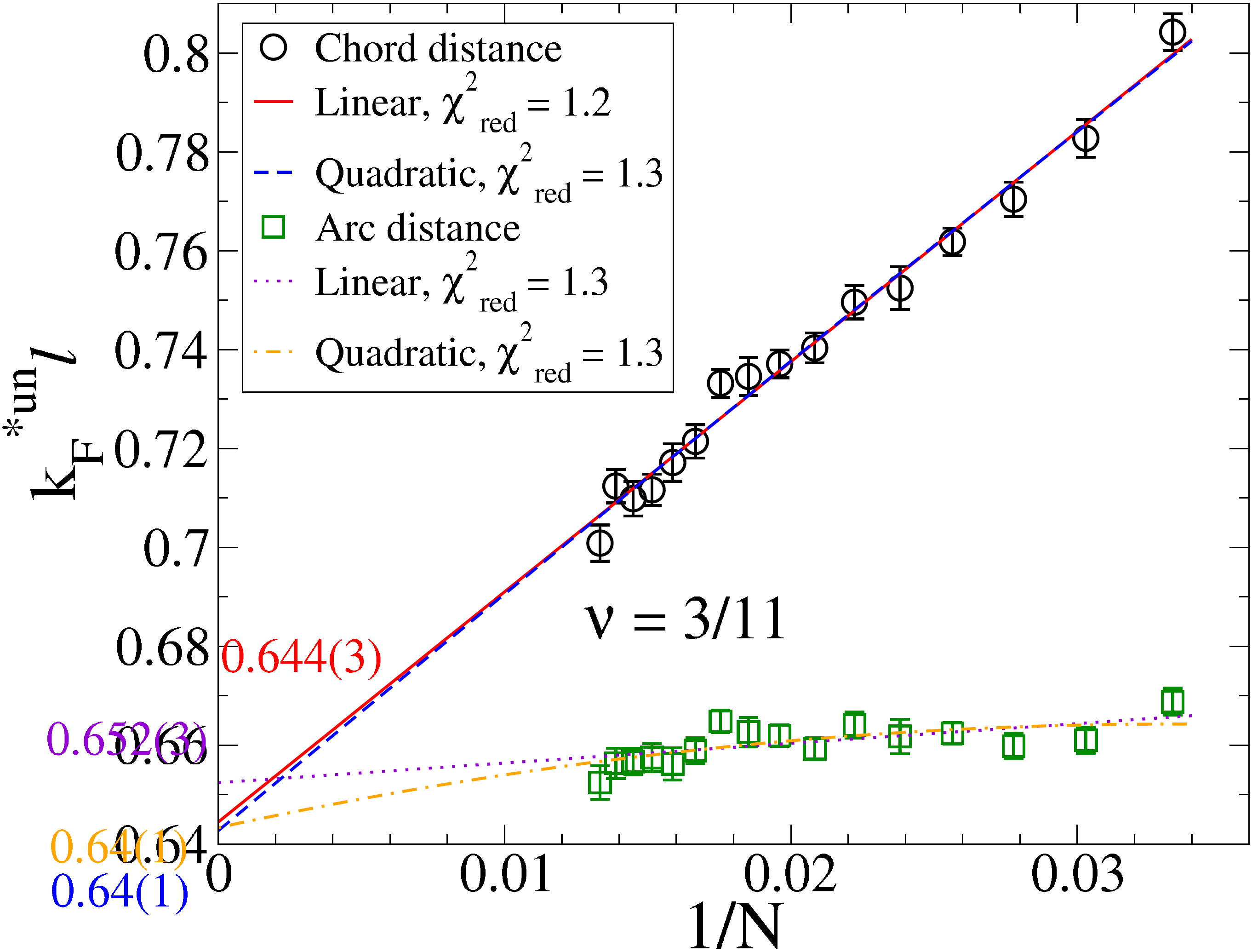}
\includegraphics[width=0.4\textwidth,height=0.25\textwidth]{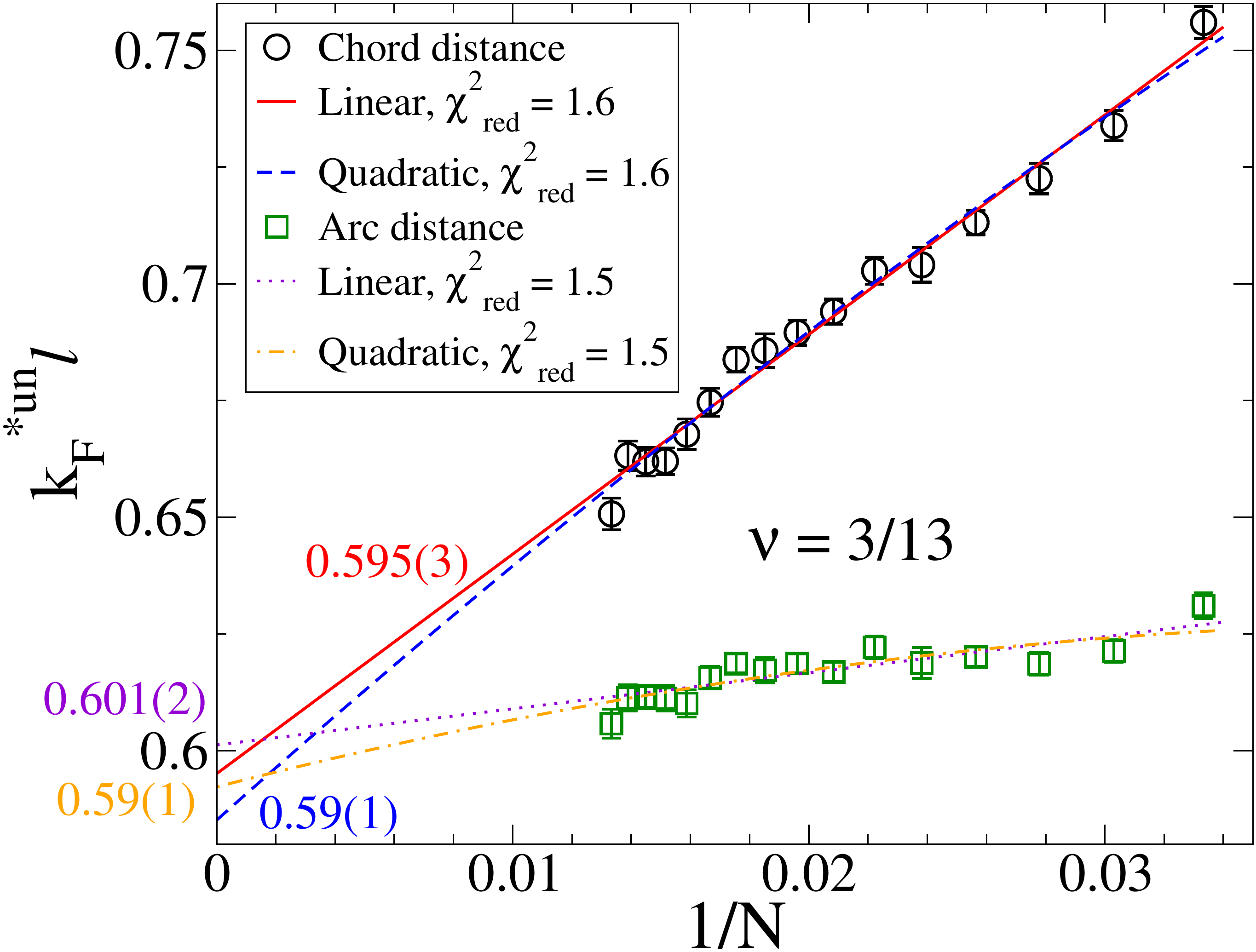}

\includegraphics[width=0.4\textwidth,height=0.25\textwidth]{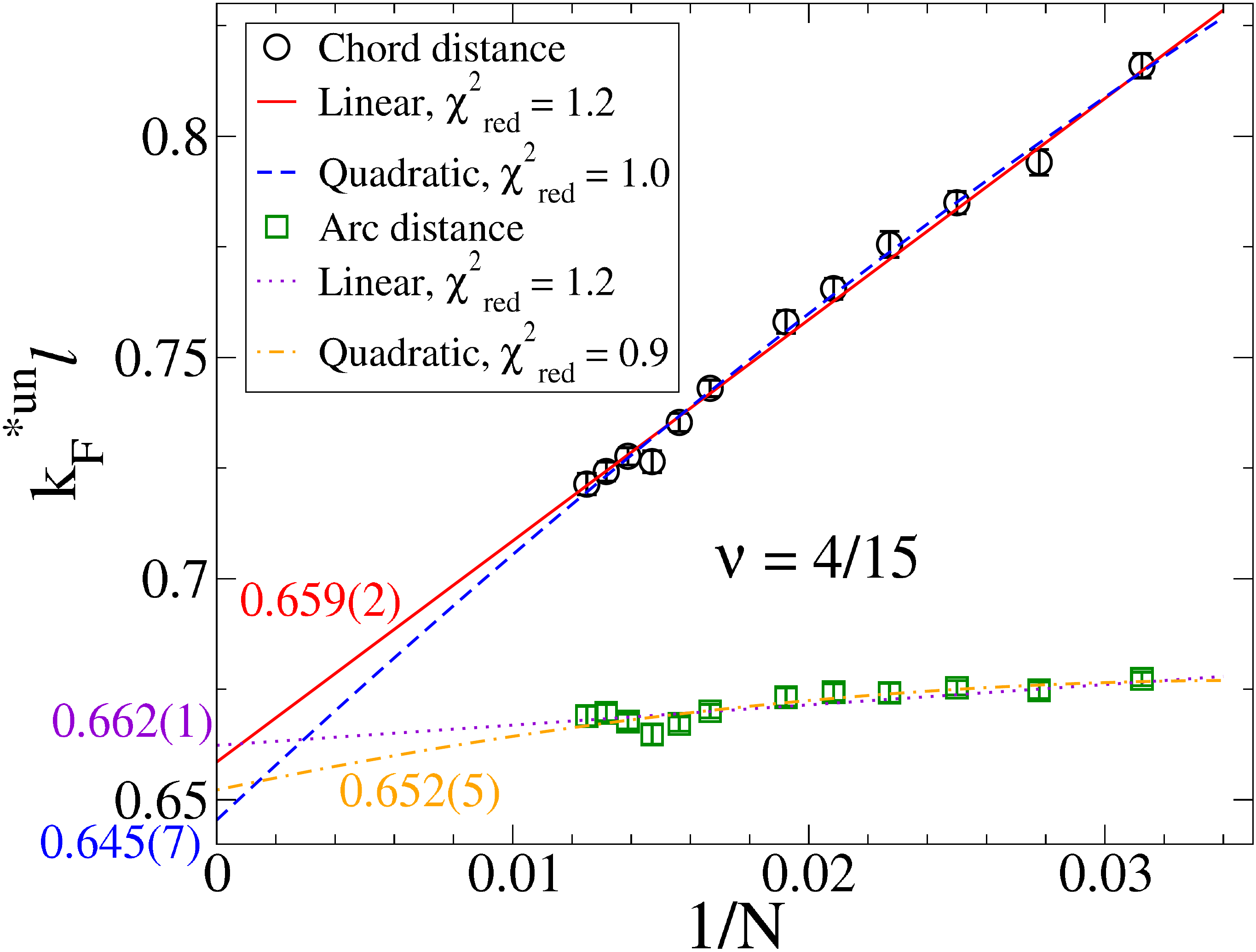}
\includegraphics[width=0.4\textwidth,height=0.25\textwidth]{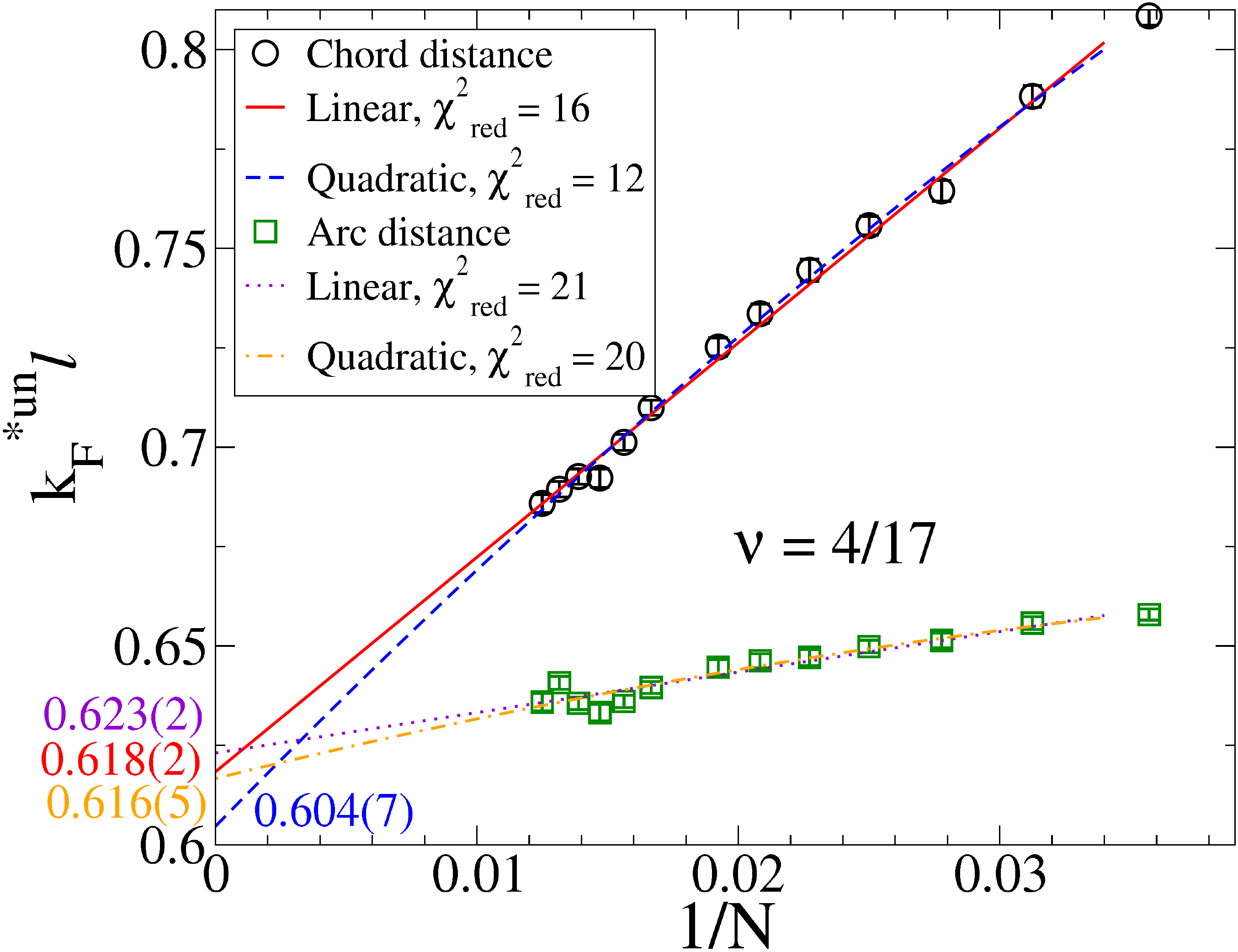}

\includegraphics[width=0.4\textwidth,height=0.25\textwidth]{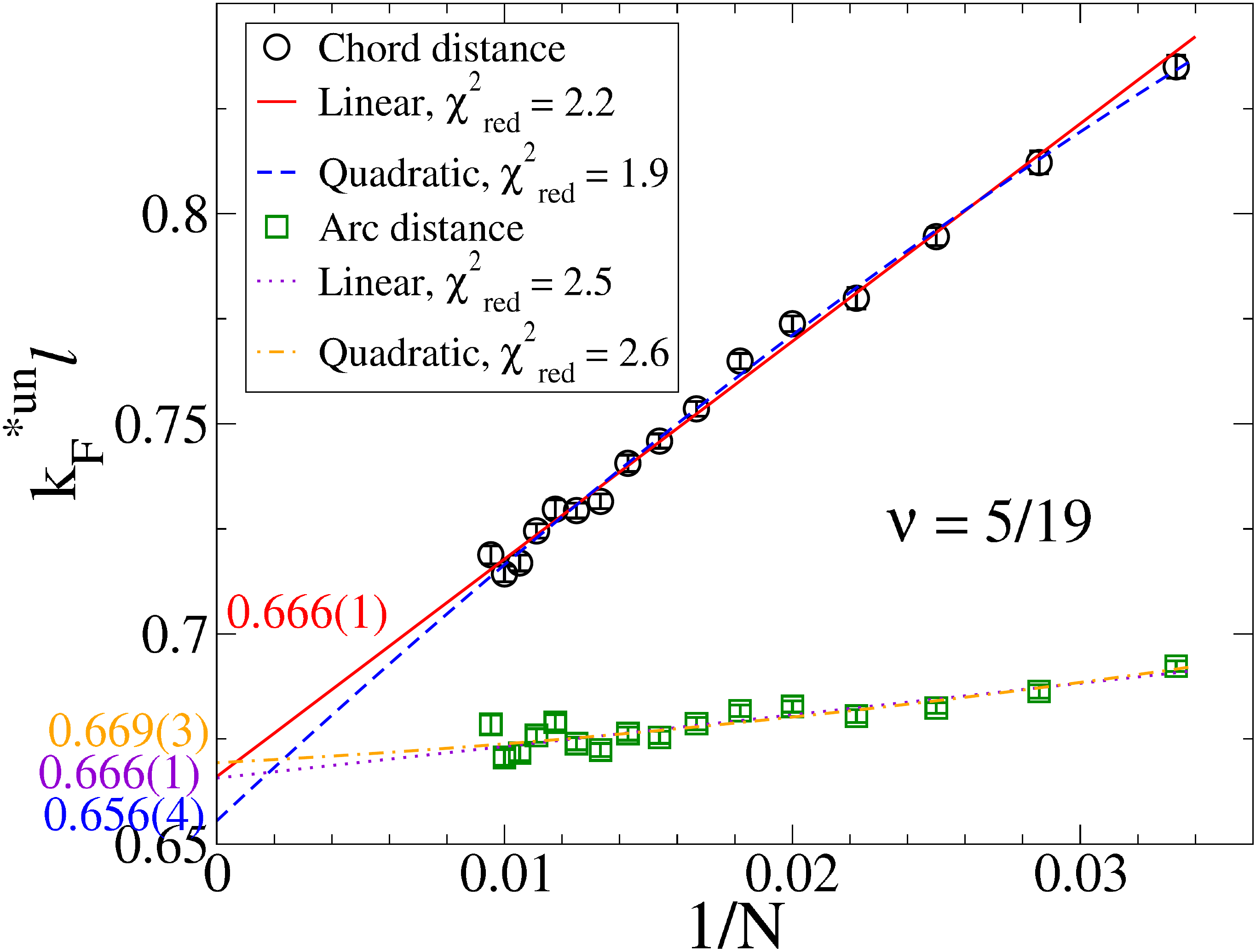}
\includegraphics[width=0.4\textwidth,height=0.25\textwidth]{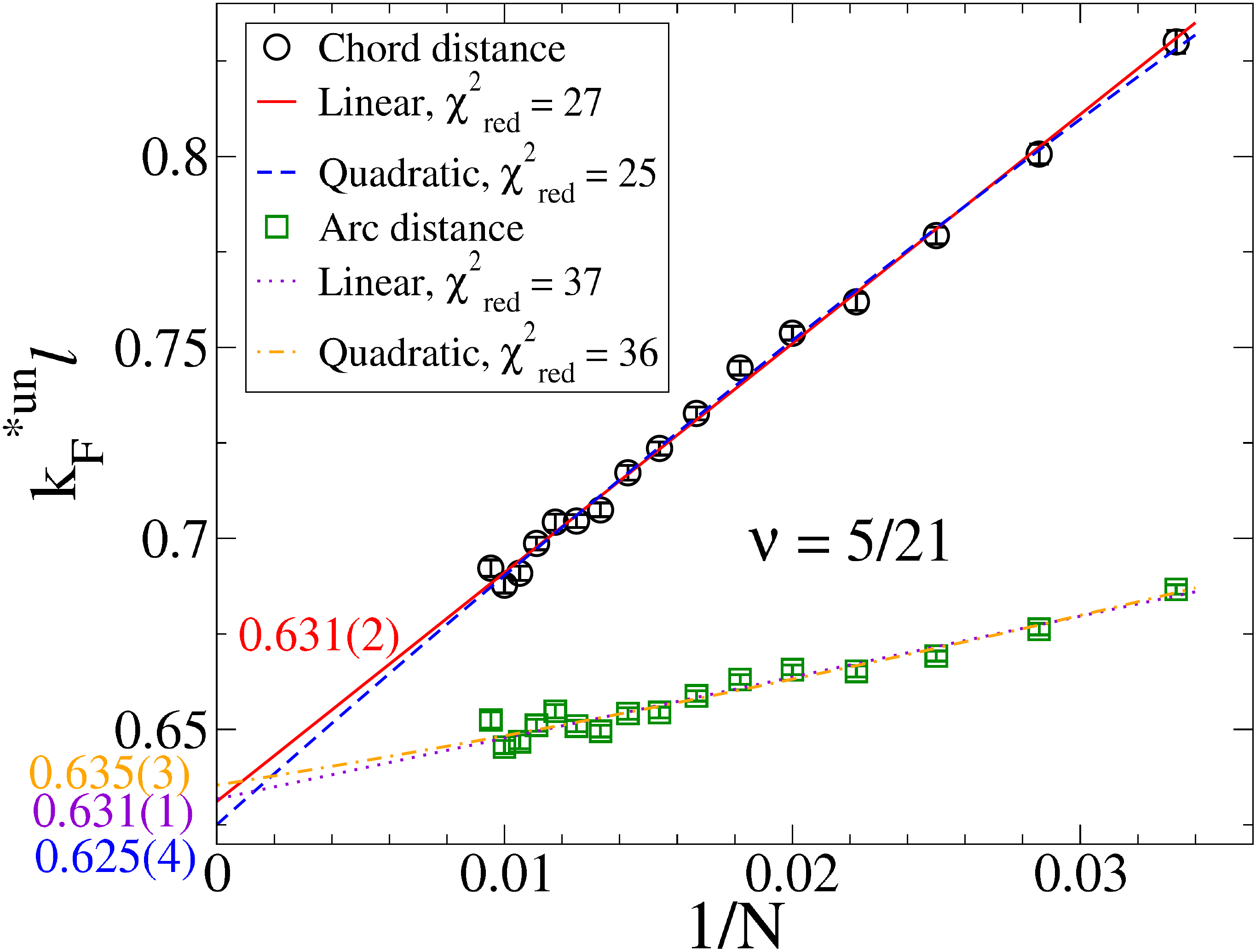}

\includegraphics[width=0.4\textwidth,height=0.25\textwidth]{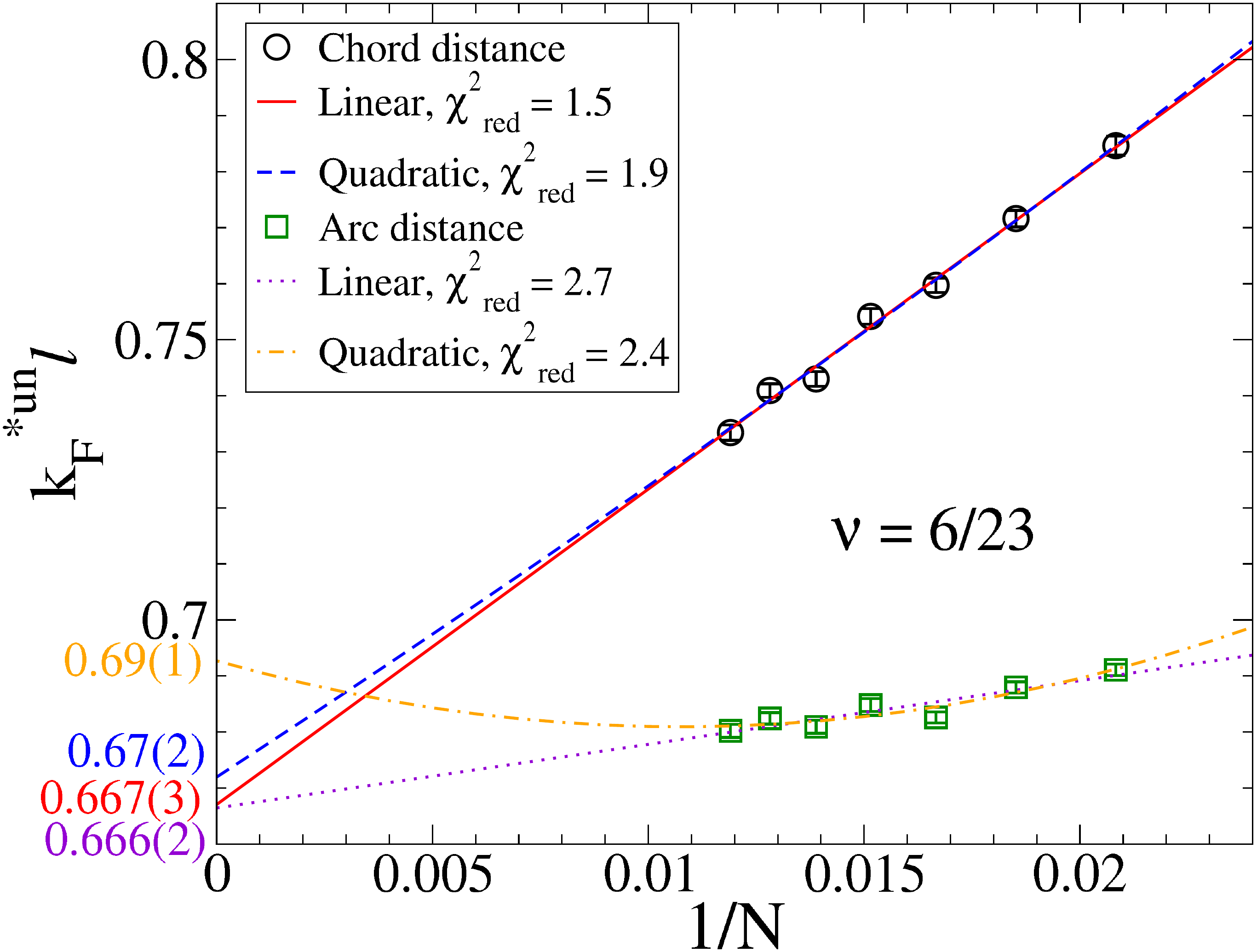}
\includegraphics[width=0.4\textwidth,height=0.25\textwidth]{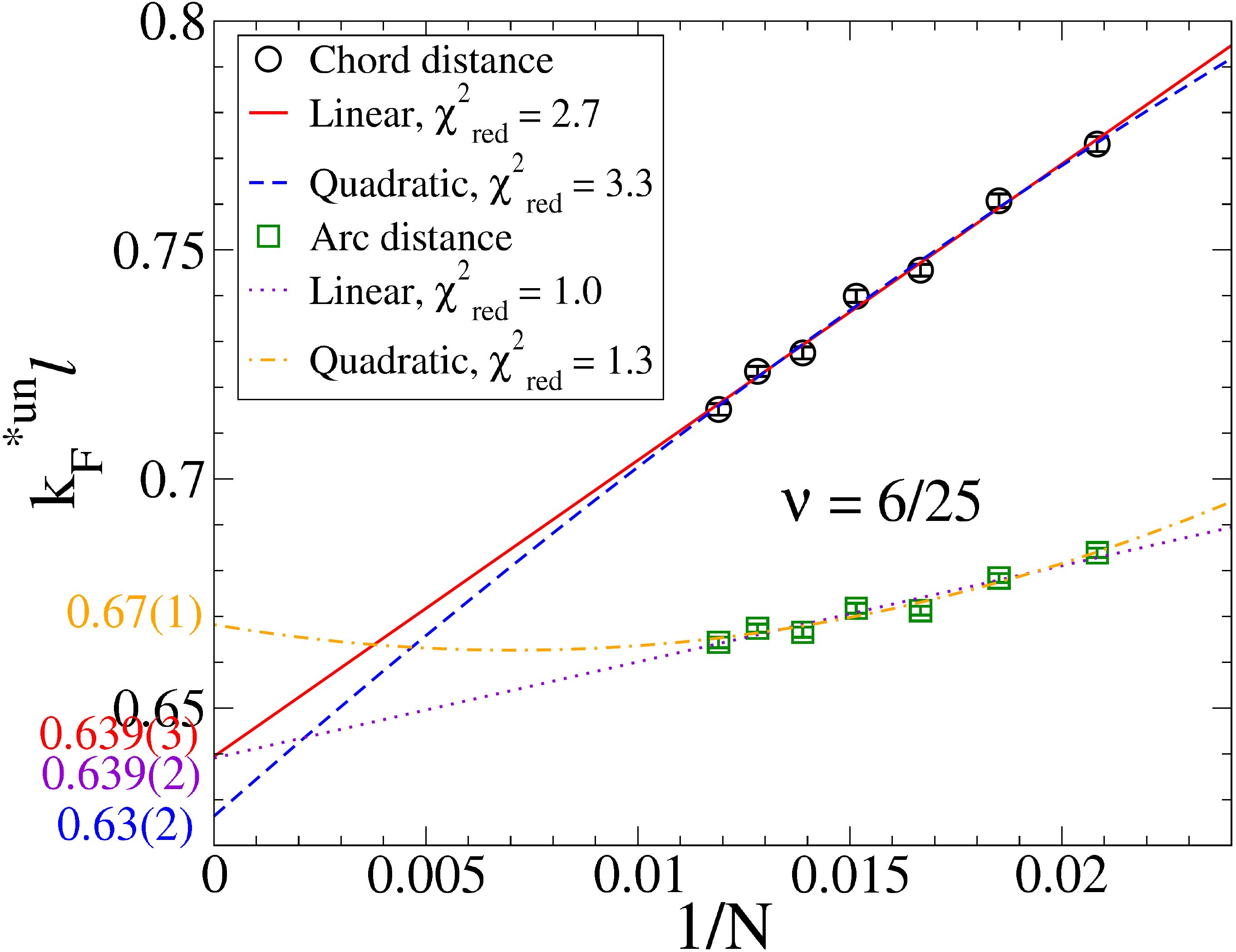}

\includegraphics[width=0.4\textwidth,height=0.25\textwidth]{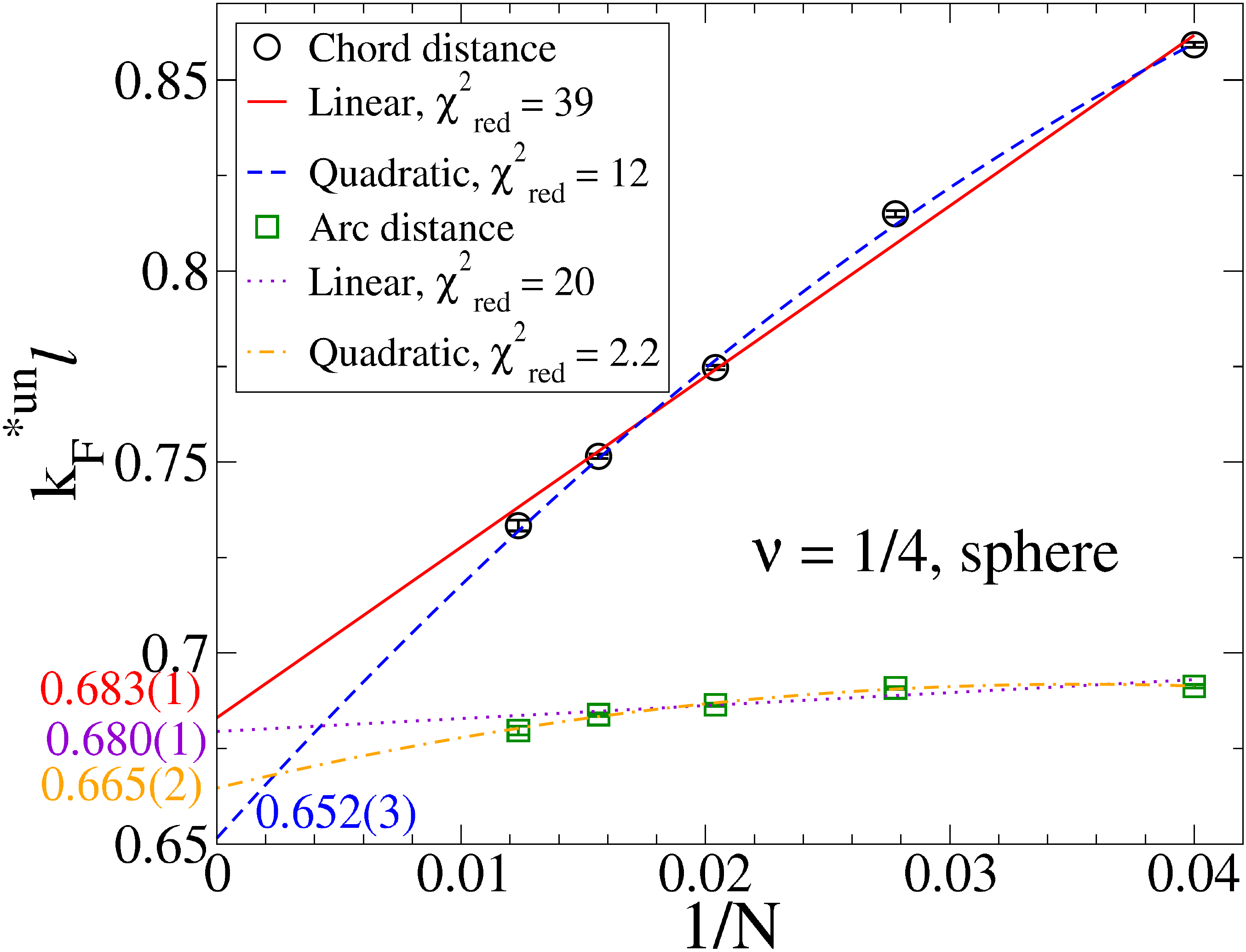}
\includegraphics[width=0.4\textwidth,height=0.25\textwidth]{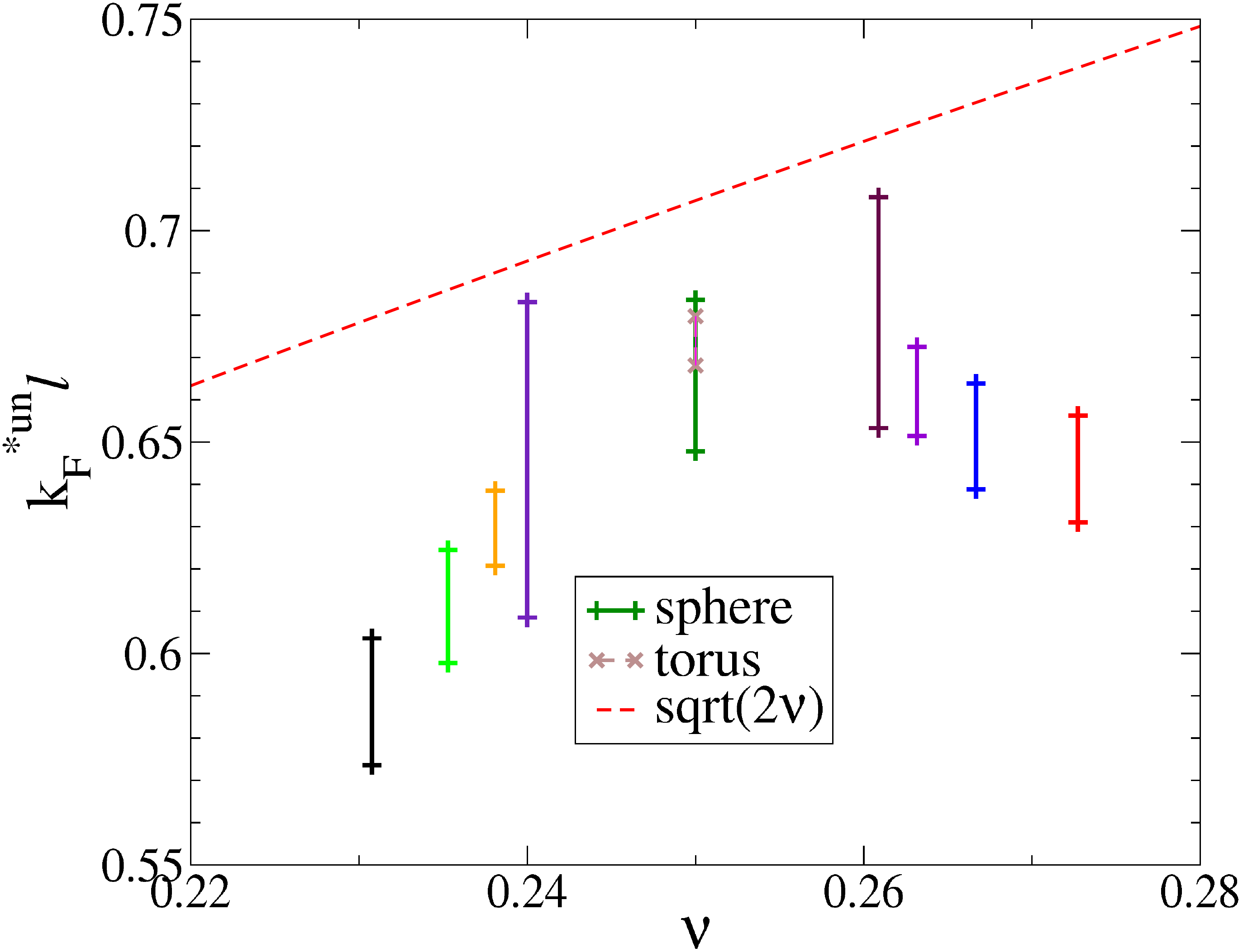}
\end{center}
\caption{Thermodynamic extrapolation of the Fermi wave vector $k_{\rm F}^{\rm * un}\ell$ obtained using the {\em unprojected}  wave functions $\Psi^{\rm un}_{n/(4n\pm 1)}$.}
\label{kFunsm_4CFs}
\end{figure*}

\begin{figure*}[htpb]
\begin{center}
\includegraphics[width=0.4\textwidth,height=0.25\textwidth]{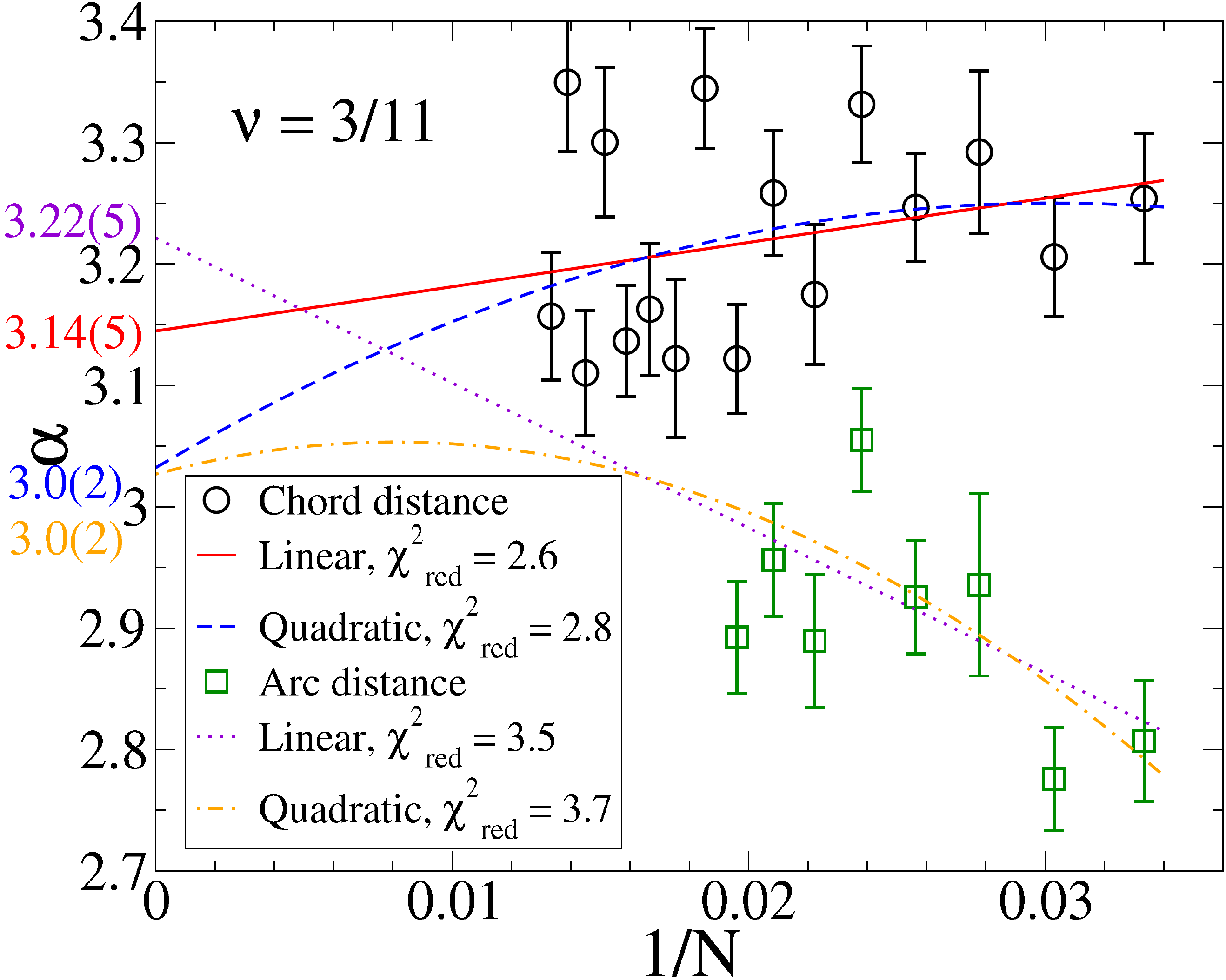}
\includegraphics[width=0.4\textwidth,height=0.25\textwidth]{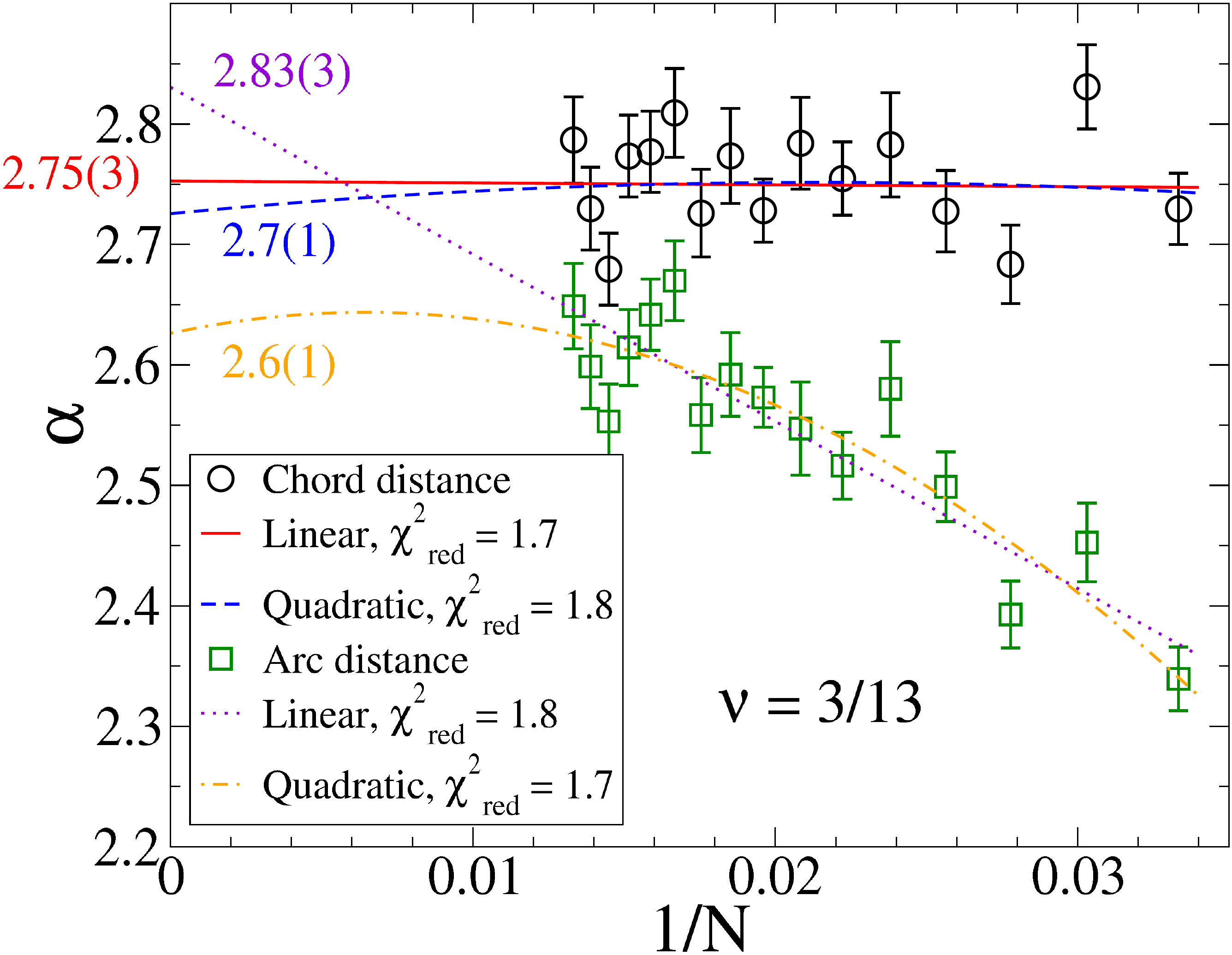}

\includegraphics[width=0.4\textwidth,height=0.25\textwidth]{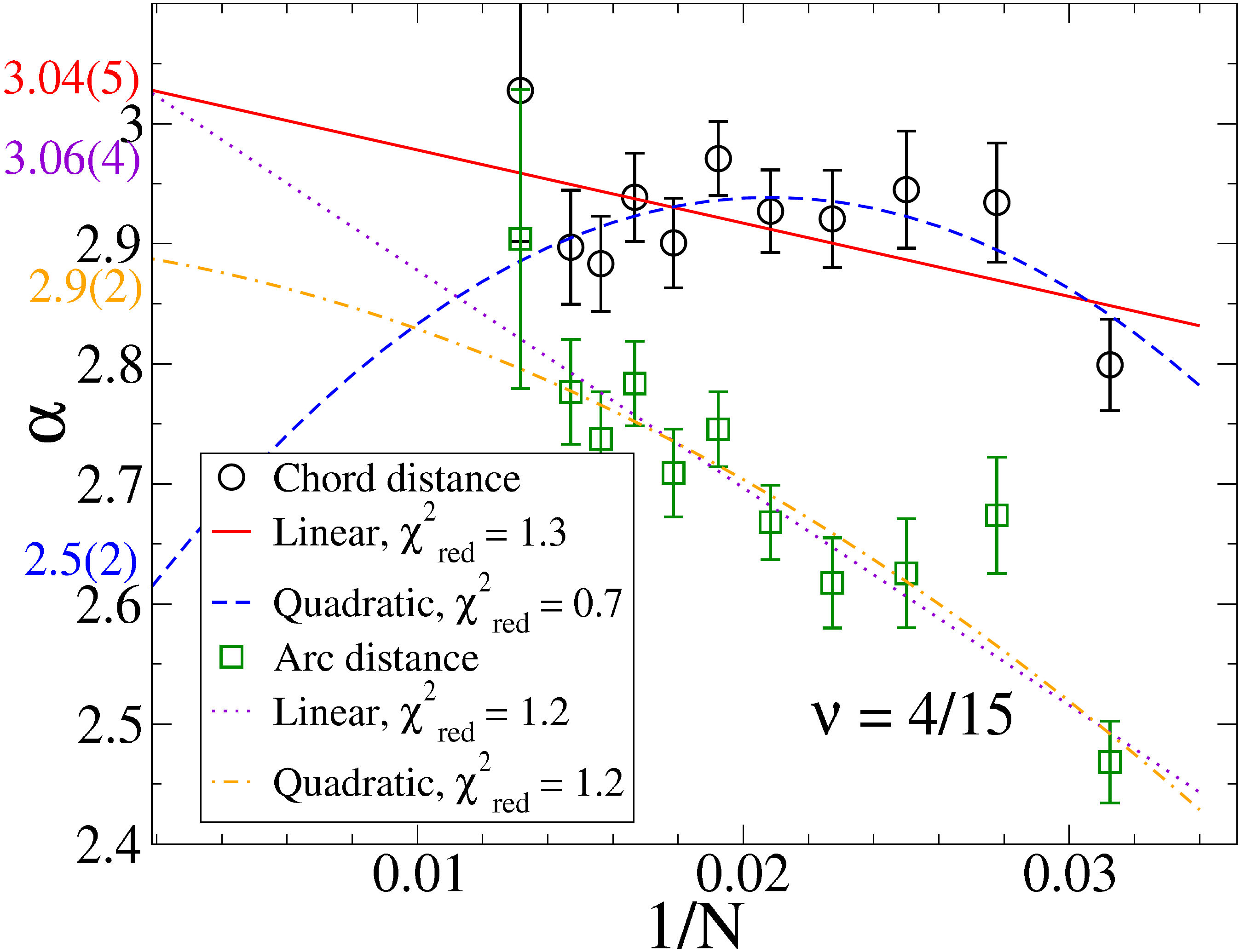}
\includegraphics[width=0.4\textwidth,height=0.25\textwidth]{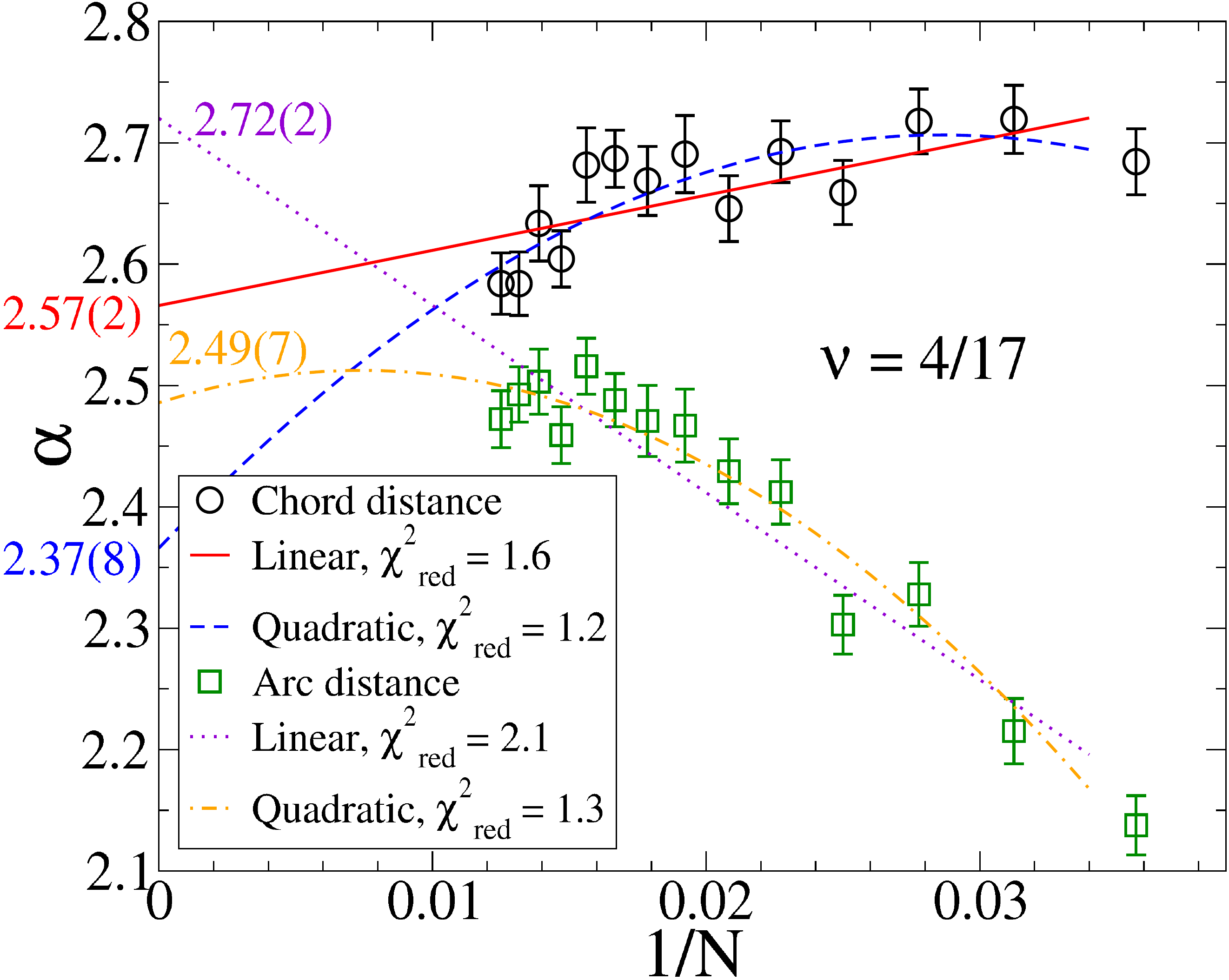}

\includegraphics[width=0.4\textwidth,height=0.25\textwidth]{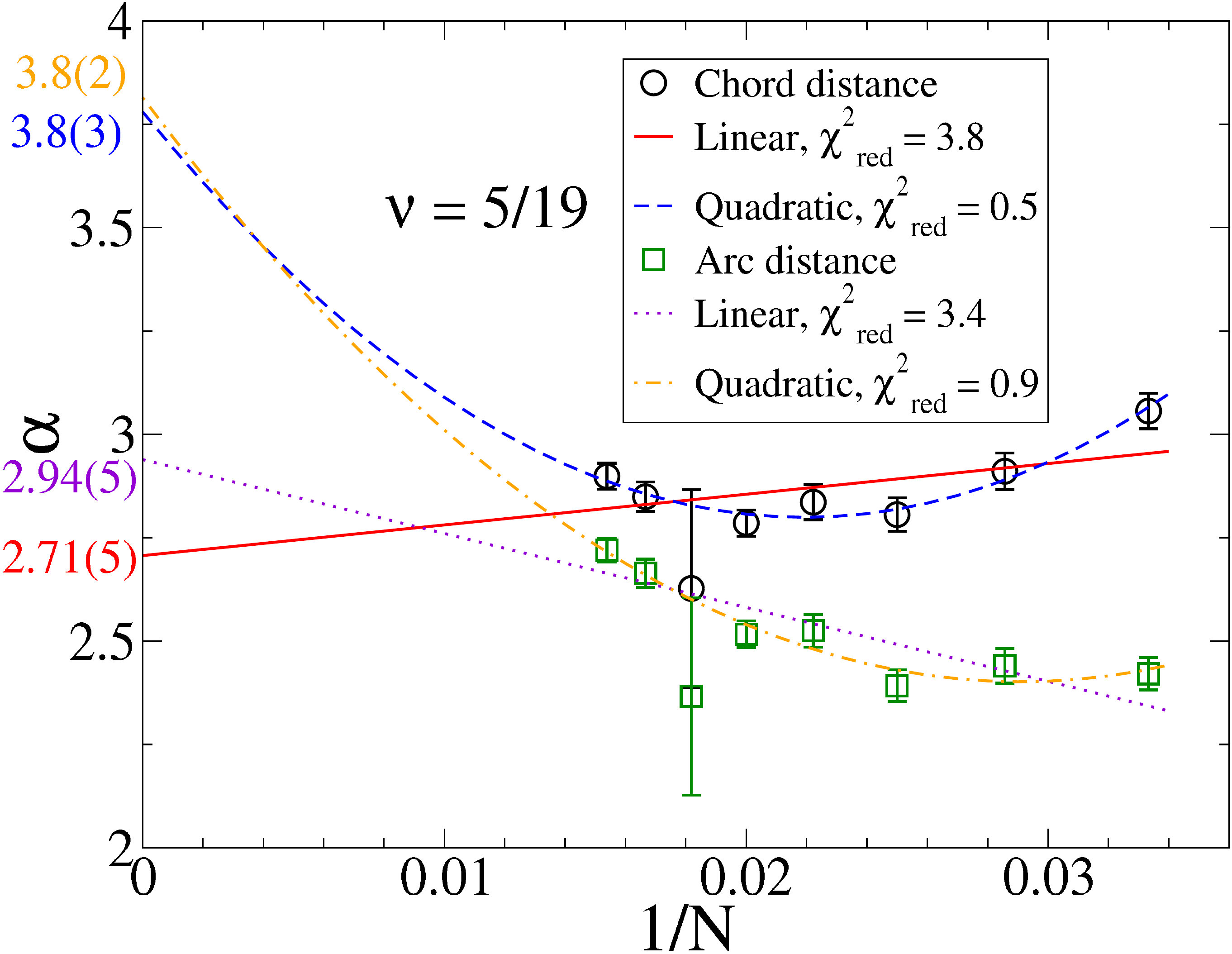}
\includegraphics[width=0.4\textwidth,height=0.25\textwidth]{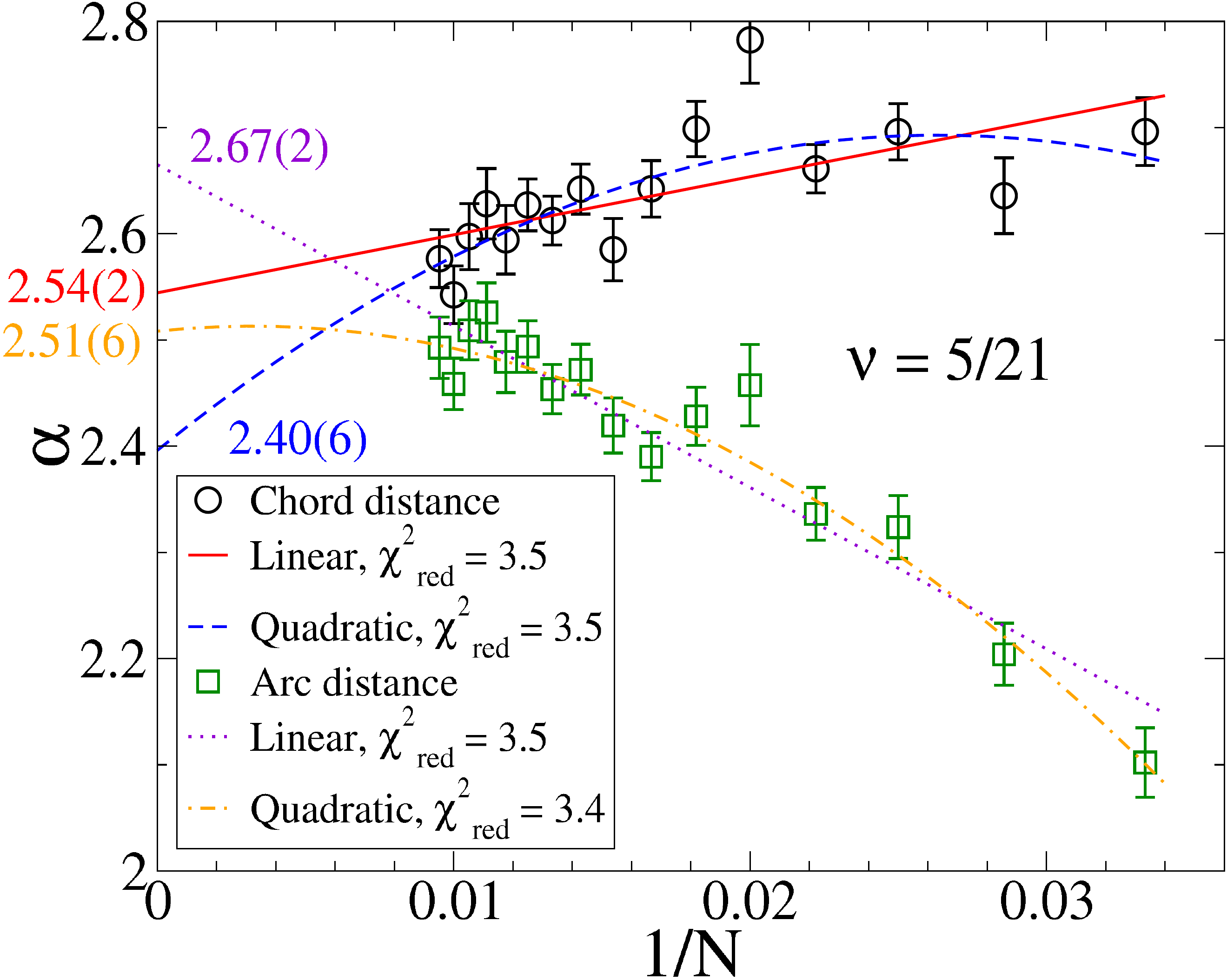}

\includegraphics[width=0.4\textwidth,height=0.25\textwidth]{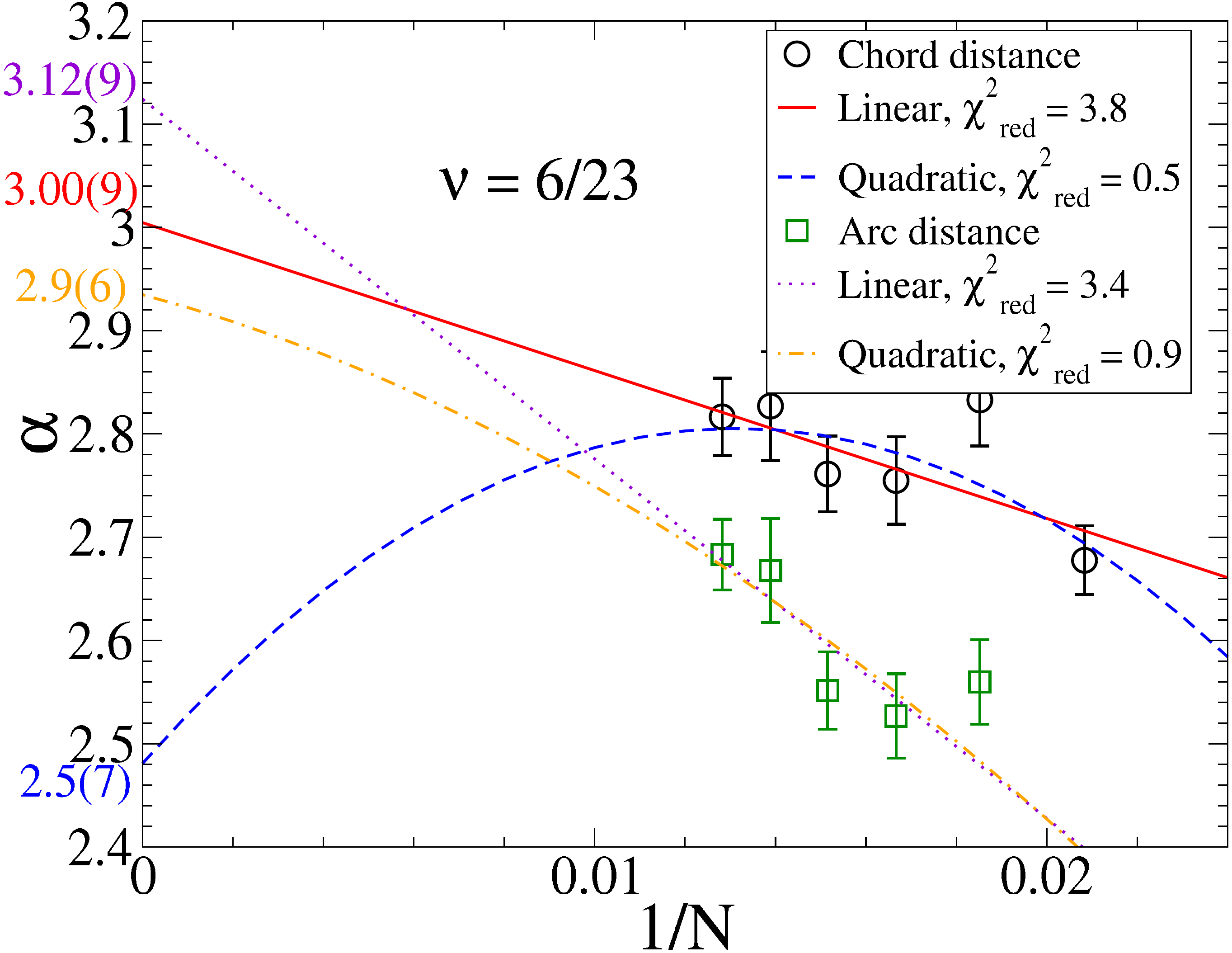}
\includegraphics[width=0.4\textwidth,height=0.25\textwidth]{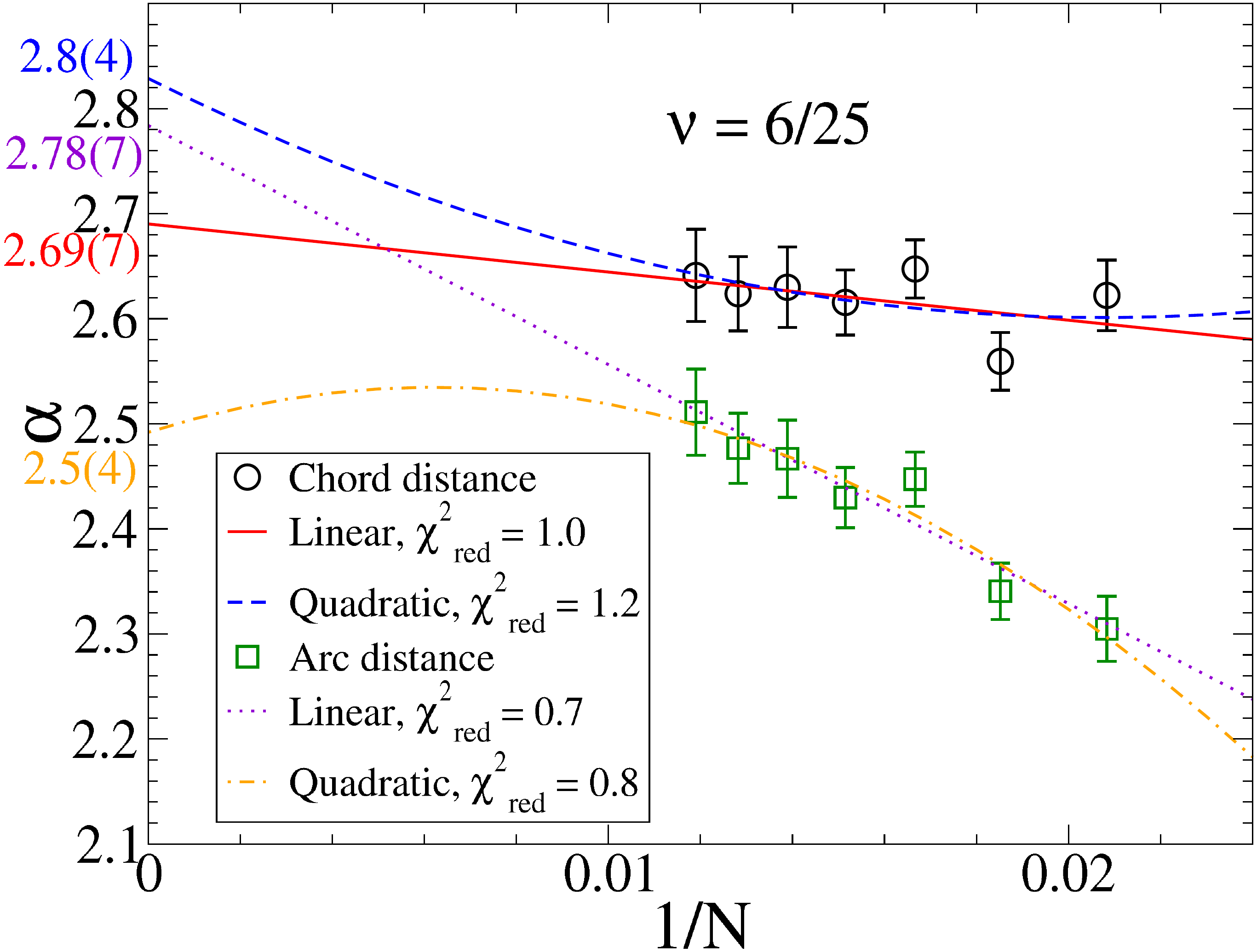}

\includegraphics[width=0.4\textwidth,height=0.25\textwidth]{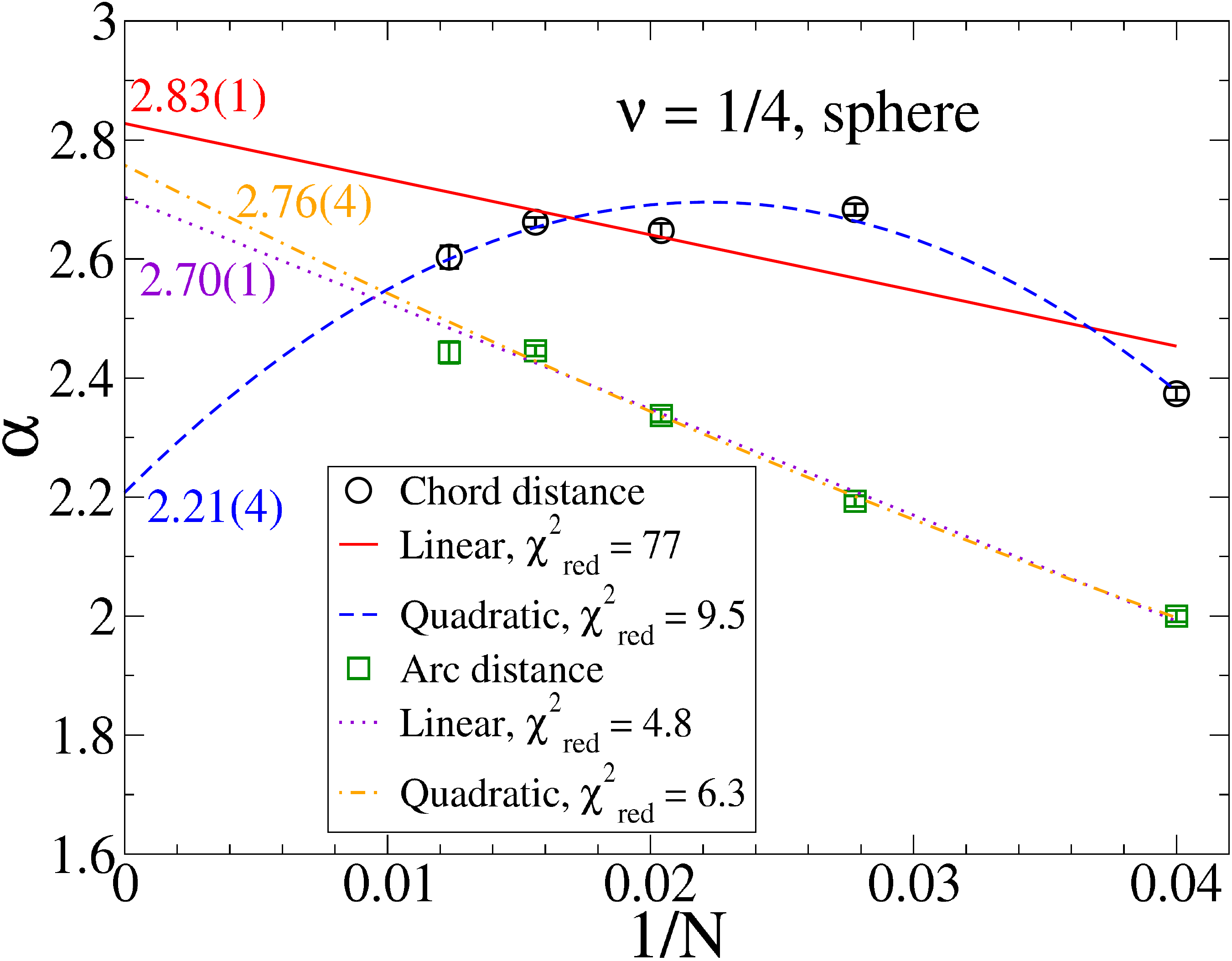}
\includegraphics[width=0.4\textwidth,height=0.25\textwidth]{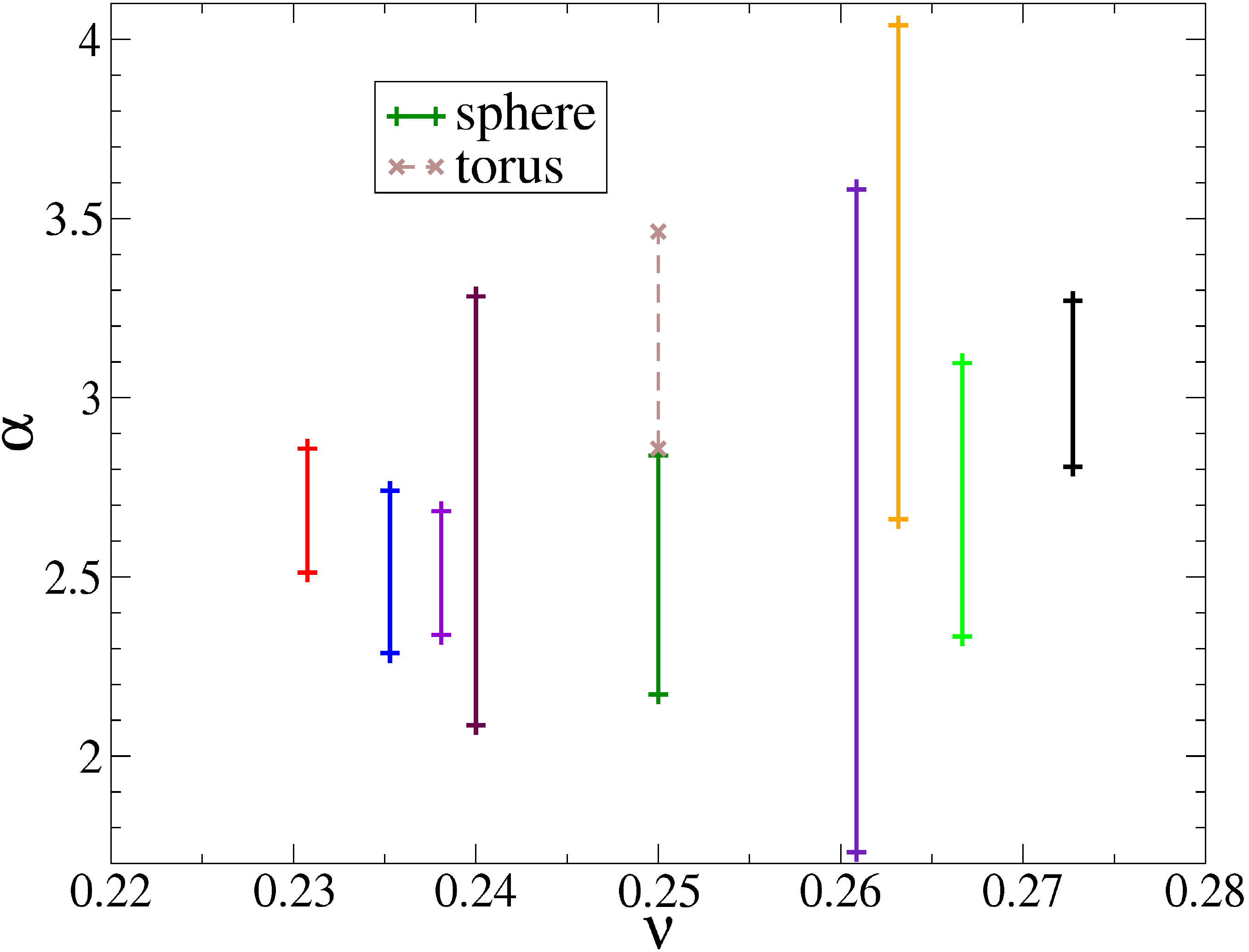}
\end{center}
\caption{Thermodynamic extrapolation of the power law exponent $\alpha$ obtained using the projected Jain wave functions at fillings $n/(4n\pm1)$. The bottom-right most panel shows thermodynamic values of $\alpha$ as a function of the filling factor $\nu$.}
\label{alpha_sm_4CFs}
\end{figure*}

\begin{figure*}[htpb]
\begin{center}
\includegraphics[width=0.4\textwidth,height=0.25\textwidth]{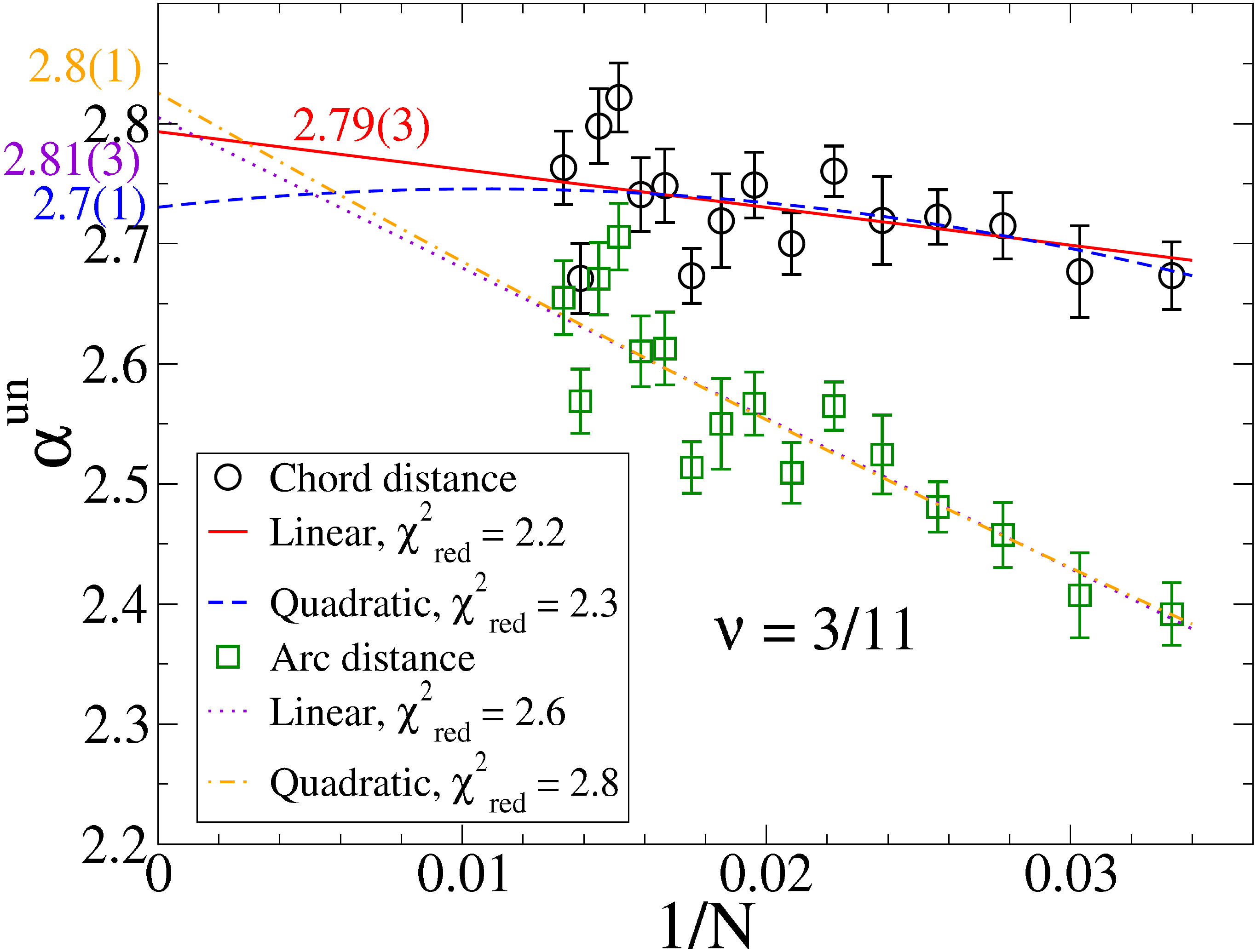}
\includegraphics[width=0.4\textwidth,height=0.25\textwidth]{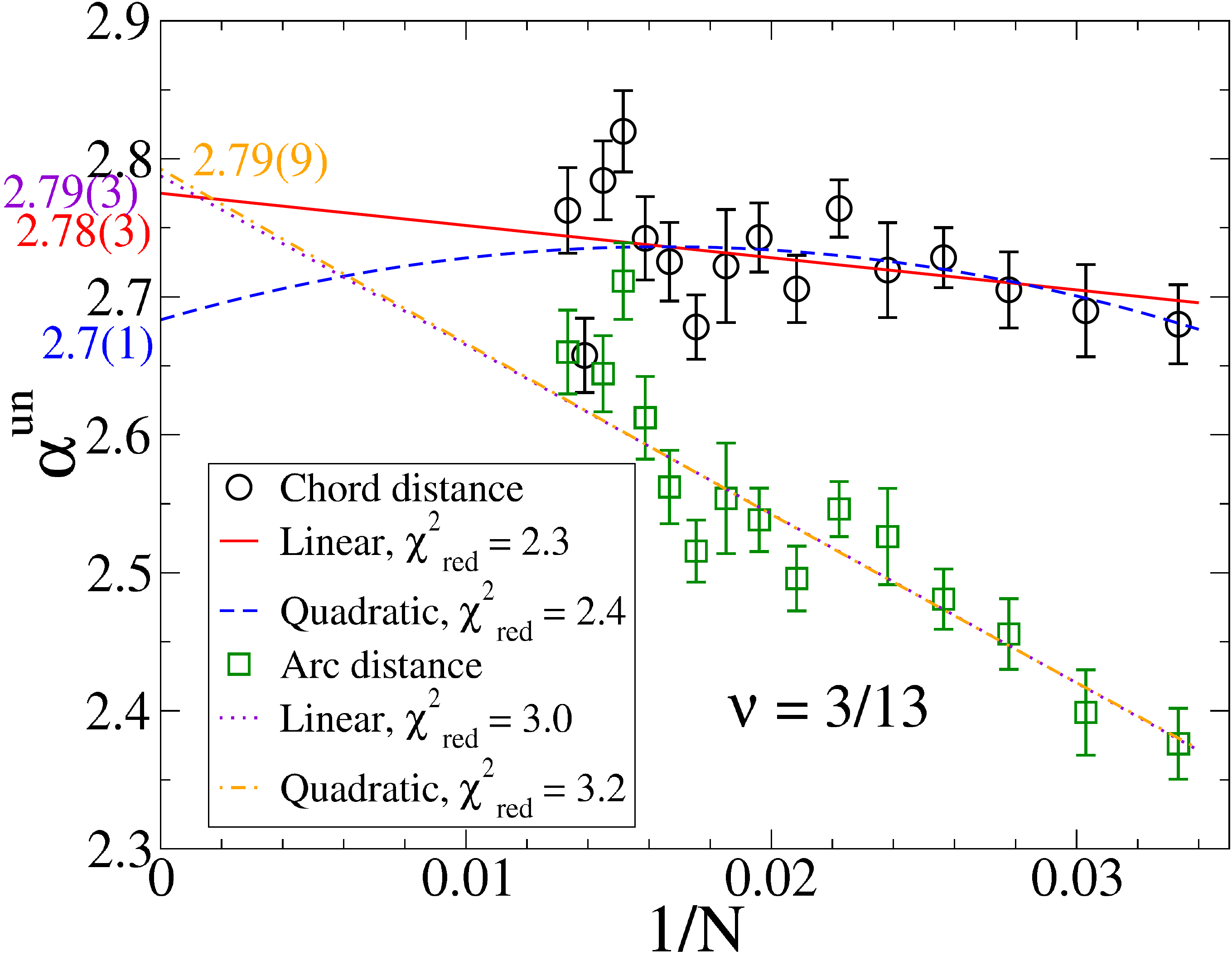}

\includegraphics[width=0.4\textwidth,height=0.25\textwidth]{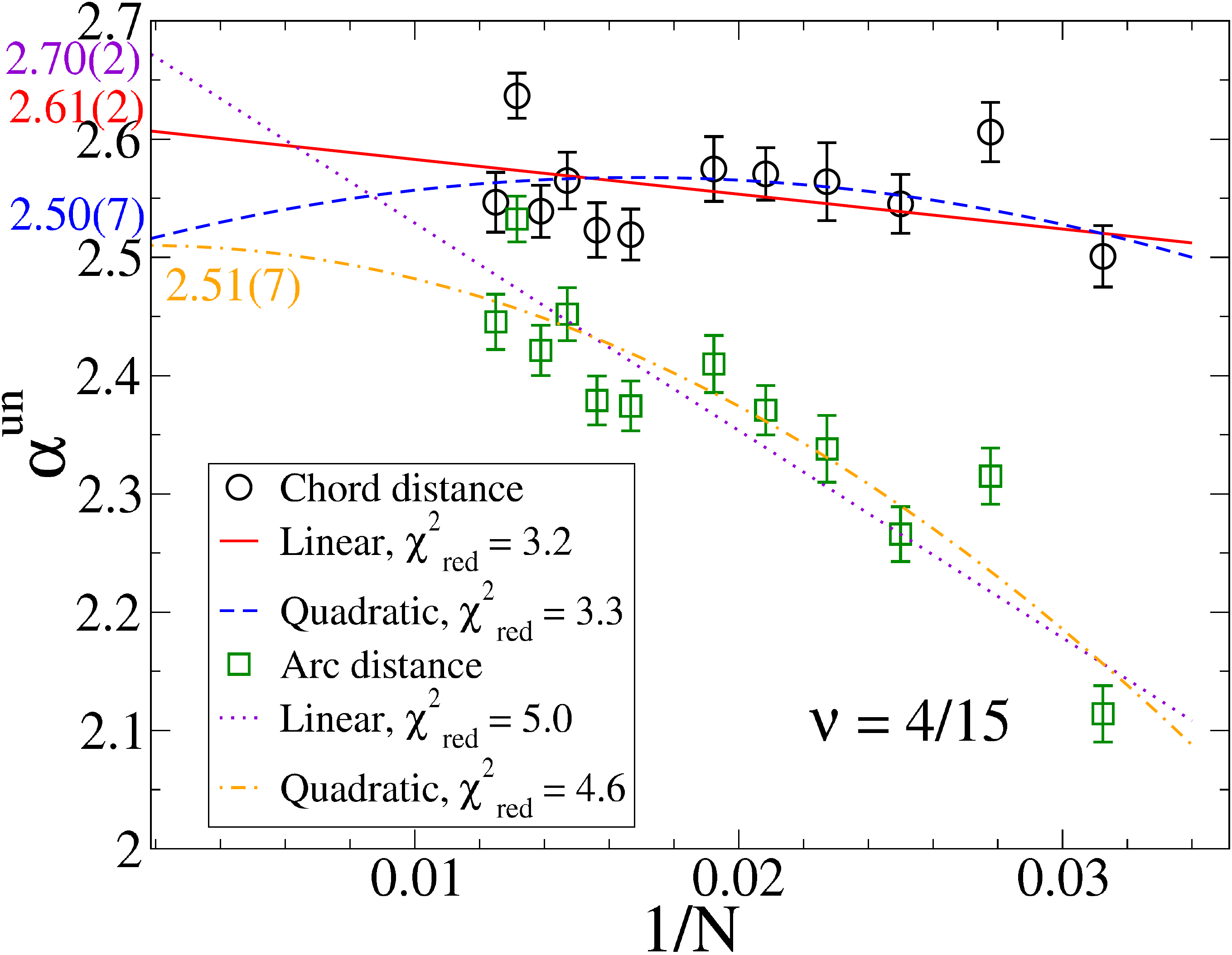}
\includegraphics[width=0.4\textwidth,height=0.25\textwidth]{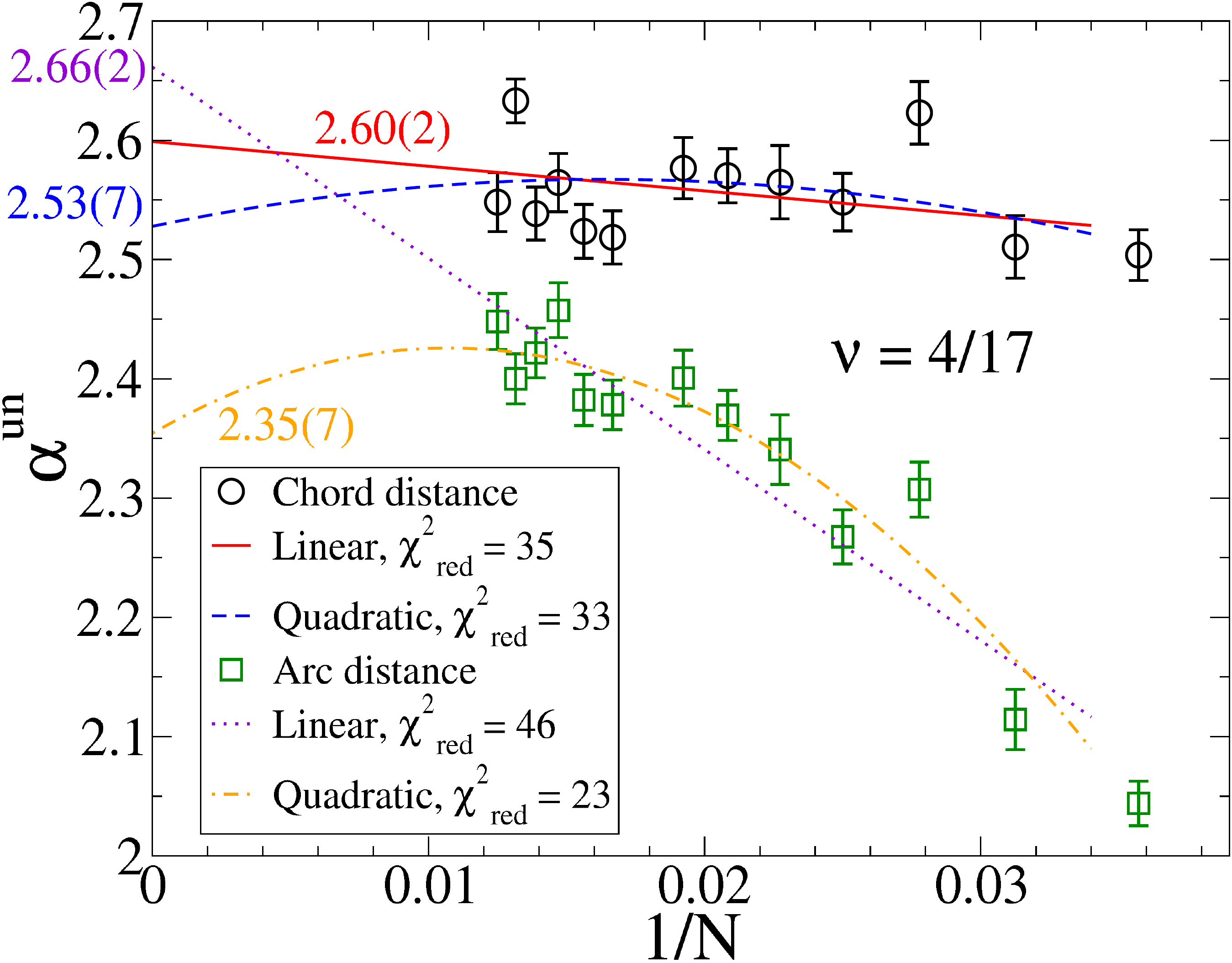}

\includegraphics[width=0.4\textwidth,height=0.25\textwidth]{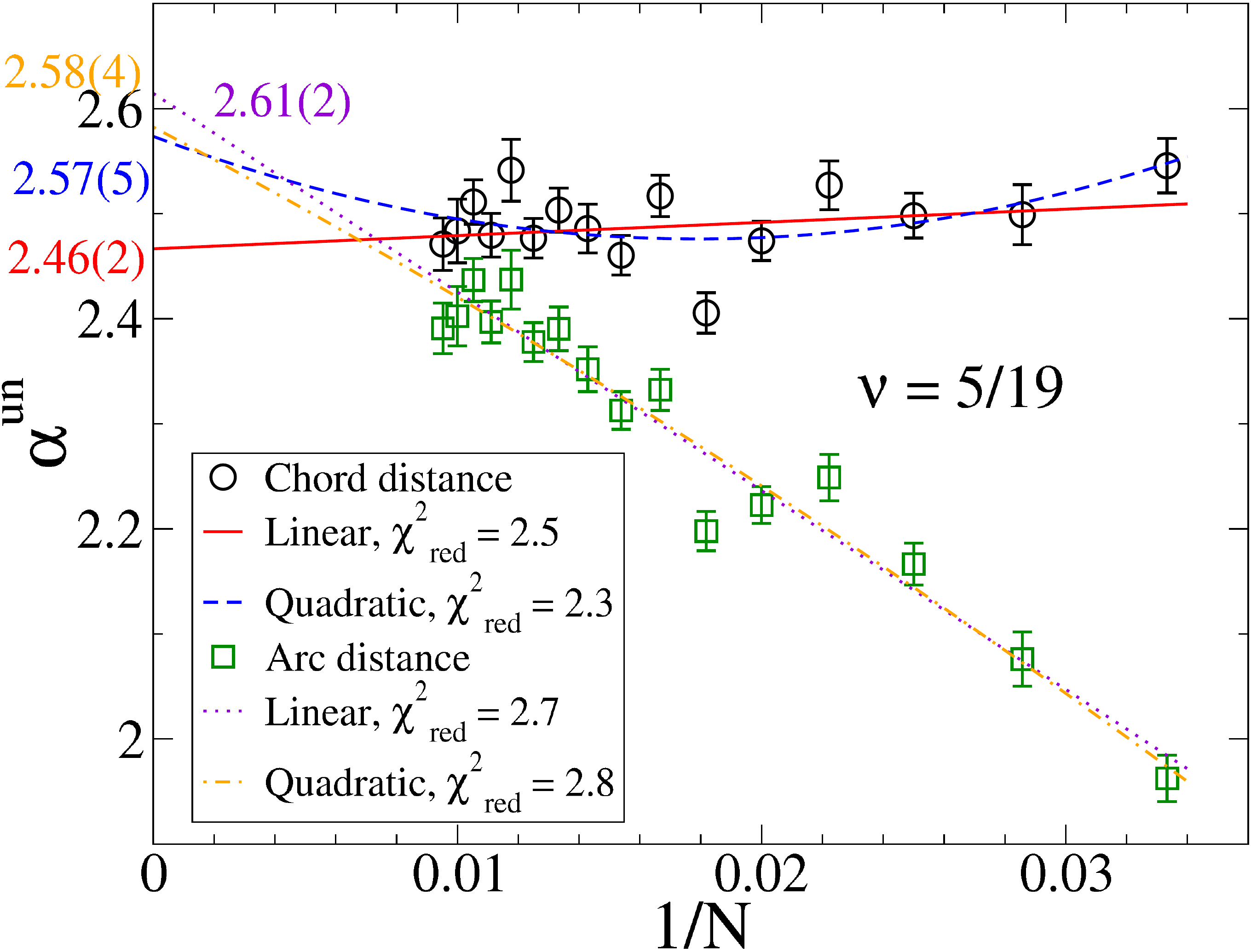}
\includegraphics[width=0.4\textwidth,height=0.25\textwidth]{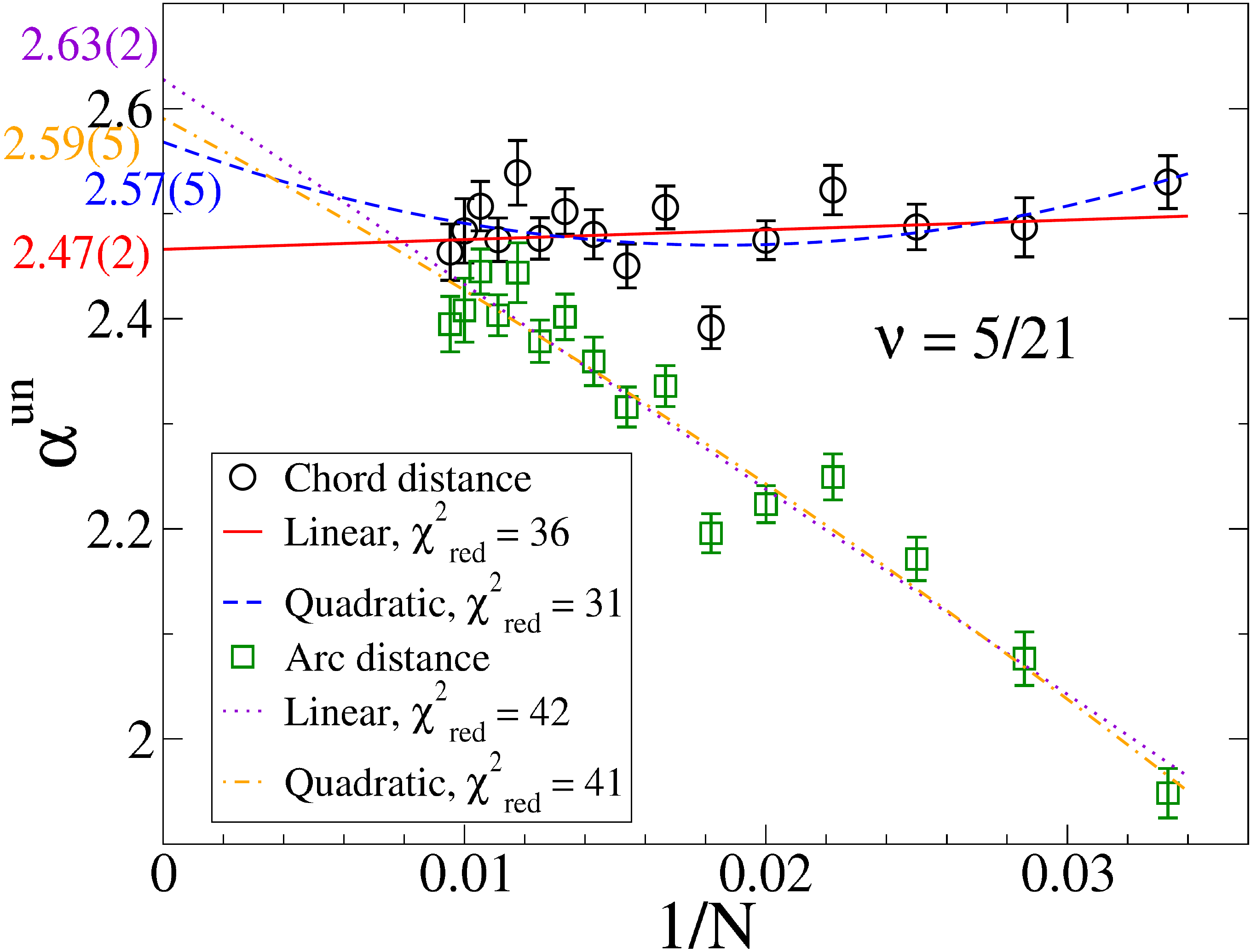}

\includegraphics[width=0.4\textwidth,height=0.25\textwidth]{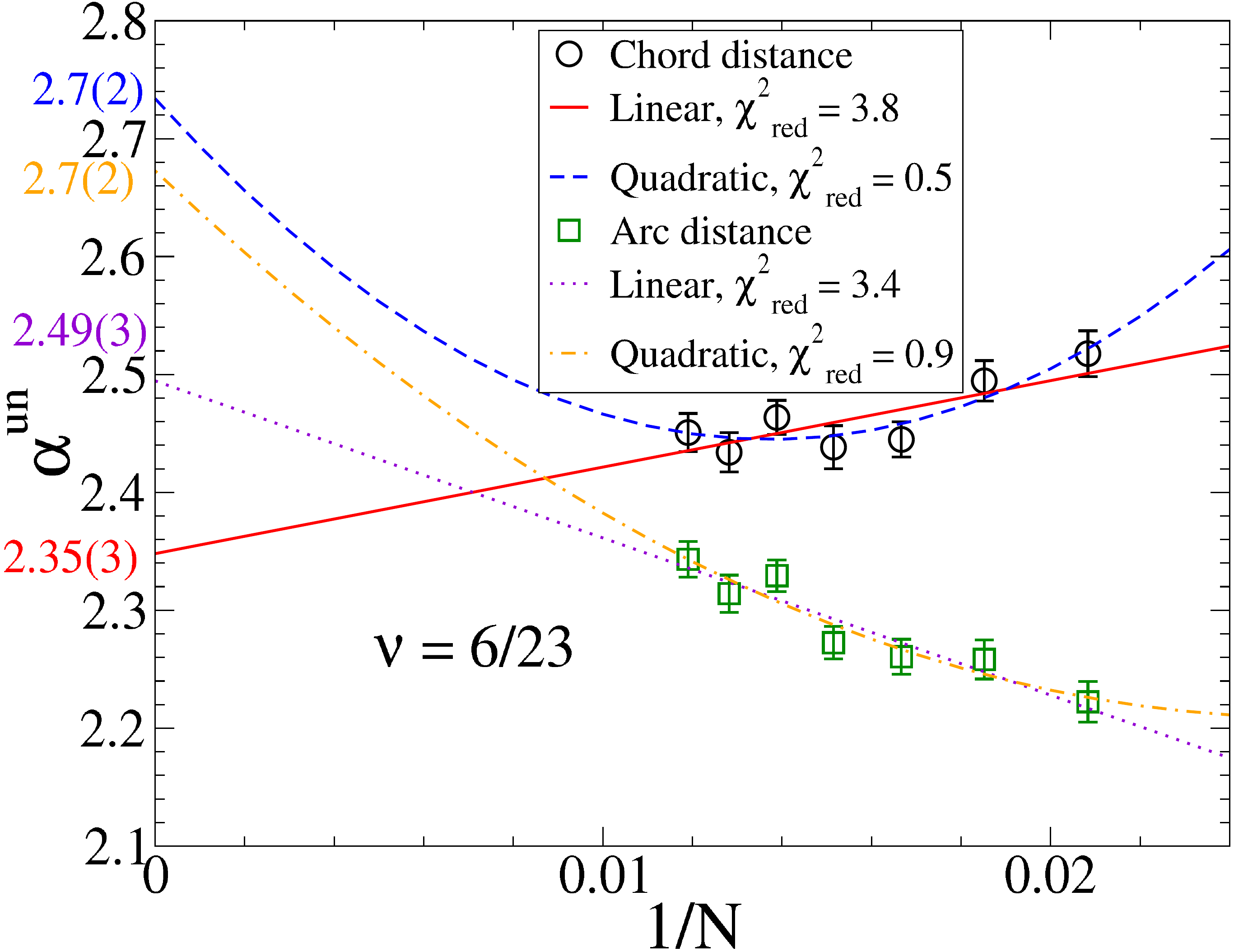}
\includegraphics[width=0.4\textwidth,height=0.25\textwidth]{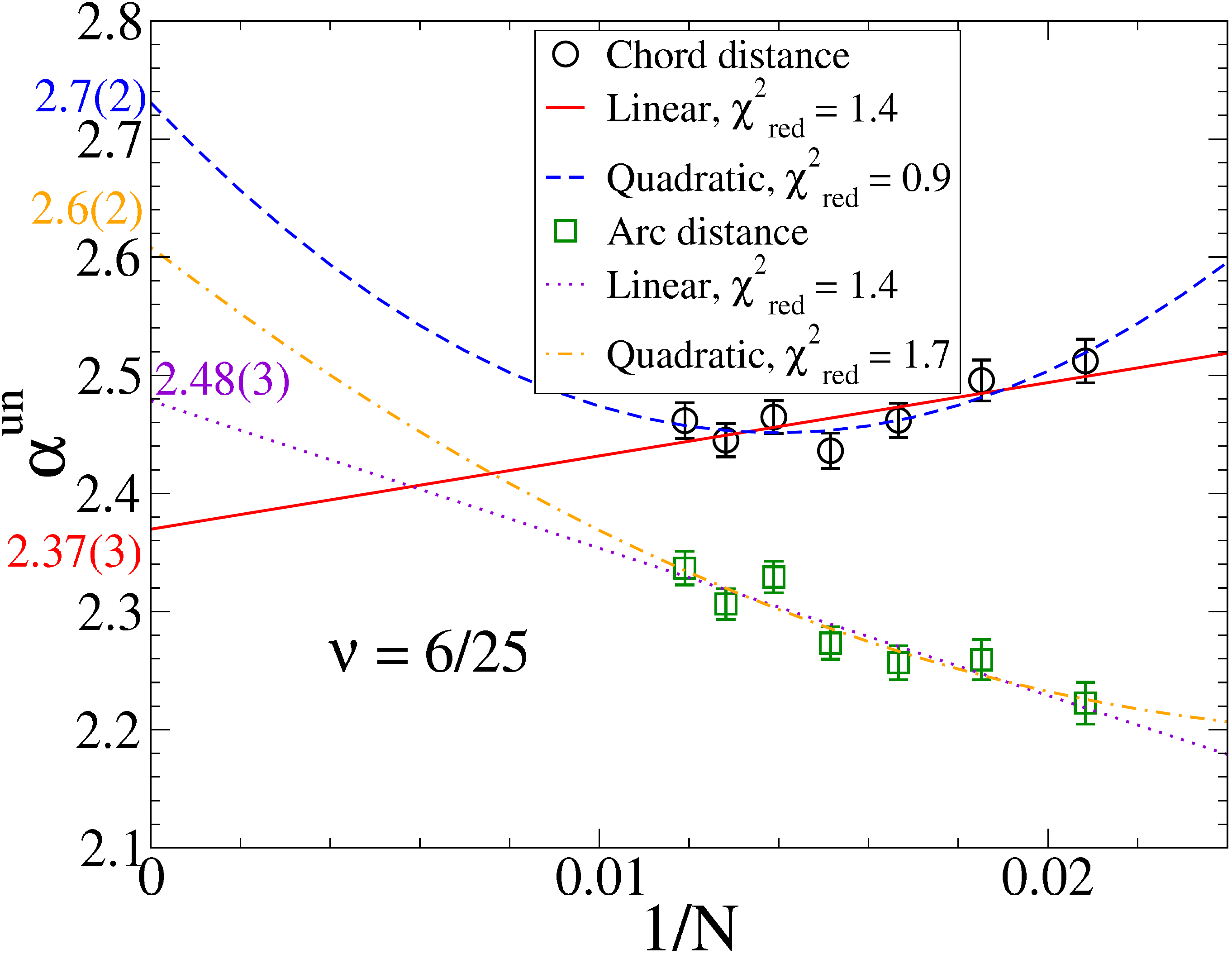}

\includegraphics[width=0.4\textwidth,height=0.25\textwidth]{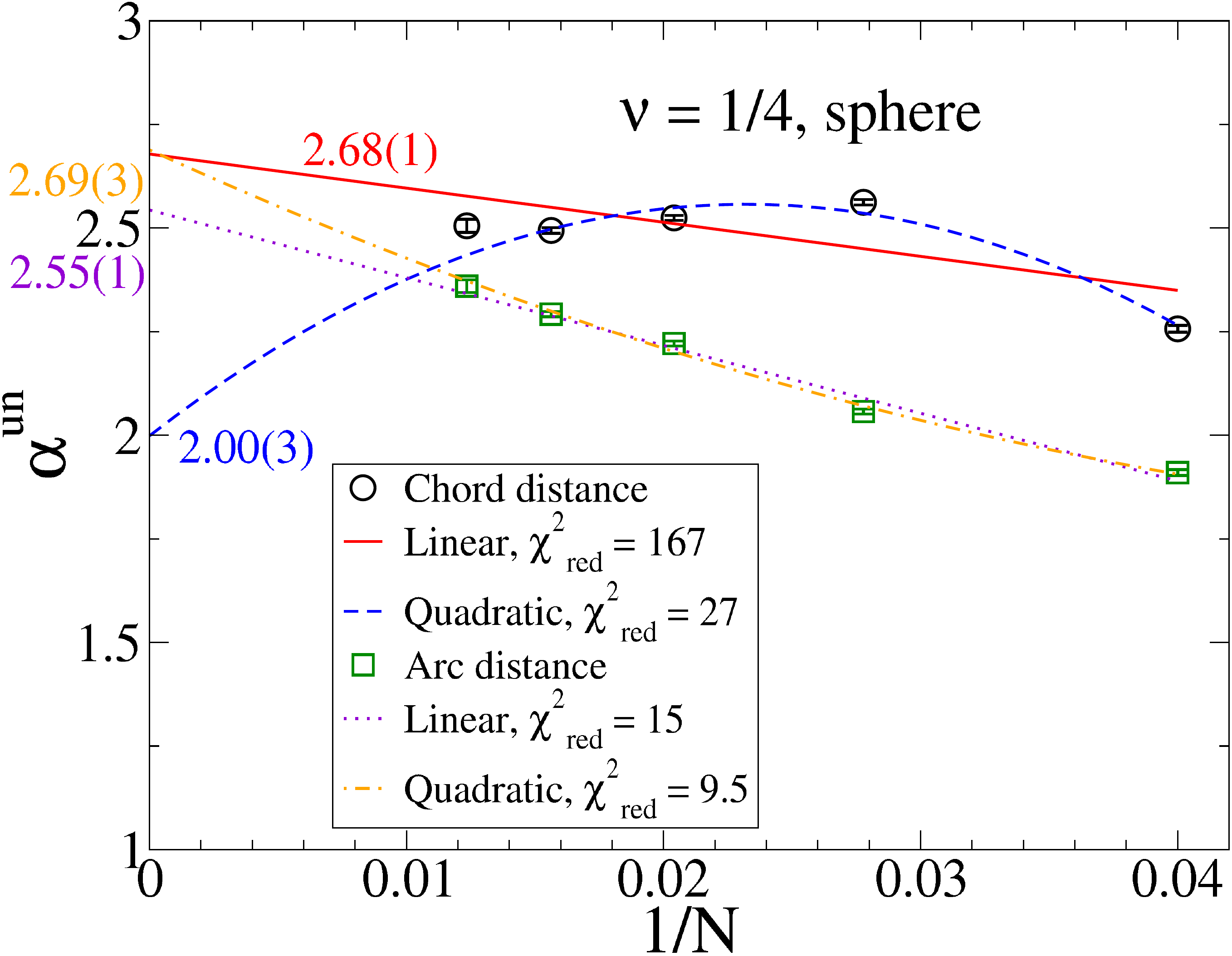}
\includegraphics[width=0.4\textwidth,height=0.25\textwidth]{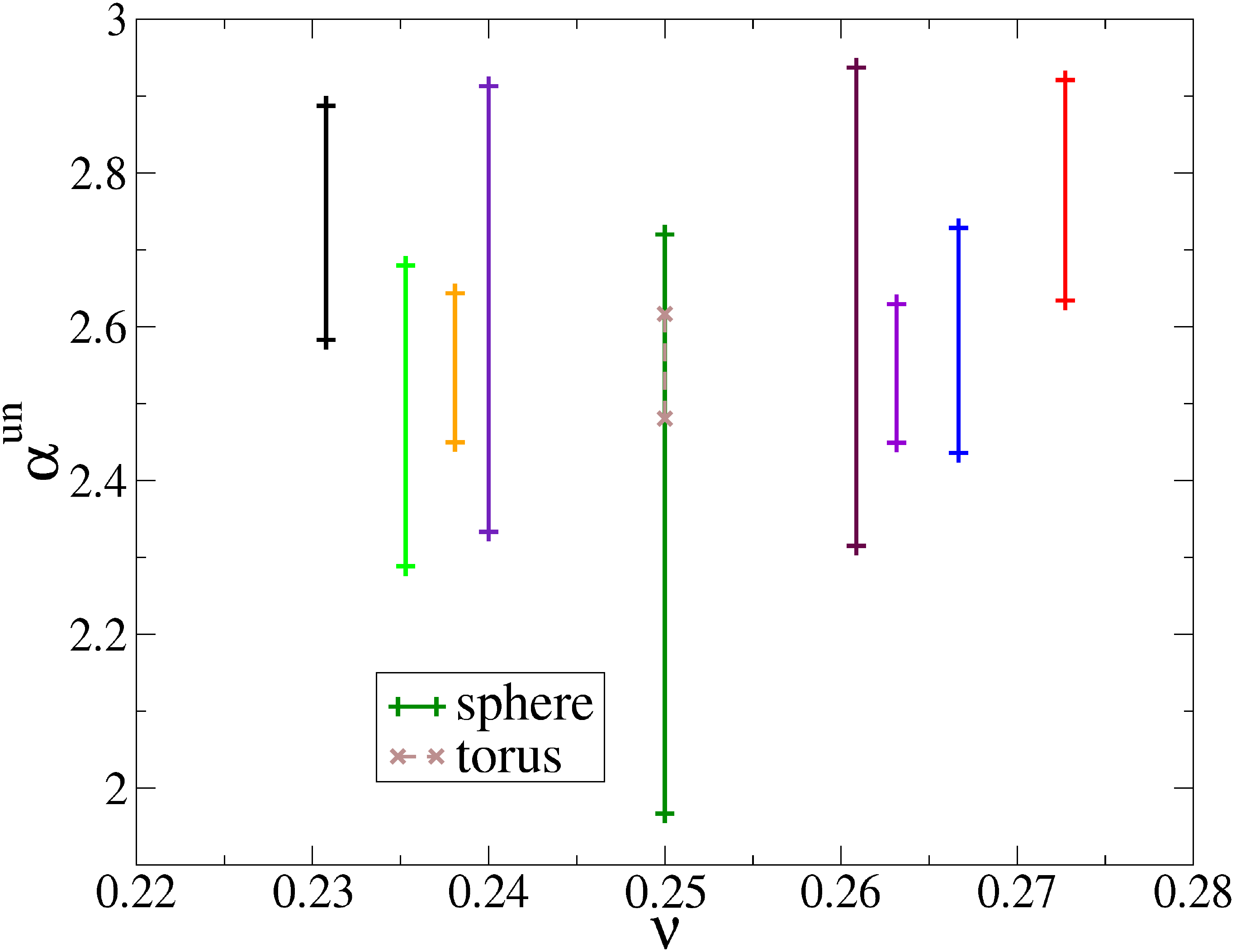}
\end{center}
\caption{Thermodynamic extrapolation of the power law exponent $\alpha^{\rm un}$ obtained using the {\em unprojected}  wave functions $\Psi^{\rm un}_{n/(4n\pm 1)}$.}
\label{alphaun_sm_4CFs}
\end{figure*}

Our results in Fig.~\ref{kFsm_4CFs} (\ref{kFunsm_4CFs}) show the extrapolations of $k_{\rm F}^*\ell$ ($k_{\rm F}^{*un}\ell$) to the thermodynamic limit for the projected (unprojected)  wave functions. The extrapolated values are shown in the bottom-most panels of the respective figures. To obtain these plots we follow exactly the same procedure as illustrated above for the sequence $n/(2n\pm 1)$. We also show the extrapolations for the power law exponent $\alpha$ ($\alpha^{un}$) as defined in Eq. 1 of the main text, obtained from the projected (unprojected) wave functions at various filling factors of composite fermions carrying four vortices in Fig. \ref{alpha_sm_4CFs} (\ref{alphaun_sm_4CFs}).

We find that the obtained Fermi wave vector in the vicinity of $\nu=1/4$ is close to the value $\sqrt{4\pi\rho_{e}}$ for both the projected and unprojected  wave functions. At precisely $\nu=1/4$ we find that the Fermi wave vector obtained from the projected wave functions is close to but slightly different from the mean field value of $1/\sqrt{2}\ell$, suggesting a slight violation of the Luttinger theorem. For the unprojected state at $1/4$ we find that the Fermi wave vector is consistent with the mean field value of $1/\sqrt{2}\ell$. 

The situation close to $\nu=3/4$ is analogous. The CF Fermi wave vector will be close to $k_{\rm F}^{*}\ell=\sqrt{4\pi \rho_{\rm h}}=\sqrt{2(1-\nu)}$ for situations where p-h symmetry is valid.

Since we are not able to go beyond 81 particles (81 particles) for $\nu=1/2$ ($\nu=1/4$) in the spherical geometry, we have also considered the CFFS in the torus geometry, where we have studied the CFFS for up to 153 particles (121 particles). The calculation in the torus geometry is described next.

\section{CF Fermi wave vector in the torus geometry}
A brief outline for the calculation on the torus geometry for $\nu=1/2p$, where $p$ is an integer, is as follows: We consider a periodic geometry in the plane spanned by two nonparallel vectors, $\mathbf L_1$ an $\mathbf L_2$. Using the symmetric gauge $\mathbf A=(-By/2,Bx/2,0)$ for a homogeneous magnetic field $\mathbf B=(0,0,B)$, the wave functions in the lowest Landau level have the form
\begin{equation}
\psi(x,y)=\exp\left(-\frac{1}{2}|z|^2\right)f(z),
\end{equation}
where $z=(x+iy)/\sqrt2\ell$ and $f(z)$ is an entire function.
Consequently, the many-body states have the form
\begin{equation}
\Psi=\exp\left(-\frac{1}{2}\sum_{i=1}^N|z_i|^2\right)F(z_1,z_2,\dots,z_N),
\end{equation}
where $F$ is entire.
We will refer to the first factor as the nonanalytic part.
We require periodic boundary conditions in each coordinate,\cite{Haldane85}
\begin{equation}
\label{pbc}
t_m(\mathbf L_j) \Psi = \exp(i\phi_j)\Psi,
\end{equation}
where $j\in\{1,2\}$, $\phi_j$ are fixed constants, and $t_m(\mathbf L)$ is the magnetic translation operator
acting on the complex coordinate $z_m$ defined in terms of the Cartesian ones $\mathbf r_m=(x_m,y_m)$:
\begin{equation}
t_m(\mathbf L) = \exp\left[\mathbf L\cdot\left(\nabla - (ie/\hbar)\mathbf A\right)
- \frac{i}{\ell^2}(\mathbf L\times\mathbf r_m)\cdot\hat{e}_{z} \right].
\end{equation}
The two conditions in Eq.~(\ref{pbc}) are compatible only if
\begin{equation}
|\mathbf L_1\times\mathbf L_2|=2\pi \ell^2 N_\phi,
\end{equation}
where $N_\phi$ is an integer; i.e., the parallelogram spanned by $\mathbf L_1$ and $\mathbf L_2$
encloses an integer number of flux quanta.

The unprojected CF Fermi sea is simply the product of a Slater determinant of plane waves and the Jastrow factor that binds an even number $2p$ of flux quanta of the many-body wave function to
each electron:
\begin{equation}
\label{unprojcffs}
\text{Det}\left(e^{i\mathbf k_j\cdot\mathbf r_i}\right)\Psi_{1/2p},
\end{equation}
where $\Psi_{1/2p}$ is the bosonic Laughlin state\cite{Rezayi94,Rezayi00} at filling factor $\nu=1/2p$,
which is a function of $N=N_\phi/2p$ electron coordinates, and $\{\mathbf k_j\}$ are
the momenta of the filled CF states, taken from reciprocal lattice that corresponds to the direct lattice
spanned by $\mathbf L_1$ and $\mathbf L_2$.

When the lowest Landau level projection is implemented in the periodic geometry,
differentiation with respect to $z_m$ becomes a (magnetic) single-body translation operation.
In particular, the CF Fermi sea state is\cite{Shao15}
\begin{equation}
\label{fullcffs}
\text{Det}\left(t_i(-\mathbf d_j)\right)\Psi_{1/2p},
\end{equation}
Here, $\mathbf d = (m\mathbf L_1 + n\mathbf L_2)/N_\phi$, and the momenta of the filled CF states
are generated as $\mathbf k=(d_y/\ell^2,-d_x/\ell^2)$.
With the determinant spelled out, there are $N!$ terms in Eq.~(\ref{fullcffs}), making it
computationally intractable.
We will use the convenient approximate projection scheme elaborated by Shao \textit{et al.}\cite{Shao15}
This leads to the wave function
\begin{multline}
\Psi_\text{CFFS}=\text{Det}_{i,j}\left[
e^{d_j^\ast z_i}\prod_{k\neq i}\sigma^{2p}(z_i-z_k-2(d_j-\overline d))\right]\\
\times F_\text{c.m.}\left(\sum_{i=1}^N(z_i-\overline d)\right)
\exp\left(-\frac{1}{2}\sum_{i=1}^N|z_i|^2\right)
\end{multline}
where $d_j=(d_{j,x} + id_{j,y})/\sqrt2\ell$, $\overline d=(1/N)\sum_{i=1}^N d_j$,
$F_\text{c.m.}$ in an analytic center-of-mass wave function, and $\sigma(z)$ is the Weierstass' sigma function
appropriate to the lattice spanned by the complex numbers $L_1=(L_{1,x} + iL_{1,y})/\sqrt2\ell$
and $L_2=(L_{2,x} + iL_{2,y})/\sqrt2\ell$:
\begin{equation}
\sigma(z)=\frac{\vartheta_1(\pi z/L_1|\tau)}{(\pi/L_1)\vartheta'_1(0|\tau)}\exp\left(
\frac{i(\pi z/L_1)^2}{\pi(\tau-\tau^\ast)}
\right)
\end{equation}
with $\tau=L_2/L_1$.
Using the properties of the $\sigma$ function, one obtains by elementary steps that the periodic boundary
conditions in Eq.~(\ref{pbc}) imply the following requirements ($j\in\{1,2\}$) for the center-of-mass wave function:
\begin{equation}
\label{pbccom}
\frac{F_\text{c.m.}(Z+L_2)}{F_\text{c.m.}(Z)}=\exp\left(
i\phi_j+\frac{p|L_j|^2}{N_\phi}+\frac{2pL_j^\ast}{N_\phi}Z\right).
\end{equation}
As the pair correlation should not depend on the center-of-mass motion, we are content with finding a particular
solution to Eq.~(\ref{pbccom}).
Seeking it in the form that is structurally analogous to the $\sigma$ function,
\begin{equation}
F_\text{c.m.}(Z)=\exp\left(\frac{\alpha Z^2}{L_1}\right)
\vartheta_1\left(\frac{\pi\beta Z}{L_1}|\beta\tau\right),
\end{equation}
the boundary conditions fix $\beta=2p$ and $\alpha=L_1^\ast/N_\phi$,
if $\phi_1=0$ and $\phi_2=\pi$.

\begin{figure*}[t]
\begin{center}
\includegraphics[width=0.48\textwidth]{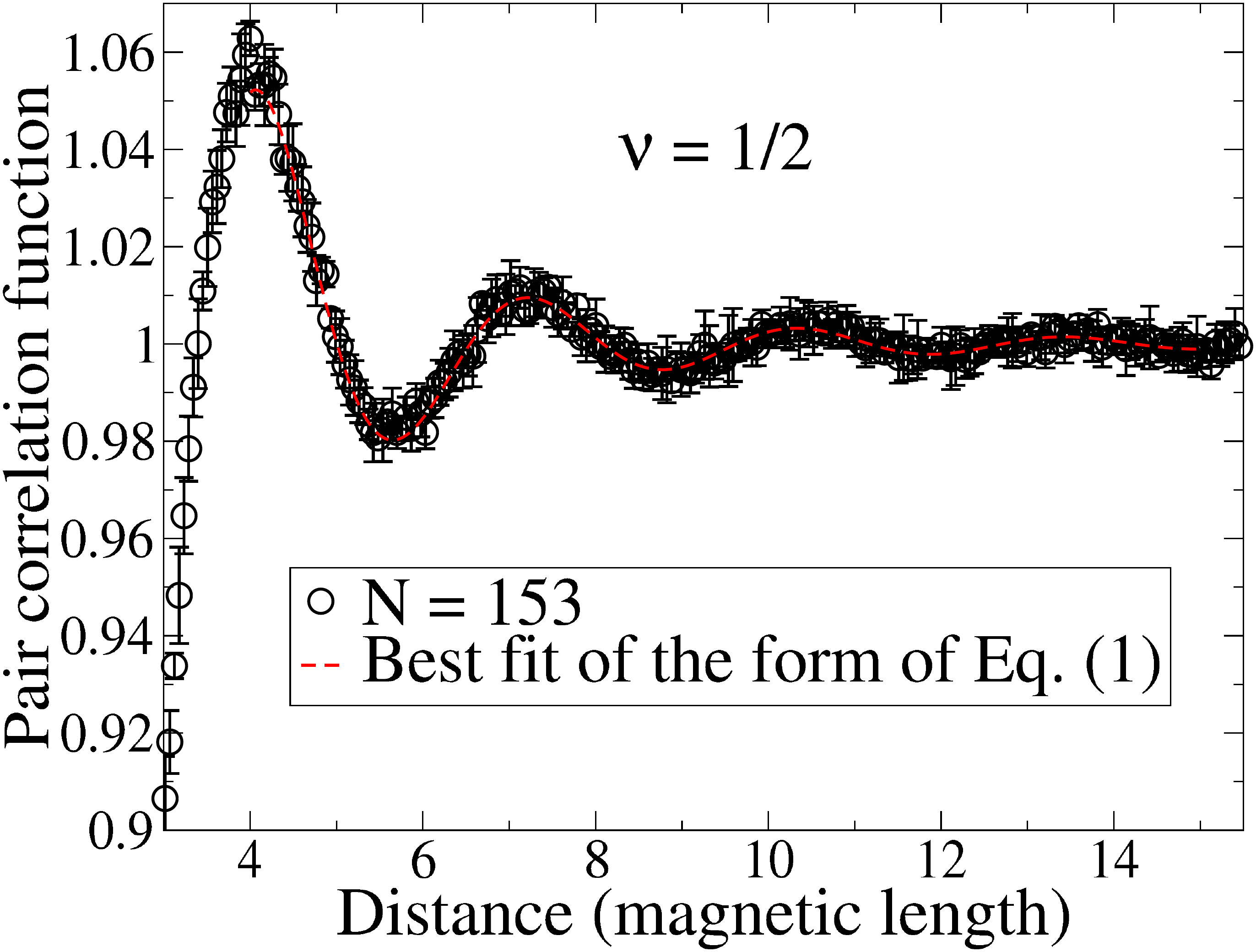}
\includegraphics[width=0.48\textwidth]{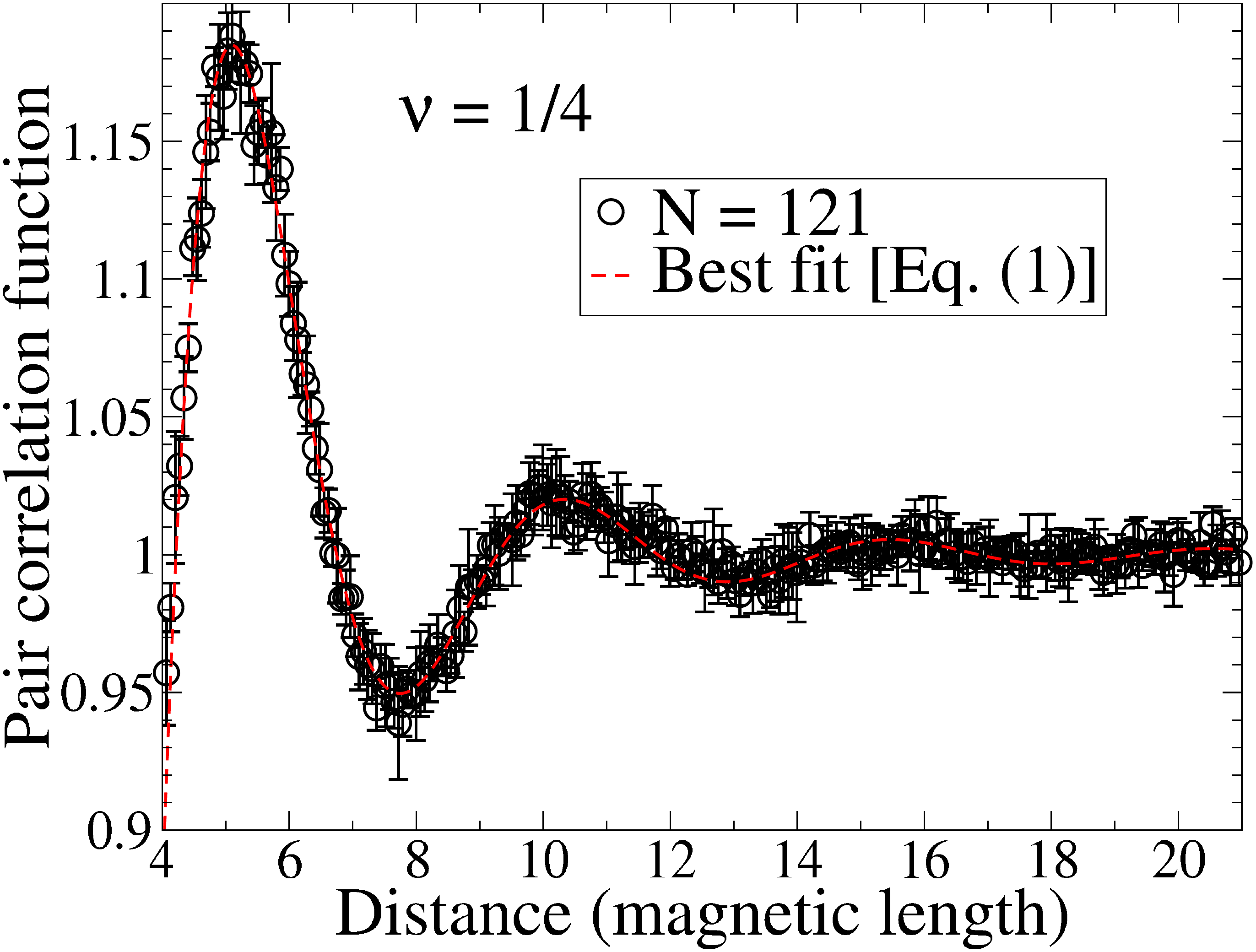}
\end{center}
\caption{
Fit of the Friedel oscillations in the pair correlation function of the composite fermion Fermi sea state at $\nu=1/2$ (left) and $\nu=1/4$ (right) in the torus geometry. The results shown are for the largest systems considered in this work. }
\label{fsfit}
\end{figure*}

\begin{figure*}[htpb]
\begin{center}
\includegraphics[width=8cm,height=5cm]{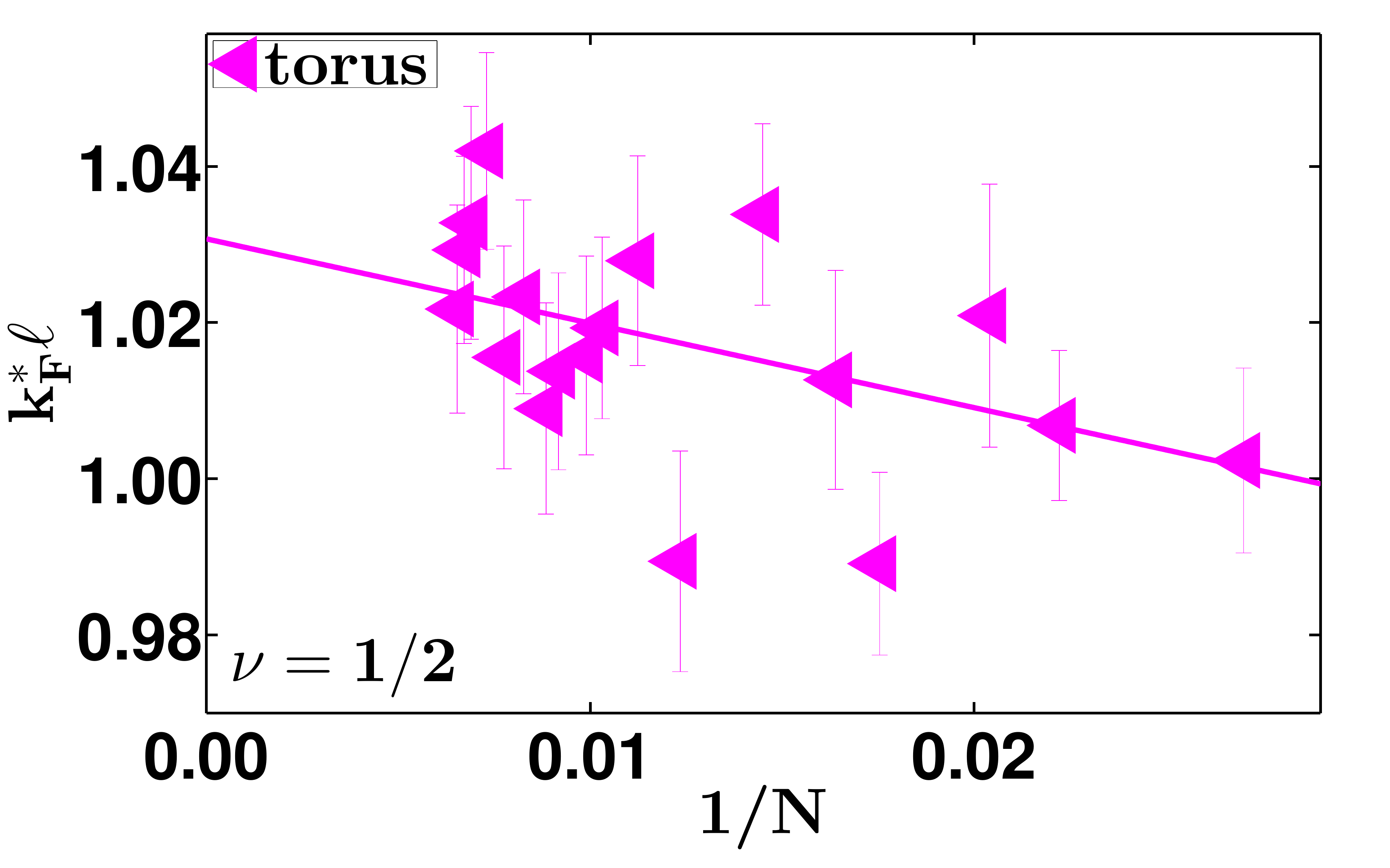}
\includegraphics[width=8cm,height=5cm]{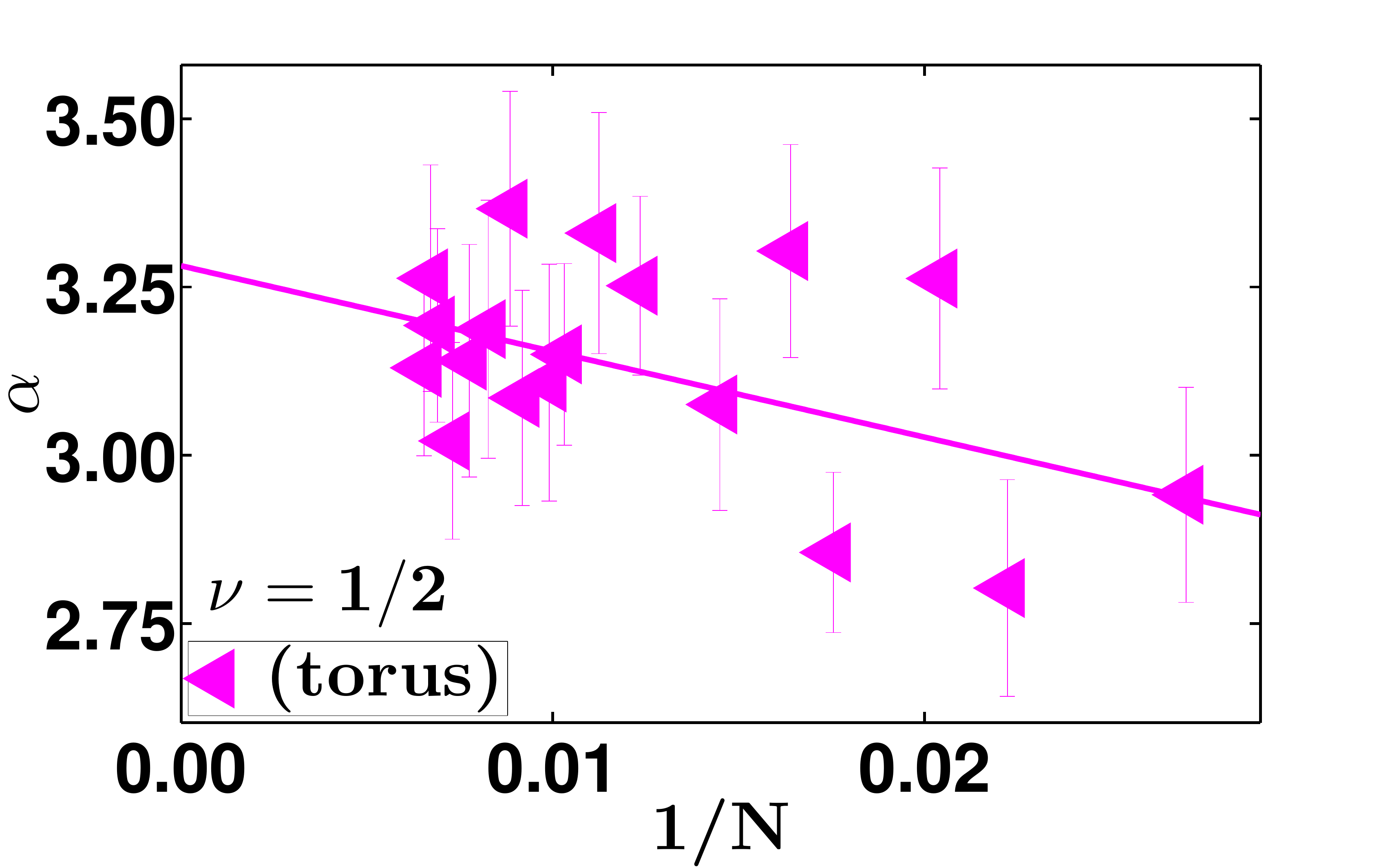}
\end{center}
\caption{
The Fermi wave vector (left panel) and the power law exponent $\alpha$ (right panel) of the projected composite fermion Fermi sea state at $\nu=1/2$ on the torus geometry as a function of $1/N$. The linear extrapolation to the thermodynamic limit is also shown.}
\label{torus_fs_p}
\end{figure*}

Specializing now for a square unit cell, $\mathbf L_1\cdot\mathbf L_2=0$ and $|\mathbf L_1|=|\mathbf L_2|$,
the pair correlation was obtained up to $N\leq 153$ at $\nu=1/2$ and up to $N\leq 121$ at $\nu=1/4$.
The functional form in Eq.~(1) of the paper was fitted to obtain the Fermi wave vector.
Examples are shown in Fig.~\ref{fsfit}. The extrapolation to the thermodynamic limit is shown in Fig.~\ref{torus_fs_p}.
At $\nu=1/2$, we get $k_{\rm F}^*l=1.03(1)$ for a linear fit in $1/N$. When we allow both linear and quadratic fits, the range is given by $k_{\rm F}^*l=1.02-1.07$, which is consistent with the results from the spherical geometry quoted in the main text.
At $\nu=1/4$, we get $k_{\rm F}^*l=0.66(2)$ for a linear fit; with both linear and quadratic fits we get the range $k_{\rm F}^*l=0.61-0.69$, which is again consistent with the results from the spherical geometry.

\begin{figure*}[htpb]
\begin{center}
\includegraphics[width=0.48\textwidth,keepaspectratio]{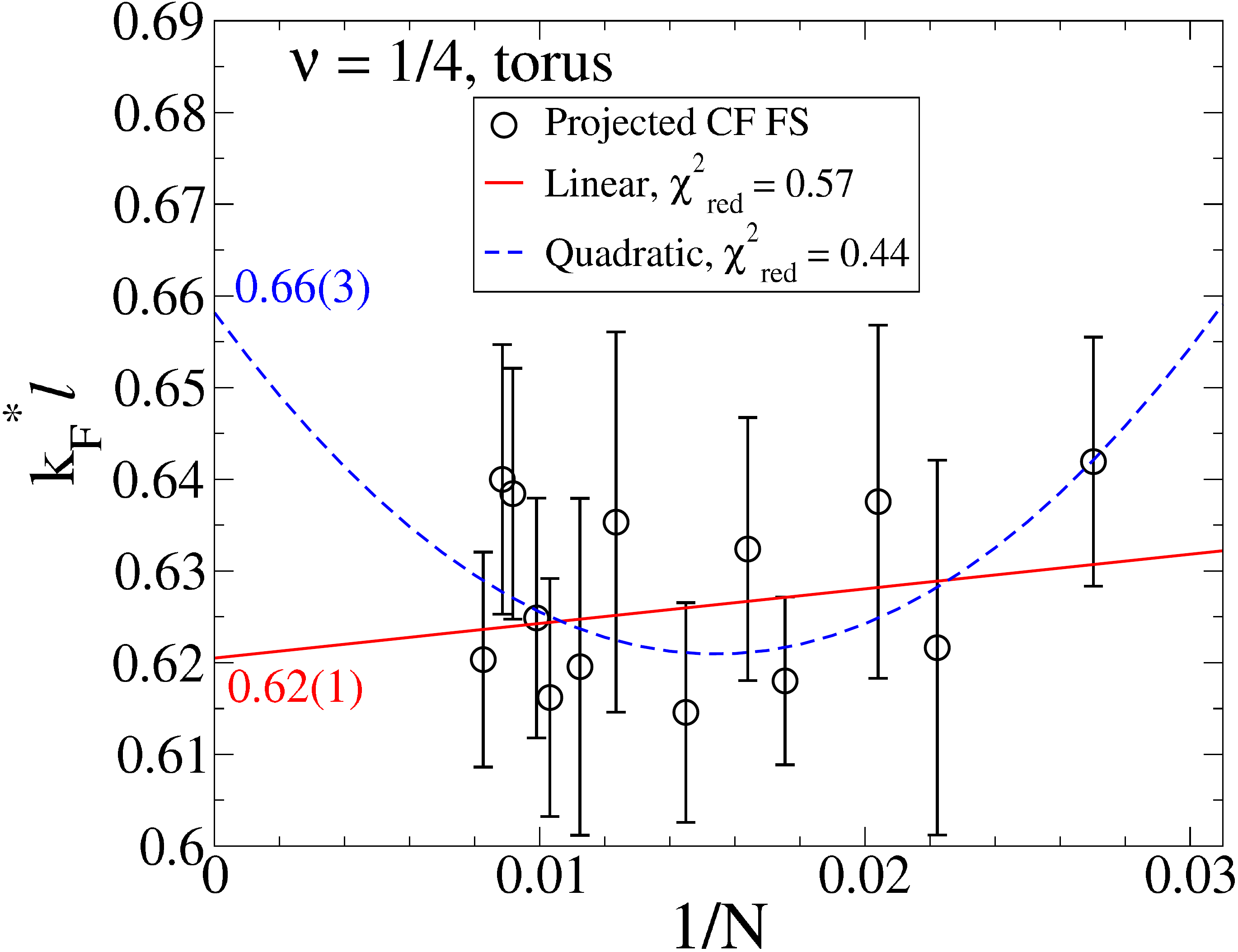}
\includegraphics[width=0.48\textwidth,keepaspectratio]{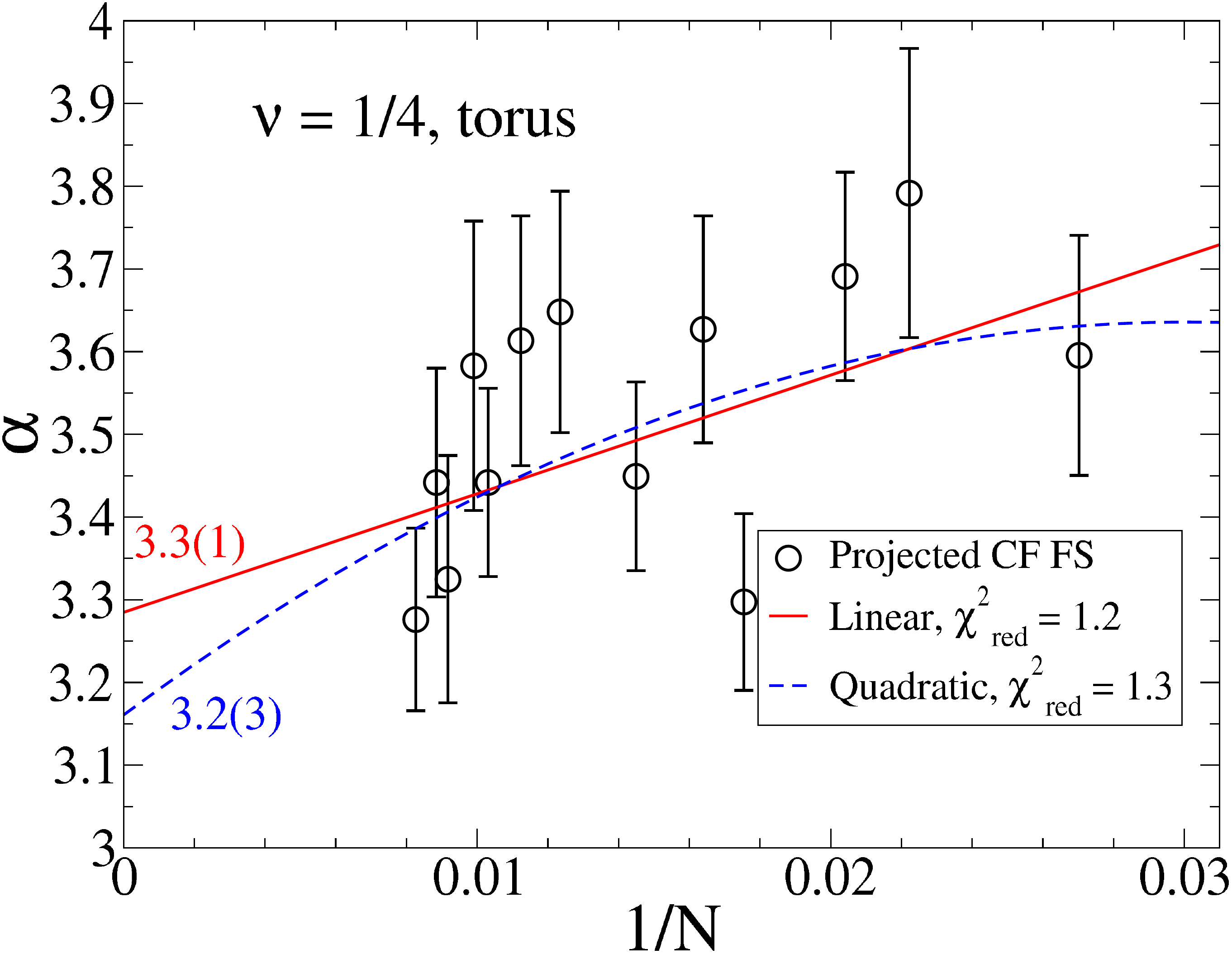}
\end{center}
\caption{
The Fermi wave vector (left panel) and the power law exponent $\alpha$ (right panel) of the projected composite fermion Fermi sea state at $\nu=1/4$ on the torus geometry as a function of $1/N$. We show both the linear and the quadratic extrapolations to the thermodynamic limit.}
\label{torus_fs4_p}
\end{figure*}

\begin{figure*}[htpb]
\begin{center}
\includegraphics[width=8cm,height=5cm]{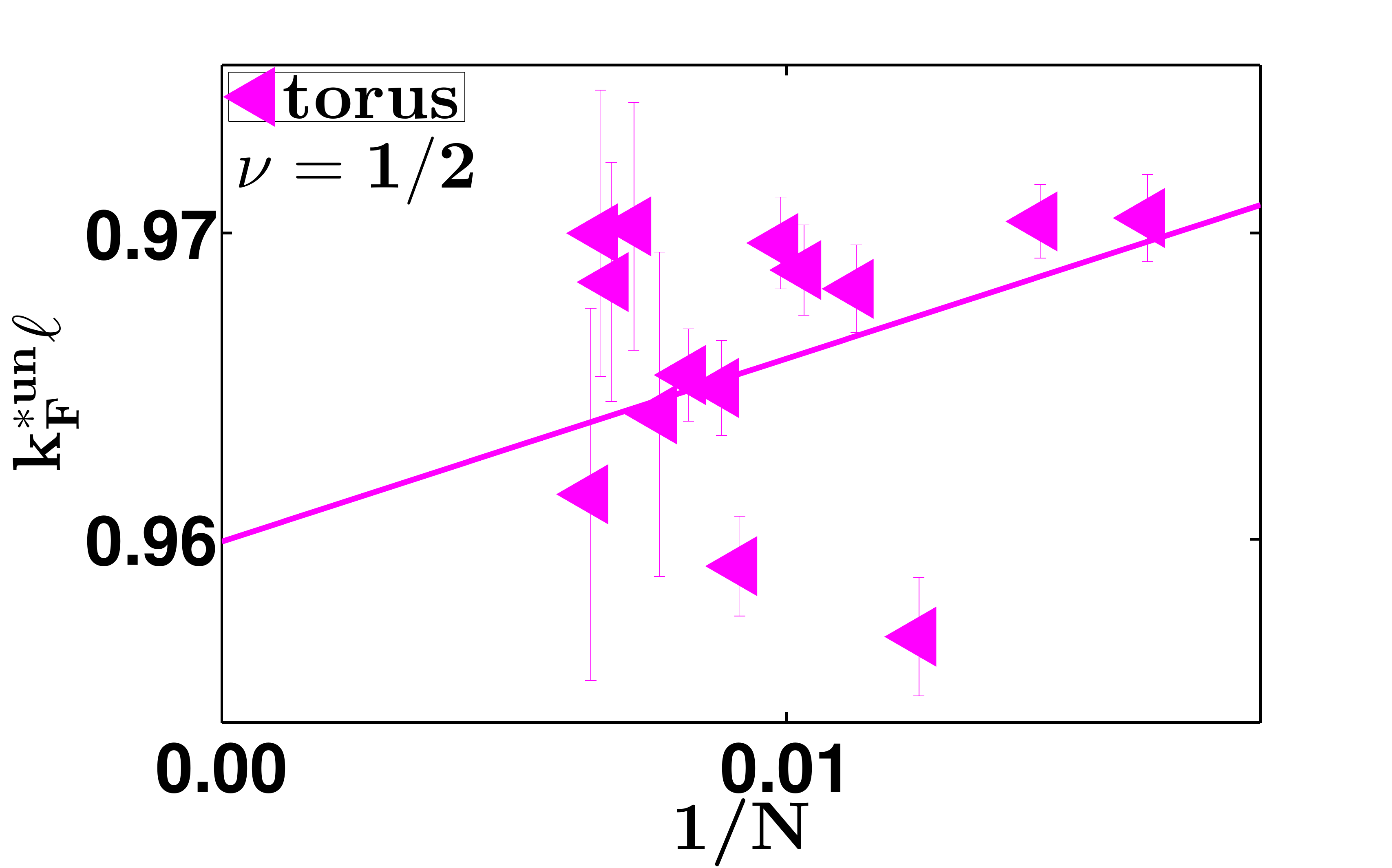}
\includegraphics[width=8cm,height=5cm]{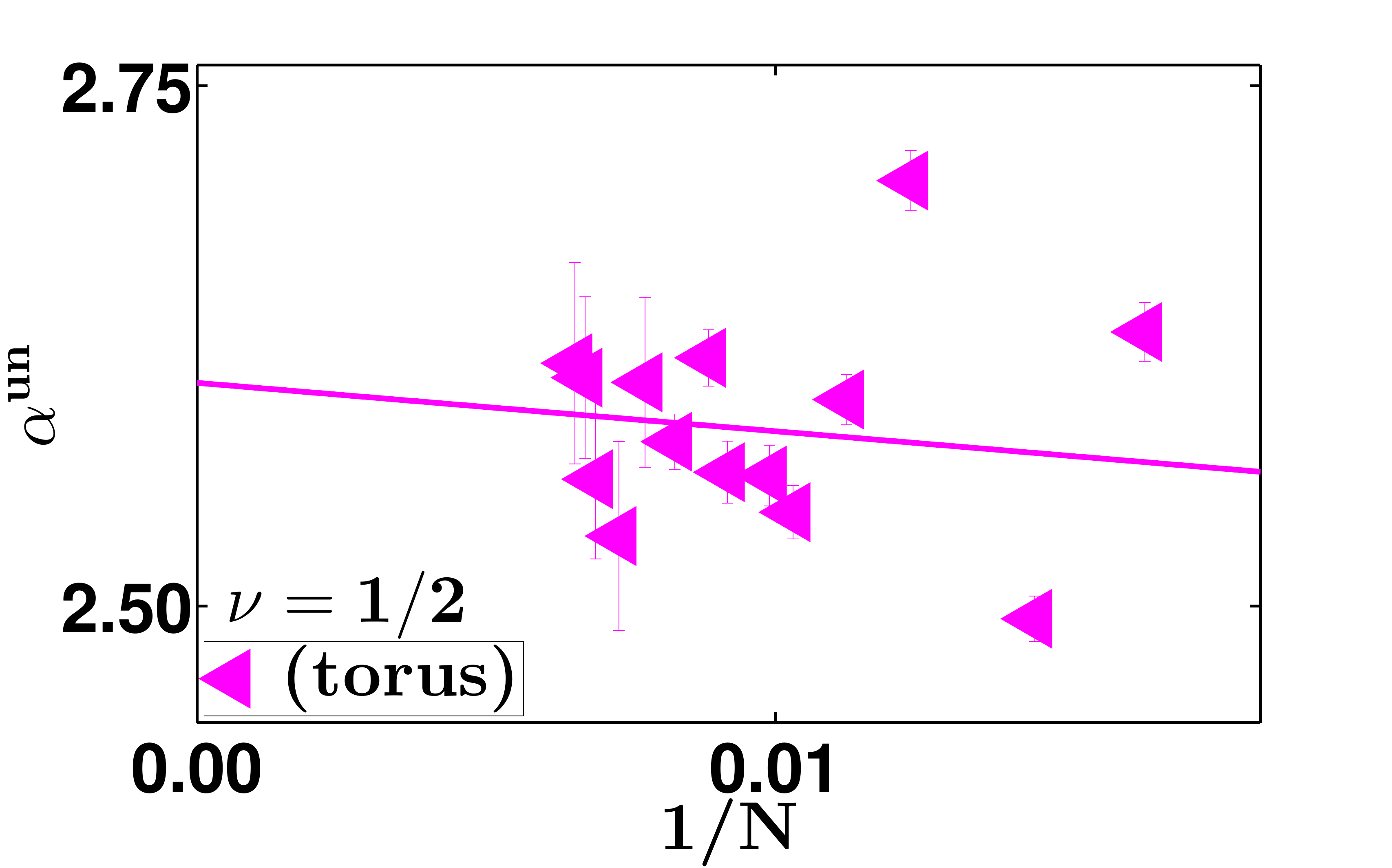}
\end{center}
\caption{
The Fermi wave vector (left panel) and the power law exponent $\alpha^{\rm un}$ (right panel) of the unprojected composite fermion Fermi sea state at $\nu=1/2$ on the torus geometry as a function of $1/N$. The linear extrapolation to the thermodynamic limit is also shown.}
\label{torus_fs_un}
\end{figure*}

\begin{figure*}[htpb]
\begin{center}
\includegraphics[width=0.48\textwidth,keepaspectratio]{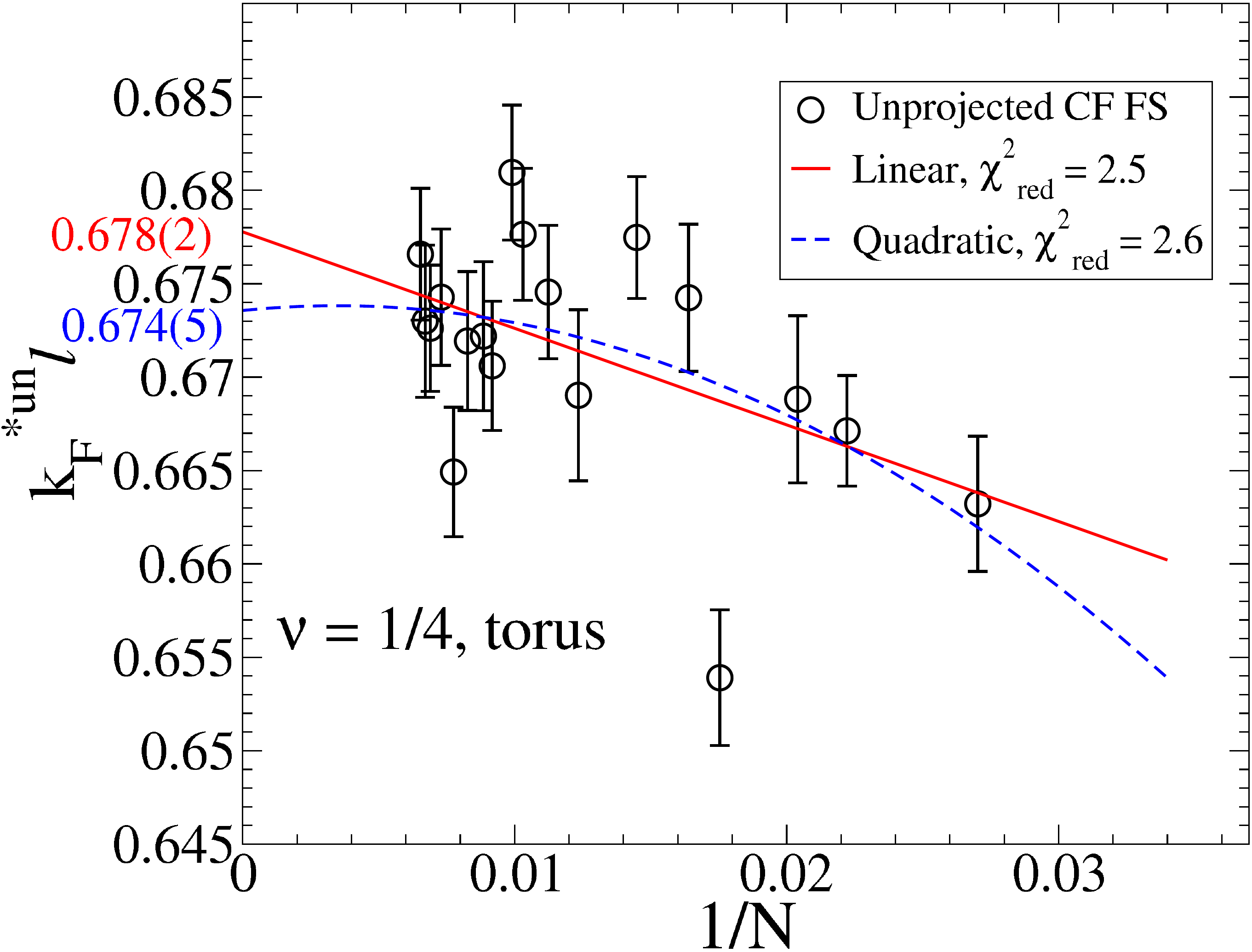}
\includegraphics[width=0.48\textwidth,keepaspectratio]{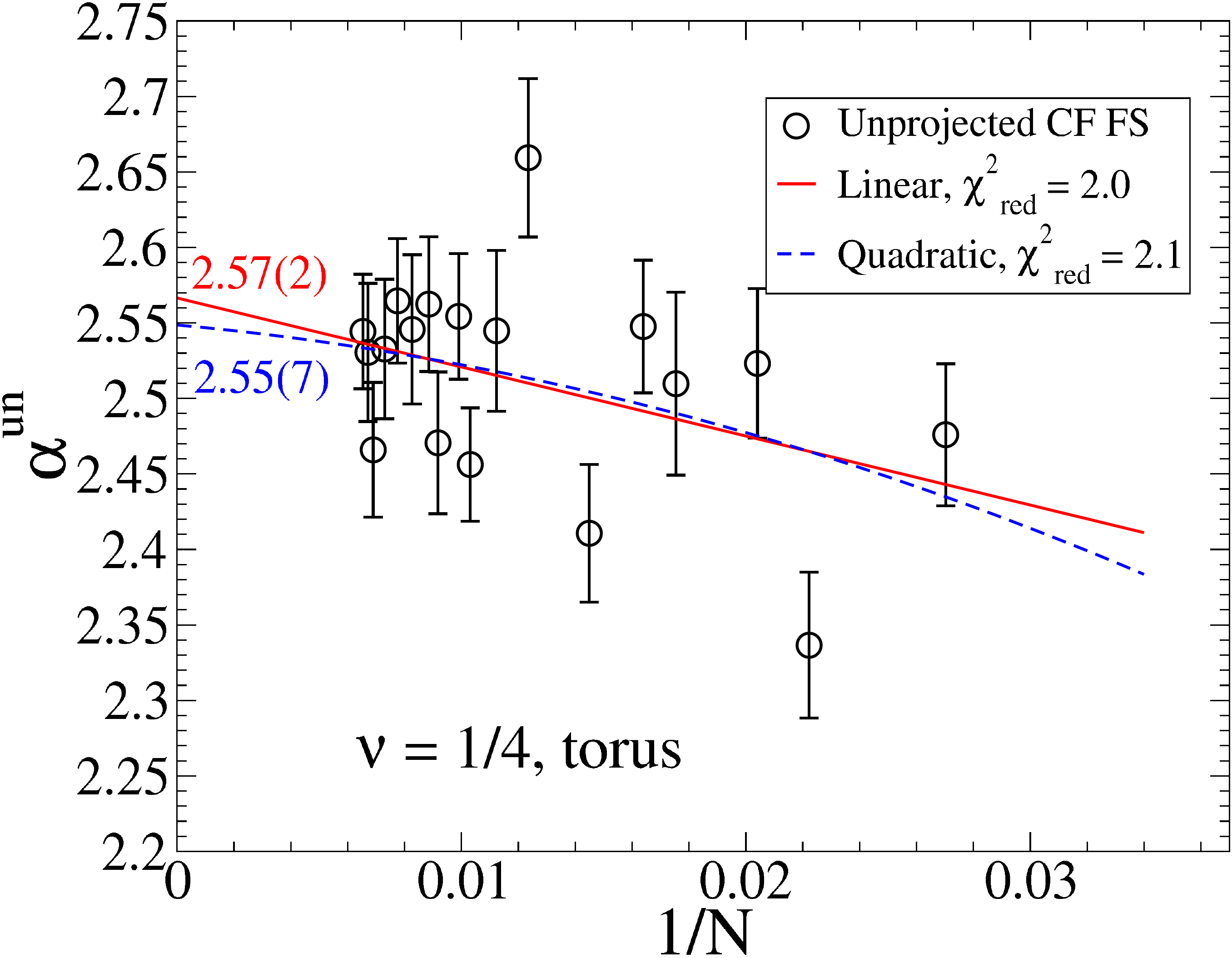}
\end{center}
\caption{
The Fermi wave vector (left panel) and the power law exponent $\alpha^{\rm un}$ (right panel) of the {\em unprojected} composite fermion Fermi sea state at $\nu=1/4$ on the torus geometry as a function of $1/N$. We show both the linear and the quadratic extrapolations to the thermodynamic limit.}
\label{torus_fs4_un}
\end{figure*}

A note regarding the scatter in our data is in order. The main problem of CF Fermi sea on the torus is that one has
to select the wave vectors from the reciprocal space (i.e., wave vectors compatible with the boundary condition) that are
occupied. Shao {\em et al.}\cite{Shao15} propose that the choice that gives the lowest energy Fermi sea for non-interacting particles, which is a roughly circular disk in k-space, is appropriate. However, for relatively smaller system sizes, the CF Fermi sea is not quite circular, which is a source of finite size effects. 

We have performed a similar analysis for the unprojected CF Fermi sea.
The results are shown in Fig.~\ref{torus_fs_un} for $\nu=1/2$.
For a linear fit we find $k_{\rm F}^*l=0.96(1)$.
When we allow both linear and quadratic fits, the range is given by $k_{\rm F}^*l=0.96-1.00$,
which again is consistent with the range quoted in the main text.
At $\nu=1/4$, we get $k_{\rm F}^*l=0.678(2)$ by linear extrapolation, or the range $k_{\rm F}^*l=0.67-0.69$ with both linear and quadratic fits.

\end{document}